\documentclass[aps,prd,preprint,nofootinbib,11pt]{revtex4}
\usepackage{amsfonts}
\usepackage{mathrsfs}
\usepackage{graphicx}
\usepackage{amsmath}
\usepackage{amssymb}
\usepackage{subfigure}
\usepackage{epsfig}
\usepackage{graphicx}
\usepackage{color}
\newcommand{\bqa}{\begin{eqnarray}}
\newcommand{\eqa}{\end{eqnarray}}
\newcommand{\beq}{\begin{equation}}
\newcommand{\eeq}{\end{equation}}
\allowdisplaybreaks[1]
\graphicspath{{fig/}{dia/}} \DeclareGraphicsExtensions{.eps}
\begin{document}

\title{Gluonic Hidden-charm Tetraquark States \\[0.7cm]}

\author{Bing-Dong Wan$^{1,2}$ \footnote{wanbd@lnnu.edu.cn} and Shuo Yang$^{1,2}$ \footnote{shuoyang@lnnu.edu.cn, corresponding author} \vspace{+3pt}}

\affiliation{$^1$Department of Physics, Liaoning Normal University, Dalian 116029, China\\
$^2$ Center for Theoretical and Experimental High Energy Physics, Liaoning Normal University, Dalian 116029, China
}
\author{~\\~\\}

\begin{abstract}
\vspace{0.5cm}
In this paper, a new type of hybrid state, which consists of two valence quarks and two valence antiquarks together with a valence gluon, the gluonic tetraquark states, are investigated. Twenty-four currents of the the gluonic hidden-charm tetraquark states in $[\bar{3}_c]_{c q}\otimes[8_c]_{G}\otimes[3_c]_{\bar{c} \bar{q^\prime}}$ configuration are constructed, and their mass spectrum are evaluated in the framework of QCD sum rules with quantum numbers of $J^P=0^{+}$, $0^{-}$, $1^{-}$, and $1^{+}$. The nonperturbative contributions up to dimension 8 are taken into account. The results indicate that there may be exist 14 gluonic hidden-charm tetraquark states, and their corresponding hidden-bottom partners are also evaluated. The possible production and decay modes of the gluonic tetraquark states are analyzed, which are hopefully measurable in BESIII, BELLEII, PANDA, Super-B, and LHCb experiments.
\end{abstract}
 \maketitle
\newpage
\section{Introduction}
Quantum Chromodynamics (QCD) \cite{Gross:1973id,Politzer:1973fx,Wilson:1974sk} is widely regarded as the fundamental theory of strong interactions. It is generally believed that the properties of hadrons are governed by the non-perturbative aspects of QCD. While perturbative QCD, which is relatively well understood, provides insights into high-energy processes, we still lack reliable and effective methods for addressing the non-perturbative effects of QCD. As a result, gaining a deeper understanding of the physics associated with non-perturbative QCD remains one of the most important challenges in high-energy physics. To tackle this, researchers often rely on phenomenological models to calculate hadronic quantities, such as hadron spectra, hadronic transition matrix elements, parton distributions, and fragmentation functions.

In the framework of QCD and the quark model (QM) \cite{GellMann:1964nj,Zweig}, various hadronic structures beyond traditional mesons ($q\bar{q}$) and baryons ($qqq$) are possible. These structures, referred to as novel hadronic states, include multiquark states, glueballs, and hybrids. Exploring these novel hadronic states could significantly expand our understanding of the hadron family and deepen our insights into QCD. With the advent of the new millennium, advancements in experimental technology in high-energy physics have led to the gradual discovery of novel hadronic states, such as the XYZ states \cite{Choi:2003ue, Aubert:2005rm, Belle:2011aa, Ablikim:2013mio, Liu:2013dau}. Today, more than 40 novel hadronic states or candidates have been reported, reminiscent of the particle zoo phase seen in the previous century. There is growing anticipation that many more novel hadronic states will emerge soon, signaling a resurgence in hadron physics. Understanding the hadronic structure behind these new experimental findings is one of the most exciting and important areas of research in hadron physics.

Building on the successes of the $XYZ$ and $P_c$ states, a search for hybrid states in the charmonium and bottomonium sectors has been proposed~\cite{Olsen:2009gi,Olsen:2009ys,Godfrey:2008nc,Close:2007ny}. A hybrid state typically refers to a state that consists of a pair of constituent quarks and a dynamic gluon. One of the main objectives of many experimental facilities, such as BESIII, GlueX, PANDA, and LHCb, is to detect the existence of hybrids. Although the existence of hybrid states has not yet been experimentally confirmed, several promising candidates have been observed, both recently and in the past. For instance, the BESIII collaboration recently observed a $1^{-+}$ structure at 1.855 GeV, named $\eta_1(1855)$, in the $\eta \eta^\prime$ invariant mass spectrum with a $19\sigma$ significance~\cite{BESIII:2022riz,BESIII:2022iwi}. In previous experiments, three candidates with the exotic quantum numbers $I^G J^{PC} = 1^-1^{-+}$ have been observed: $\pi_1(1400)$\cite{IHEP-Brussels-LosAlamos-AnnecyLAPP:1988iqi}, $\pi_1(1600)$\cite{E852:2001ikk,COMPASS:2009xrl}, and $\pi_1(2015)$~\cite{E852:2004gpn}.

Compared to normal hybrid states, a new type of hybrid state—consisting of two valence quarks, two valence antiquarks, and a valence gluon—has been proposed to explain some exotic properties of $X(6900)$~\cite{Wan:2020fsk}. The narrow structure $X(6900)$ was first reported by LHCb collaboration in 2020. It was observed in the di-$J/\psi$ invariant mass spectrum around $6.9$ GeV with a significance greater than $5\; \sigma$ by using proton-proton collision at the center-of-mass energies of $\sqrt{s}=7$, 8 and 13 TeV~\cite{Aaij:2020fnh}. This structure is for the first time that clear structure in the $J/\psi$-pair mass spectrum were observed in the experiment, thus it's considered as a huge breakthrough in the exploration of hadron spectroscopy. In 2023, $X(6900)$ was confirmed by CMS collaboration, and two other structures named $X(6600)$ and $X(7100)$ are also reported~\cite{CMS:2023owd}. The mass of $X(6900)$ is higher than the double $J/\psi$ threshold by approximately $700$ MeV, which is larger than the typical energy gap between ground and excited states. Within the tetraquark hybrid model, the large energy gap can be naturally attributed to the dynamic gluon, and the $X(7100)$ was also be predicted.

Similar to the tetracharm hybrid state, gluonic charmonium-like states that have a same configuration but with a charm quark and an anti-charm quark replaced by an up quark and an anti-down quark. The $cq$ and $\bar{c}\bar{q}^\prime$ diquark pairs are configured in color $\bar{3}_c$ and $3_c$ respectively within the SU(3) gauge group. With the dynamic gluon, the gluonic charmonium-like states have the configuration of $[\bar{3}_c]_{c q}\otimes[8_c]_{G}\otimes[3_c]_{\bar{c} \bar{q^\prime}}$.

In this paper, we evaluate the gluonic charmonium-like states in the framework of QCD sum rules (QCDSR) \cite{Shifman}. 
Rather than phenomenological models, QCDSR, as a QCD-based theoretical framework, incorporates nonperturbative effects universally order by order, offering unique advantages in exploring hadron properties involving nonperturbative QCD and has already achieved a lot in the study of hadron spectroscopy~\cite{Albuquerque:2013ija,Wang:2013vex,Govaerts:1984hc,Reinders:1984sr,P.Col,Narison:1989aq,Tang:2021zti,Qiao:2014vva,Qiao:2015iea,Tang:2019nwv,Wan:2019ake,Wan:2020oxt,Wan:2021vny,Wan:2022xkx,Zhang:2022obn,Wan:2022uie,Wan:2023epq,Wan:2024dmi,Tang:2024zvf,Li:2024ctd,Zhao:2023imq,Yin:2021cbb,Yang:2020wkh,Wan:2024cpc}. To establish the QCDSR, the first step is to construct appropriate interpolating currents corresponding to the hadron of interest. Using these currents, one can then construct the two-point correlation function, which has two representations: the QCD representation and the phenomenological representation. By equating these two representations, the QCDSR is formally established, allowing for the deduction of the masses of hadrons.

The rest of the paper is arranged as follows. After the Introduction, a brief interpretation of QCD sum rules and some primary formulas in our calculation are presented in Sec.~\ref{formalism}. The numerical analysis are given in Sec.~\ref{numerical}, and the possible hybrids production and decay modes are offered in Sec.~\ref{decay}. The last part is left for a brief summary.

\section{Formalism}\label{formalism}
The procedure of QCDSR begin with the establishing of the two-point correlation function:
\begin{eqnarray}
\Pi(q^2) &=& i \int d^4 x e^{i q \cdot x} \langle 0 | T \{ j (x),\;  j^\dagger (0) \} |0 \rangle\;,\label{two-points-a}\\
\Pi_{\mu\nu}(q^2) &=& i \int d^4 x e^{i q \cdot x} \langle 0 | T \{ j_\mu (x).\;  j_\nu^\dagger (0) \} |0 \rangle \;.\label{two-points-b}
\end{eqnarray}
Here, $j(x)$ and $j_\mu(x)$ are the relevant hadronic interpolating currents with $J = 0$ and $1$, respectively, $\mu$ stands for the the Lorentz index, and for $j_\mu(x)$, the correlation function has the following Lorentz covariance form:
\begin{eqnarray}
\Pi_{\mu\nu}(q^2) &=&-\Big( g_{\mu \nu} - \frac{q_\mu q_\nu}{q^2}\Big) \Pi_1(q^2)+ \frac{q_\mu q_\nu}{q^2}\Pi_0(q^2)\;.
\end{eqnarray}
Here, the subscripts $1$ and $0$ denote the quantum numbers of the spin $1$ and $0$ mesons, respectively.

With the correlation function, the interpolating currents for the gluonic tetraquark states have to be constructed for evaluating their mass spectrum. 
The explanation on constructing of the currents can be found in Ref.~\cite{Wan:2020fsk}.
For the $0^+$ state, the interpolating currents can take the following forms:
\begin{eqnarray}
j^{A}_{0^{+}} (x)&=& g_s \epsilon_{ikl}\epsilon_{jmn} [q_k^T C \gamma_\mu c_l]\frac{\lambda_{ij}^a}{2} G_{\mu\nu}^a [\bar{c}_m \gamma_\nu C \bar{q}_n^{\prime\,T}]\, , \label{current-0+A} \\
j^{B}_{0^{+}}(x) &=& g_s \epsilon_{ikl}\epsilon_{jmn} [q_k^T C \gamma_\mu\gamma_5 c_l]\frac{\lambda_{ij}^a}{2} G_{\mu\nu}^a [\bar{c}_m \gamma_\nu\gamma_5 C \bar{q}_n^{\prime\,T}]\, , \label{current-0+B} \\
j^{C}_{0^{+}} (x)&=& g_s \epsilon_{ikl}\epsilon_{jmn}[q_k^T C \gamma_\mu c_l]\frac{\lambda_{ij}^a}{2} \tilde{G}_{\mu\nu}^a [\bar{c}_m \gamma_\nu \gamma_5 C \bar{q}_n^{\prime\,T}]\, , \label{current-0+C} \\
j^{D}_{0^{+}} (x)&=& g_s \epsilon_{ikl}\epsilon_{jmn}[q_k^T C \gamma_\mu \gamma_5 c_l]\frac{\lambda_{ij}^a}{2} \tilde{G}_{\mu\nu}^a [\bar{c}_m \gamma_\nu  C \bar{q}_n^{\prime\,T}]\, , \label{current-0+D}
\end{eqnarray}
where $q$ and $q^\prime$ stand for up quark and down quark, respectively, $G_{\mu\nu}^a$ and $\tilde{G}_{\mu\nu}^a$ denote gluon field strength and dual field strength, respectively, where $\tilde{G}_{\mu\nu}^a=\frac{1}{2}\epsilon_{\mu\nu\alpha\beta} G^{a,\;\alpha\beta}$, $g_s$ is the QCD coupling constant, $\lambda^a$ are the Gell-Mann matrices, $C$ represents the charge conjugation matrix, and the indices $i$, $j$, $k$, $\cdots$ are color indices. 

For the $0^-$ state, the interpolating currents may be constructed exclusively as:
\begin{eqnarray}
j^{A}_{0^{-}} (x)&=& g_s \epsilon_{ikl}\epsilon_{jmn} [q_k^T C \gamma_\mu c_l]\frac{\lambda_{ij}^a}{2} \tilde{G}_{\mu\nu}^a [\bar{c}_m \gamma_\nu C \bar{q}_n^{\prime\,T}]\, , \label{current-0-A} \\
j^{B}_{0^{-}}(x) &=& g_s \epsilon_{ikl}\epsilon_{jmn} [q_k^T C \gamma_\mu\gamma_5 c_l]\frac{\lambda_{ij}^a}{2} \tilde{G}_{\mu\nu}^a [\bar{c}_m \gamma_\nu\gamma_5 C \bar{q}_n^{\prime\,T}]\, , \label{current-0-B}\\
j^{C}_{0^{-}} (x)&=& g_s \epsilon_{ikl}\epsilon_{jmn}[q_k^T C \gamma_\mu c_l]\frac{\lambda_{ij}^a}{2} G_{\mu\nu}^a [\bar{c}_m \gamma_\nu \gamma_5 C \bar{q}_n^{\prime\,T}]\, , \label{current-0-C} \\
j^{D}_{0^{-}} (x)&=& g_s \epsilon_{ikl}\epsilon_{jmn}[q_k^T C \gamma_\mu\gamma_5 c_l]\frac{\lambda_{ij}^a}{2} G_{\mu\nu}^a [\bar{c}_m \gamma_\nu  C \bar{q}_n^{\prime\,T}]\, . \label{current-0-D}
\end{eqnarray}

The currents for the $1^-$ state are found to be in forms:
\begin{eqnarray}
j^{A}_{1^{-}\;,\alpha} (x)&=& g_s \epsilon_{ikl}\epsilon_{jmn}[q_k^T C \sigma_{\alpha\mu} c_l]\frac{\lambda_{ij}^a}{2} G_{\mu\nu}^a [\bar{c}_m \gamma_\nu C \bar{q}_n^{\prime\,T}]\, , \label{current-1-A} \\
j^{B}_{1^{-}\;,\alpha} (x)&=& g_s \epsilon_{ikl}\epsilon_{jmn}[q_k^T C \sigma_{\alpha\mu}\gamma_5 c_l]\frac{\lambda_{ij}^a}{2} G_{\mu\nu}^a [\bar{c}_m \gamma_\nu\gamma_5 C \bar{q}_n^{\prime\,T}]\, , \label{current-1-B} \\
j^{C}_{1^{-}\;,\alpha} (x)&=& g_s \epsilon_{ikl}\epsilon_{jmn}[q_k^T C \sigma_{\alpha\mu} c_l]\frac{\lambda_{ij}^a}{2} \tilde{G}_{\mu\nu}^a [\bar{c}_m \gamma_\nu\gamma_5 C \bar{q}_n^{\prime\,T}]\, , \label{current-1-C} \\
j^{D}_{1^{-}\;,\alpha} (x)&=& g_s \epsilon_{ikl}\epsilon_{jmn}[q_k^T C \sigma_{\alpha\mu}\gamma_5 c_l]\frac{\lambda_{ij}^a}{2} \tilde{G}_{\mu\nu}^a [\bar{c}_m \gamma_\nu C \bar{q}_n^{\prime\,T}]\, ,\label{current-1-D}\\
j^{E}_{1^{-}\;,\alpha} (x)&=& g_s \epsilon_{ikl}\epsilon_{jmn}[q_k^T C \gamma_\mu c_l]\frac{\lambda_{ij}^a}{2} G_{\mu\nu}^a [\bar{c}_m \sigma_{\alpha\nu} C \bar{q}_n^{\prime\,T}]\, , \label{current-1-E} \\
j^{F}_{1^{-}\;,\alpha} (x)&=& g_s \epsilon_{ikl}\epsilon_{jmn}[q_k^T \gamma_\mu\gamma_5 C  c_l]\frac{\lambda_{ij}^a}{2} G_{\mu\nu}^a [\bar{c}_m \sigma_{\alpha\nu}\gamma_5 C \bar{q}_n^{\prime\,T}]\, , \label{current-1-F} \\
j^{G}_{1^{-}\;,\alpha} (x)&=& g_s \epsilon_{ikl}\epsilon_{jmn}[q_k^T C \gamma_\mu\gamma_5 c_l]\frac{\lambda_{ij}^a}{2} \tilde{G}_{\mu\nu}^a [\bar{c}_m \sigma_{\alpha\nu} C \bar{q}_n^{\prime\,T}]\, , \label{current-1-G} \\
j^{H}_{1^{-}\;,\alpha} (x)&=& g_s \epsilon_{ikl}\epsilon_{jmn}[q_k^T C \gamma_\mu c_l]\frac{\lambda_{ij}^a}{2} \tilde{G}_{\mu\nu}^a [\bar{c}_m \sigma_{\alpha\nu}\gamma_5 C \bar{q}_n^{\prime\,T}]\, . \label{current-1-H} 
\end{eqnarray}

The currents of the $1^+$ gluonic tetraquark state are constructed as:
\begin{eqnarray}
j^{A}_{1^{+}\;,\alpha} (x)&=& g_s \epsilon_{ikl}\epsilon_{jmn}[q_k^T C \sigma_{\alpha\mu} c_l]\frac{\lambda_{ij}^a}{2} \tilde{G}_{\mu\nu}^a [\bar{c}_m \gamma_\nu C \bar{q}_n^{\prime\,T}]\, , \label{current-1+A} \\
j^{B}_{1^{+}\;,\alpha} (x)&=& g_s \epsilon_{ikl}\epsilon_{jmn}[q_k^T C \sigma_{\alpha\mu}\gamma_5 c_l]\frac{\lambda_{ij}^a}{2} \tilde{G}_{\mu\nu}^a [\bar{c}_m \gamma_\nu\gamma_5 C \bar{q}_n^{\prime\,T}]\, , \label{current-1+B} \\
j^{C}_{1^{+}\;,\alpha} (x)&=& g_s \epsilon_{ikl}\epsilon_{jmn}[q_k^T C \sigma_{\alpha\mu} c_l]\frac{\lambda_{ij}^a}{2} G_{\mu\nu}^a [\bar{c}_m \gamma_\nu\gamma_5 C \bar{q}_n^{\prime\,T}]\, , \label{current-1+C} \\
j^{D}_{1^{+}\;,\alpha} (x)&=& g_s \epsilon_{ikl}\epsilon_{jmn}[q_k^T C \sigma_{\alpha\mu}\gamma_5 c_l]\frac{\lambda_{ij}^a}{2} G_{\mu\nu}^a [\bar{c}_m \gamma_\nu C \bar{q}_n^{\prime\,T}]\, , \label{current-1+D} \\
j^{E}_{1^{+}\;,\alpha} (x)&=& g_s \epsilon_{ikl}\epsilon_{jmn}[q_k^T C \gamma_\mu c_l]\frac{\lambda_{ij}^a}{2} \tilde{G}_{\mu\nu}^a [\bar{c}_m \sigma_{\alpha\nu} C \bar{q}_n^{\prime\,T}]\, , \label{current-1+E} \\
j^{F}_{1^{+}\;,\alpha} (x)&=& g_s \epsilon_{ikl}\epsilon_{jmn}[q_k^T C \gamma_\mu\gamma_5 c_l]\frac{\lambda_{ij}^a}{2} \tilde{G}_{\mu\nu}^a [\bar{c}_m \sigma_{\alpha\nu}\gamma_5 C \bar{q}_n^{\prime\,T}]\, , \label{current-1+F} \\
j^{G}_{1^{+}\;,\alpha} (x)&=& g_s \epsilon_{ikl}\epsilon_{jmn}[q_k^T C \gamma_\mu\gamma_5 c_l]\frac{\lambda_{ij}^a}{2} G_{\mu\nu}^a [\bar{c}_m \sigma_{\alpha\nu} C \bar{q}_n^{\prime\,T}]\, , \label{current-1+G} \\
j^{H}_{1^{+}\;,\alpha} (x)&=& g_s \epsilon_{ikl}\epsilon_{jmn}[q_k^T C \gamma_\mu c_l]\frac{\lambda_{ij}^a}{2} G_{\mu\nu}^a [\bar{c}_m \sigma_{\alpha\mu}\gamma_5 C \bar{q}_n^{\prime\,T}]\, . \label{current-1+H} 
\end{eqnarray}

\begin{figure}
\includegraphics[width=12cm]{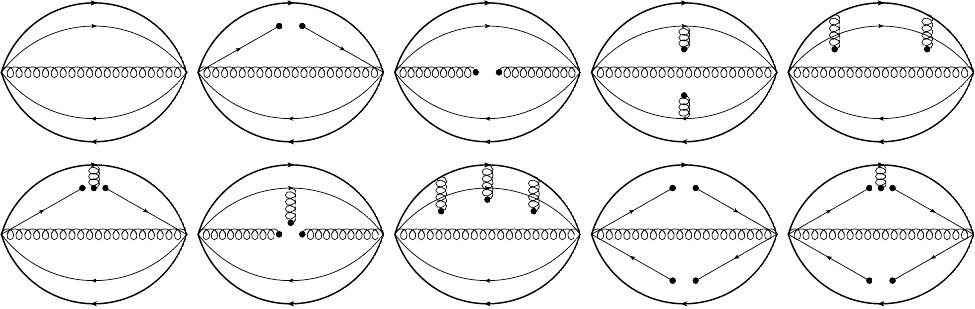}
\caption{The typical Feynman diagrams related to the correlation function, where the thick solid line represents the heavy quark, the thin solid line stands for the light quark, and the spiral line denotes the gluon.} \label{figfeyn}
\end{figure}

With the currents (\ref{current-0+A})$-$(\ref{current-1+H}), the two-point correlation function (\ref{two-points-a}) and (\ref{two-points-b}) can be calculated on both operator product expansion (OPE) side and phenomenological side. On OPE side, one can express the correlation in terms of a dispersion relation:
\begin{eqnarray}
\Pi^{OPE}(q^2) &=& \int_{s_{min}}^{\infty} d s
\frac{\rho^{OPE} (s)}{s - q^2} \; .
\label{OPE-hadron}
\end{eqnarray}
Here, $\rho^{OPE}(s) = \text{Im} [\Pi^{OPE}(s)] / \pi$ is the spectral density on the OPE side and $s_{min}$ is a kinematic limit, which usually corresponds to the square of the sum of the current quark masses of the hadron \cite{Albuquerque:2013ija}, i.e., $s_{min}=(2 m_c+m_q+m_q^\prime )^2$. Up to dimension $8$ of the condensates, $\rho^{OPE}(s)$ can be express as:
\begin{eqnarray}
\rho^{OPE}(s)  =  \rho^{pert}(s) + \rho^{\langle \bar{q} q
\rangle}(s) +\rho^{\langle G^2 \rangle}(s) + \rho^{\langle \bar{q} G q \rangle}(s)
+ \rho^{\langle \bar{q} q \rangle^2}(s) + \rho^{\langle G^3 \rangle}(s) 
+ \rho^{\langle \bar{q} q \rangle\langle \bar{q} G q \rangle}(s)  . \label{rho-OPE}
\end{eqnarray}
To calculate the spectral density of the OPE side, Eq. (\ref{rho-OPE}), 
the full propagators of the light quark $S^q_{i j}(x)$ and the heavy quark $S^Q_{ij}(p)$ are employed:
\begin{eqnarray}
S^q_{j k}(x) \! \! & = & \! \! \frac{i \delta_{j k} x\!\!\!\slash}{2 \pi^2
x^4} - \frac{\delta_{jk} m_q}{4 \pi^2 x^2} - \frac{i t^a_{j k} G^a_{\alpha\beta}}{32 \pi^2 x^2}(\sigma^{\alpha \beta} x\!\!\!\slash
+ x\!\!\!\slash \sigma^{\alpha \beta}) - \frac{\delta_{jk}}{12} \langle\bar{q} q \rangle + \frac{i\delta_{j k}
x\!\!\!\slash}{48} m_q \langle \bar{q}q \rangle - \frac{\delta_{j k} x^2}{192} \langle g_s \bar{q} \sigma \cdot G q \rangle \nonumber \\ &+& \frac{i \delta_{jk} x^2 x\!\!\!\slash}{1152} m_q \langle g_s \bar{q} \sigma \cdot G q \rangle - \frac{t^a_{j k} \sigma_{\alpha \beta}}{192}
\langle g_s \bar{q} \sigma \cdot G q \rangle
+ \frac{i t^a_{jk}}{768} (\sigma_{\alpha \beta} x \!\!\!\slash + x\!\!\!\slash \sigma_{\alpha \beta}) m_q \langle
g_s \bar{q} \sigma \cdot G q \rangle \;,
\end{eqnarray}
\begin{eqnarray}
S^Q_{j k}(p) \! \! & = & \! \! \frac{i \delta_{j k}(p\!\!\!\slash + m_Q)}{p^2 - m_Q^2} - \frac{i}{4} \frac{t^a_{j k} G^a_{\alpha\beta} }{(p^2 - m_Q^2)^2} [\sigma^{\alpha \beta}
(p\!\!\!\slash + m_Q)
+ (p\!\!\!\slash + m_Q) \sigma^{\alpha \beta}] \nonumber \\ &+& \frac{i\delta_{jk}m_Q  \langle g_s^2 G^2\rangle}{12(p^2 - m_Q^2)^3}\bigg[ 1 + \frac{m_Q (p\!\!\!\slash + m_Q)}{p^2 - m_Q^2} \bigg] \nonumber \\ &+& \frac{i \delta_{j k}}{48} \bigg\{ \frac{(p\!\!\!\slash +
m_Q) [p\!\!\!\slash (p^2 - 3 m_Q^2) + 2 m_Q (2 p^2 - m_Q^2)](p\!\!\!\slash + m_Q) }{(p^2 - m_Q^2)^6}
\bigg\} \langle g_s^3 G^3 \rangle \; .
\end{eqnarray}
The vacuum condensates are clearly displayed in $S^q_{i j}(x)$ and $S^Q_{ij}(p)$. For more explanation on above propagator, readers may refer to Refs.~\cite{Wang:2013vex, Albuquerque:2013ija}. 
Furthermore, the perturbative gluon propagator employed in our analytical calculation is considered in coordinate space, which can be expressed as~\cite{Govaerts:1984hc}:
\begin{eqnarray}
 S_{\mu\nu,\rho\sigma}^{ab}(x)&=& \frac{\delta^{ab}}{2\pi^{2}} \times\frac{1}{x^{6}}\big\{(g_{\mu\rho}x^{2} - 4x_{\mu}x_{\rho})g_{\nu\sigma} - (g_{\mu\sigma}x^{2} - 4x_{\mu}x_{\sigma})g_{\rho\nu}\nonumber\\
 &-& (g_{\rho\nu}x^{2} - 4x_{\rho}x_{\nu})g_{\mu\sigma} + (g_{\nu\sigma}x^{2} - 4x_{\nu}x_{\sigma})g_{\rho\mu}\big\}.\label{pert-gluon}
\end{eqnarray}
The Feynman diagrams corresponding to each term of Eq. (\ref{rho-OPE}) are schematically shown in Fig. \ref{figfeyn}.

On the phenomenological side, the correlation can be expressed to the ground state with higher excited states and continuum states. We separate the ground state from the pole terms, the correlation function $\Pi(q^2)$ is obtained as a dispersion over a physical regime, i.e.,
 \begin{eqnarray}
\Pi^{phen}(q^2) & = & \frac{\lambda^2}{m^2 - q^2} + \frac{1}{\pi} \int_{s_0}^\infty d s \frac{\rho(s)}{s - q^2} \; , \label{hadron}
\end{eqnarray}
where $m$, $\rho(s)$, $\lambda$, and $s_0$ denote the mass of the hadronic state, spectral density that contains the contributions from the higher excited states, coupling constant, and the threshold of higher excited states and continuum states, respectively.

In practice, it is necessary to take control the contributions from higher order condensates in the OPE and the contributions from higher excited states and the continuum on the phenomenological side. An effective and common approach is to perform the Borel transformation on both sides of the QCDSR simultaneously. That is
\begin{eqnarray}
\hat{B}[f(Q^2)]:=\lim_{Q^2\rightarrow \infty,n\rightarrow \infty
\atop
{Q^2 / n}=
{M_B^2}}\frac{(-Q^2)^n}{(n-1)!}\left(\frac{d}{dQ^2}\right)^n\;f(Q^2)\ .
\end{eqnarray}

By performing the Borel transform on both OPE side and phenomenological side, i.e., Eqs. (\ref{OPE-hadron}) and (\ref{hadron}), and the using the quark-hadron duality principle, we can get the main function of QCDSR:
\begin{eqnarray}
\lambda^2 e^{-m^2/M_B^2} + \int_{s_0}^\infty d s \rho^{OPE}(s) e^{-s/M_B^2}= \int_{s_{min}}^\infty d s \rho^{OPE}(s) e^{-s/M_B^2}\;,
\end{eqnarray}
then the mass of the tetraquark hybrid state can be readily obtained:
\begin{eqnarray}
m(s_0, M_B^2) &=& \sqrt{-\frac{L_1(s_0, M_B^2)}{L_0(s_0, M_B^2)}} \; . \label{mass-Eq}
\end{eqnarray}
Here the moments $L_1$ and $L_0$ are defined as follows:
\begin{eqnarray}
L_0(s_0, M_B^2) & = & \int_{s_{min}}^{s_0} d s \; \rho(s) e^{-s / M_B^2} \; , \label{L0}\\
L_1(s_0, M_B^2) & = & \frac{\partial}{\partial{\frac{1}{M_B^2}}}{L_0(s_0, M_B^2)} \; .
\end{eqnarray}

\section{Numerical analysis}\label{numerical}
To evaluate the tetraquark hybrid mass numerically, one needs to give certain inputs to yield meaningful physical results. 
In this work, the broadly accepted inputs are taken\cite{Shifman, Albuquerque:2013ija, Reinders:1984sr, P.Col, Narison:1989aq}:
$m_c (m_c) = \overline{m}_c= (1.275 \pm 0.025)\; \text{GeV}$,
$m_b (m_b) = \overline{m}_b= (4.18 \pm 0.03)\; \text{GeV}$,
$m_u=2.16^{+0.49}_{-0.26}\;\text{MeV}$,
$m_d=4.67^{+0.48}_{-0.17}\;\text{MeV}$,
$\langle \bar{q} q \rangle = - (0.23 \pm 0.03)^3 \; \text{GeV}^3$,
$\langle g_s^2 G^2 \rangle = 0.88 \; \text{GeV}^4$,
$\langle g_s^3 G^3\rangle = 0.045 \; \text{GeV}^6$,
$\langle \bar{q} g_s \sigma \cdot G q \rangle = m_0^2 \langle\bar{q} q \rangle$, and $m_0^2 = (0.8 \pm 0.2) \; \text{GeV}^2$, in which the $\overline{\text{MS}}$ running heavy quark masses are adopted. Furthermore, the leading order strong coupling constant
\begin{eqnarray}
\alpha_s(M_B^2)=\frac{4\pi}{\big(11-\frac{2}{3}n_f \big) \text{ln} \big(\frac{M_B^2}{\Lambda_{\text{QCD}}^2}\big)}
\end{eqnarray}
with $\Lambda_{\text{QCD}} = 300$ MeV is taken, and $n_f$ represents the number of active quarks.

Moreover, the masses of the tetraquark hybrids depend on the continuum threshold $s_0$ and the Borel parameter $M_B^2$. Introducing these two parameters in establishing the sum rules requires meeting two criteria \cite{Shifman, Reinders:1984sr, P.Col,Albuquerque:2013ija}. First, to ensure a reasonable description of the ground-state hadron with the truncated OPE, avoiding significant errors from neglecting higher-dimensional terms, the OPE's convergence must be satisfied. Practically, we compare the relative contribution of higher-dimension condensates to the total OPE contribution and then select a reliable range for $M^2_B$ to maintain convergence. The OPE's convergence can be expressed as:
\begin{eqnarray}
  R^{OPE} = \left| \frac{L^{dim=8}_{0}(s_0, M_B^2)}{L_{0}(s_0, M_B^2)}\right|\, ,\label{RatioOPE}
\end{eqnarray}
Second, the pole contribution (PC) should be substantial enough to ensure that the primary contribution in the mass equation comes directly from the ground state. In practice, the PC should be larger than $40\%$~\cite{Tang:2021zti}, which can be formulated as:
\begin{eqnarray}
  R^{PC} = \frac{L_{0}(s_0, M_B^2)}{L_{0}(\infty, M_B^2)} \; . \label{RatioPC}
\end{eqnarray}

Furthermore, among the various $s_0$ values satisfying the two criteria, it is necessary to determine the proper one that provides an optimal window for the Borel parameter $M_B^2$. It should be noted that the Borel parameter $M_B^2$ is not a physical quantity, so within the optimal Borel window given by the chosen $s_0$, the physical quantity, namely the mass of the concerned hadron in this work, should be as independent of $M_B^2$ as possible. In practice, we may vary $s_0$ by 0.2 GeV in numerical calculations \cite{Qiao:2014vva,Qiao:2015iea}, which sets the upper and lower bounds and, consequently, the uncertainties of $s_0$.

\begin{figure}
\includegraphics[width=6.8cm]{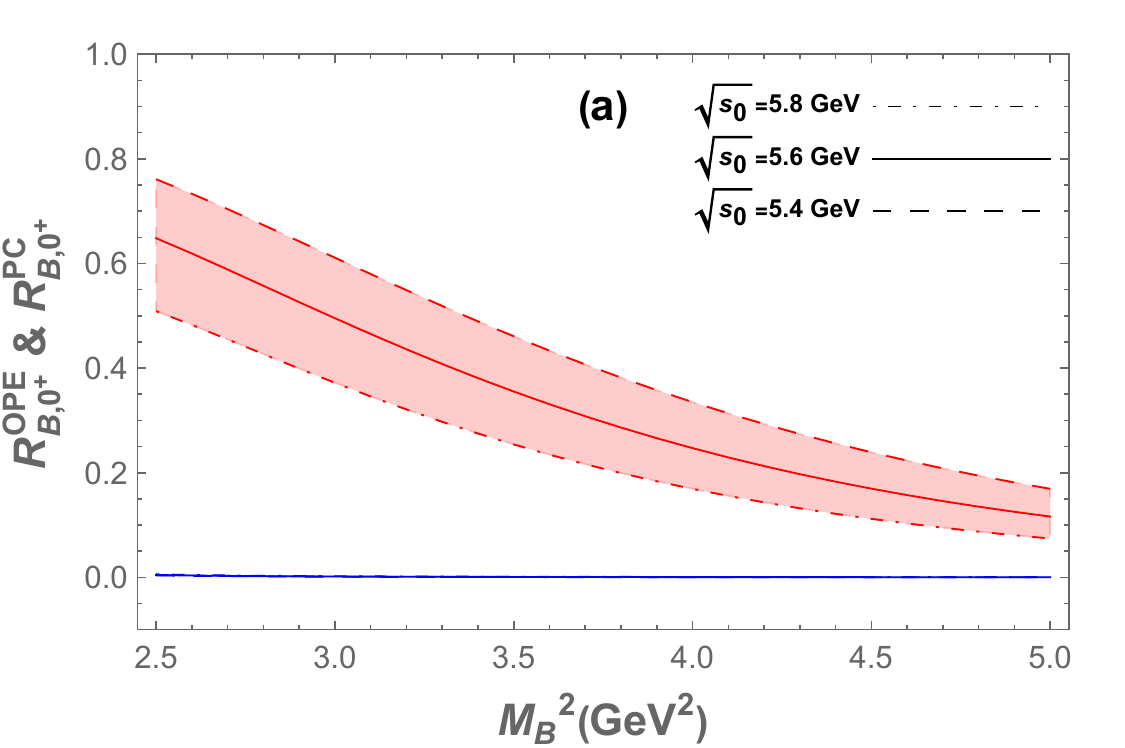}
\includegraphics[width=6.8cm]{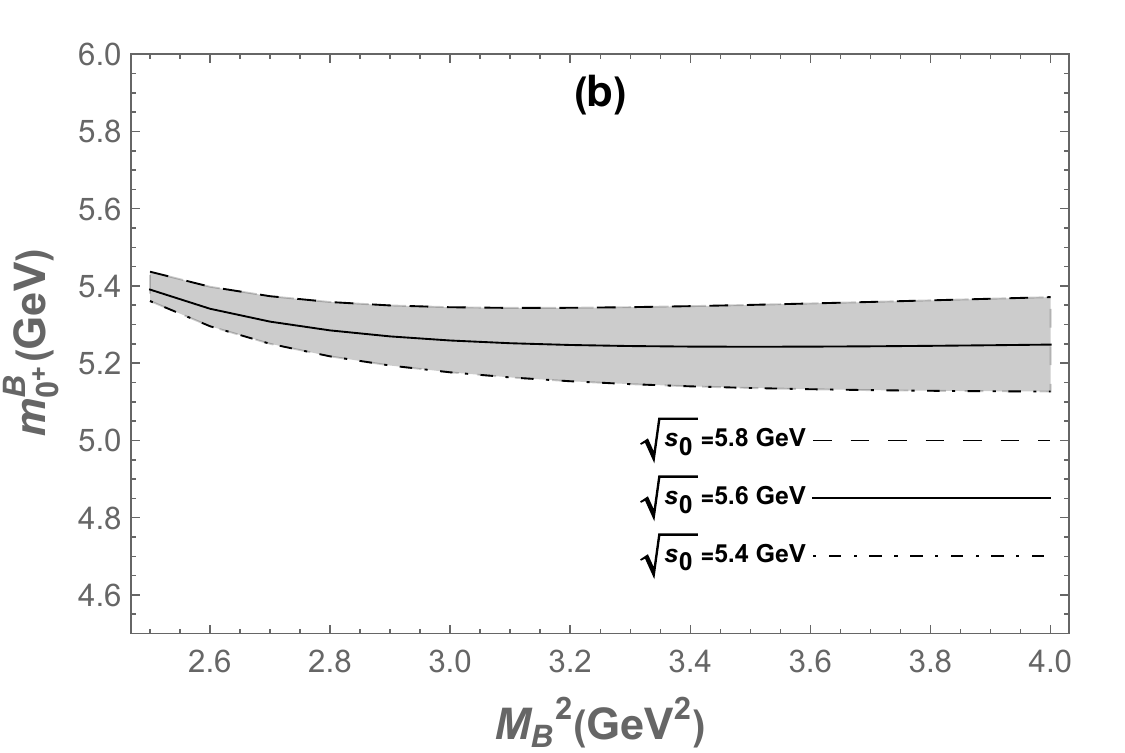}
\caption{ (a) The ratios ${R_{0^{+}}^{B,\;OPE}}$ and ${R_{0^{+}}^{B,\;PC}}$ as functions of the Borel parameter $M_B^2$ for different values of $\sqrt{s_0}$, where blue lines represent ${R_{0^{+}}^{B,\;OPE}}$ and red lines denote ${R_{0^{+}}^{B,\;PC}}$. (b) The mass $M_{0^{+}}^{B}$ as a function of the Borel parameter $M_B^2$ for different values of $\sqrt{s_0}$.} \label{fig0+B}
\end{figure}

With the above preparation we can numerically evaluate the mass spectrum of the tetraquark hybrid states. As an example, the ratios ${R_{0^{+}}^{B,\;OPE}}$ and ${R_{0^{+}}^{B,\;PC}}$ for the current in Eq.~(\ref{current-0+B}) are shown as functions of Borel parameter $M_B^2$ in Fig. \ref{fig0+B}(a) with different values of $\sqrt{s_0}$, i.e., $5.4$, $5.6$, and $5.8$ GeV. The dependence relationships between $m_{0^{+}}^{B}$ and parameter $M_B^2$ are given in Fig. \ref{fig0+B}(b). The optimal window of the Borel parameter is between $2.6 \le M_B^2 \le 3.4\; \rm{GeV}^2$, where a smooth section, the so called stable plateau, in the $m_{0^{+}}^{B}$-$M_B^2$ curve exists, which suggest the mass of the possible $0^{+}$ tetraquark hybrid state. The mass $m_{0^{+}}^{B}$ can be extracted as follows:
\begin{eqnarray}
&&m_{0^{+}}^{B} = (5.24 \pm 0.10) \, \text{GeV}  \; .\label{eq-mass-0+B}
\end{eqnarray}

With the same analyses, and the OPE, pole contribution and the masses as functions of Borel parameter $M^2_B$ can be found in the Appendix \ref{App_B}, the corresponding masses couple to Eqs. (\ref{current-0+B})–(\ref{current-1+H}) are readily obtained, and they are tabulated in Table~\ref{mass_c}. Note, for the currents (\ref{current-0+A}), (\ref{current-0-A}), (\ref{current-1-A}), (\ref{current-1-D}), (\ref{current-1-E}), (\ref{current-1-H}), (\ref{current-1+A}), (\ref{current-1+B}), (\ref{current-1+D}), (\ref{current-1+E}), (\ref{current-1+F}), (\ref{current-1+H}), we cannot find reasonable parameters $M_B^2$ that satisfy the masses independent of $M_B^2$, which implies those currents these currents do not have a strong coupling with the hidden-charm tetraquark hybrid states. It also should be noted that, the masses of the light quarks $m_u$ and $m_d$ are the same in the chiral limit, and considering the symmetries in the structures of the currents, we have $m_{0^+}^C=m_{0^+}^D$, $m_{0^-}^C=m_{0^-}^D$, $m_{1^-}^B=m_{1^-}^F$, $m_{1^-}^C=m_{1^-}^G$, and $m_{1^+}^C=m_{1^+}^G$, which meets our calculations.

\begin{table}
\begin{center}
\renewcommand\tabcolsep{10pt}
\caption{The continuum thresholds, Borel parameters, and predicted masses of the hidden-charm tetraquark hybrid states.}\label{mass_c}
\begin{tabular}{ccccc}\hline\hline
$J^P$             &Current   & $\sqrt{s_0}\;(\text{GeV})$     &$M_B^2\;(\text{GeV}^2)$ &$M^X\;(\text{GeV})$       \\ \hline
$0^+$            &$A$         & $-$                                        & $-$                                   &$-$                            \\
                      &$B$         & $5.6\pm0.2$                         &$2.6-3.4$                          &$5.24\pm0.10$          \\
                      &$C$         & $5.7\pm0.2$                         &$2.6-3.4$                          &$5.34\pm0.12$         \\
                      &$D$         & $5.7\pm0.2$                         &$2.6-3.4$                          &$5.34\pm0.12$          \\\hline
$0^-$             &$A$         & $-$                                        &$-$                                    &$-$                             \\
                      &$B$         & $5.7\pm0.2$                         &$2.6-3.5$                          &$5.32\pm0.12$.         \\
                      &$C$         & $5.9\pm0.2$                         &$2.6-3.7$                          &$5.45\pm0.13$           \\
                      &$D$         & $5.9\pm0.2$                         &$2.6-3.7$                          &$5.45\pm0.13$           \\\hline
$1^-$             &$A$         & $-$                                        &$-$                                    &$-$                              \\
                      &$B$         & $5.7\pm0.2$                         &$2.6-3.5$                          &$5.48\pm0.12$          \\
                      &$C$         & $5.3\pm0.2$                         &$2.1-3.1$                          &$4.87\pm0.12$         \\
                      &$D$         &$-$                                         &$-$                                    &$-$                            \\
                      &$E$         & $-$                                        &$-$                                    &$-$                              \\
                      &$F$         & $5.7\pm0.2$                         &$2.6-3.5$                          &$5.48\pm0.12$          \\
                      &$G$         & $5.3\pm0.2$                         &$2.1-3.1$                          &$4.87\pm0.12$         \\
                      &$H$         &$-$                                         &$-$                                    &$-$                            \\\hline
$1^+$            &$A$         & $-$                                        &$-$                                    &$-$                              \\
                      &$B$         & $-$                                        &$-$                                    &$-$                              \\
                      &$C$         & $6.2\pm0.2$                         &$3.2-4.4$                          &$5.70\pm0.13$         \\
                      &$D$         &$-$                                         &$-$                                    &$-$                            \\
                      &$E$         & $-$                                        &$-$                                    &$-$                              \\
                      &$F$         & $-$                                        &$-$                                    &$-$                              \\
                      &$G$         & $6.2\pm0.2$                         &$3.2-4.4$                          &$5.70\pm0.13$         \\
                      &$H$         &$-$                                         &$-$                                    &$-$                            \\
 \hline
 \hline
\end{tabular}
\end{center}
\end{table}

For the bottom sector, performing the same procedure but with $m_c$ replaced by $m_b$, readers can find the OPE, pole contribution and the masses as functions of Borel parameter $M^2_B$ in the Appendix \ref{App_B}, we can easily obtain the masses of the hidden-bottom tetraquark hybrid states. The relevant numerical results are tabulated in Table~\ref{mass_b}

\begin{table}
\begin{center}
\renewcommand\tabcolsep{10pt}
\caption{The continuum thresholds, Borel parameters, and predicted masses of the hidden-bottom tetraquark hybrid states.}\label{mass_b}
\begin{tabular}{ccccc}\hline\hline
$J^P$             &Current   & $\sqrt{s_0}\;(\text{GeV})$     &$M_B^2\;(\text{GeV}^2)$ &$M^X\;(\text{GeV})$       \\ \hline
$0^+$            &$A$         & $-$                                        & $-$                                   &$-$                            \\
                      &$B$         & $11.6\pm0.2$                        &$6.3-7.6$                          &$11.15\pm0.11$          \\
                      &$C$         & $11.7\pm0.2$                       &$6.2-7.4$                          &$11.24\pm0.11$         \\
                      &$D$         & $11.7\pm0.2$                       &$6.2-7.4$                          &$11.24\pm0.11$          \\\hline
$0^-$             &$A$         & $-$                                        &$-$                                    &$-$                             \\
                      &$B$         & $11.7\pm0.2$                        &$6.5-7.9$                          &$11.22\pm0.11$.         \\
                      &$C$         & $11.9\pm0.2$                       &$6.5-8.0$                          &$11.39\pm0.08$           \\
                      &$D$         & $11.9\pm0.2$                       &$6.5-8.0$                          &$11.39\pm0.08$           \\\hline
$1^-$             &$A$         & $-$                                        &$-$                                    &$-$                              \\
                      &$B$         & $11.7\pm0.2$                       &$6.5-7.8$                          &$11.32\pm0.11$          \\
                      &$C$         & $11.2\pm0.2$                       &$5.2-6.8$                          &$10.74\pm0.11$         \\
                      &$D$.        &$-$                                         &$-$                                    &$-$                            \\
                      &$E$         & $-$                                        &$-$                                    &$-$                              \\
                      &$F$         & $11.7\pm0.2$                       &$6.5-7.8$                          &$11.32\pm0.11$          \\
                      &$G$         & $11.2\pm0.2$                       &$5.2-6.8$                          &$10.74\pm0.11$         \\
                      &$H$         &$-$                                         &$-$                                    &$-$                            \\\hline
$1^+$            &$A$         & $-$                                        &$-$                                    &$-$                              \\
                      &$B$         & $-$                                        &$-$                                    &$-$                              \\
                      &$C$         & $12.2\pm0.2$                       &$7.2-9.0$                          &$11.62\pm0.12$         \\
                      &$D$.        &$-$                                         &$-$                                    &$-$                            \\
                      &$E$         & $-$                                        &$-$                                    &$-$                              \\
                      &$F$         & $-$                                        &$-$                                    &$-$                              \\
                      &$G$         & $12.2\pm0.2$                       &$7.2-9.0$                          &$11.62\pm0.12$         \\
                      &$H$         &$-$                                         &$-$                                    &$-$                            \\
 \hline
 \hline
\end{tabular}
\end{center}
\end{table}

The errors in results Tab.~\ref{mass_c} and Tab.~\ref{mass_b} mainly stem from the uncertainties in quark masses, condensates and threshold parameter $\sqrt{s_0}$. It should be noted that in each category of interpolating currents, the resulting masses for different currents are nearly identical. This can be explained by the fact that they correspond to the same particle with different fine structures. From Eqs. (\ref{current-0+A})–(\ref{current-1+H}), we can see that the spin structures of the different currents with each category are distinct.

\section{Production and decay analyses}\label{decay}
Experimentally, the hybrid states should be produced in gluon-rich processes, like hadrons collision and $\Upsilon$ decay. Thus, we can find the hidden-charm hybrid tetraquark states in these processes. Hereafter, we refer to these hidden-charm hybrid tetraquark states as $X_{J^{P}}$ in discussion. The typical production modes of these hybrid are exhibited in Tab. \ref{production}
\begin{table}
\begin{center}
\renewcommand\tabcolsep{10pt}
\caption{The typical production modes of hidden-charm tetraquark hybrid states.}\label{production}
\begin{tabular}{ccccc}\hline\hline
$J^P$             &Production modes     \\ \hline
$0^+$            &$\Upsilon\to \rho(770)X_{0^+}+c.c. $&$\Upsilon\to \rho(1450)X_{0^+}+c.c. $  \\     
                      &$\chi_{b_1}\to\gamma \rho(770)X_{0^+}+c.c. $  &$\chi_{b_1}\to\gamma \rho(1450)X_{0^+}+c.c. $    \\ \hline
$0^-$            &$\Upsilon(1S)\to a_1(1260)X_{0^-}-c.c. $,   &$\Upsilon(1S)\to b_1(1235)X_{0^-}+c.c. $ \\     
                      &$\chi_{b_1}\to \gamma a_1(1260)X_{0^-}-c.c. $  &$\chi_{b_1}\to \gamma b_1(1235)X_{0^-}+c.c. $ \\ \hline
$1^-$            &$\Upsilon(1S)\to a_0(980)X_{1^-}-c.c. $ &$\Upsilon(1S)\to a_0(1450)X_{1^-}-c.c. $   \\     
                      &$\chi_{b_1}\to \gamma a_0(980)X_{1^-}-c.c. $ & $\chi_{b_1}\to \gamma a_0(1450)X_{1^-}-c.c. $   \\ \hline
$1^+$            &$\Upsilon(1S)\to \pi X_{0^-}-c.c. $ &  $\Upsilon(1S)\to \pi(1300)X_{0^-}-c.c. $  \\     
                      &$\chi_{b_1}\to \gamma \pi X_{0^-}-c.c. $ & $\chi_{b_1}\to \gamma \pi(1300) X_{0^-}-c.c. $   \\
 \hline
 \hline
\end{tabular}
\end{center}
\end{table}

\begin{figure}
\includegraphics[width=12cm]{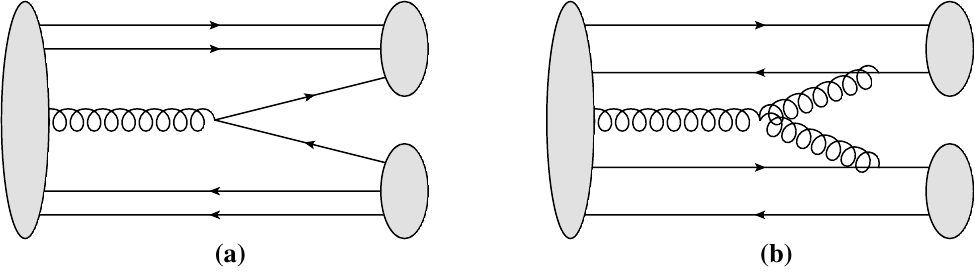}
\caption{Schematic diagram for two possible decay mechanism of hidden-charm hybrid tetraquark states.} \label{decay_mechanism}
\end{figure}

To finally ascertain these hidden-charm hybrid tetraquark states, the straightforward procedure is to reconstruct them from its decay products, though the detailed characters of them still ask for more exploration. There are two possible decay mechanism for hidden-charm hybrid tetraquark states:
\begin{enumerate}
\item  The dynamical gluon of the hybrid can split into a quark pair, then there will be three quarks and three anti-quarks in the final states, see Fig.~\ref{decay_mechanism}(a);
\item The dynamical gluon of the hybrid can split into a gluon pair, and two gluon will be absorbed by a quark and an anti-quark in the hybrid, then there will be two quarks and two anti-quarks in the final states, see Fig.~\ref{decay_mechanism}(b).
\end{enumerate}
Based on the two mechanism, we show the typical decay modes of the these hybrid in Tab. \ref{decay}. Since the there is only one QCD vertex in the first decay mechanism and three QCD vertices in the second decay mechanism, the decay modes with the second mechanism will be suppressed, and the decay modes with the first mechanism will be the primary decay modes.

\begin{table}
\begin{center}
\renewcommand\tabcolsep{10pt}
\caption{The typical decay modes of hidden-charm tetraquark hybrid states.}\label{decay}
\begin{tabular}{ccccc}\hline\hline
$J^P$             &Decay modes with the 1st mechanism       & Decay modes with the 2nd mechanism\\ \hline
$0^+$            &$X_{0^+}\to\Sigma_c^+\Sigma_c^0 $          &$X_{0^+}\to D^+D^0$  \\     
                      &$X_{0^+}\to\Sigma_c^{-}\Sigma_c^{++} $    &$X_{0^+}\to \eta_c\pi^+$    \\ 
                      &$X_{0^+}\to\Xi_c^+\Xi_c^0 $                       &$X_{0^+}\to J/\psi \rho(770)$\\\hline
$0^-$           &$X_{0^-}\to\Sigma_c^+\Sigma_c^0 $          &$X_{0^-}\to D^+D_{0}^{\ast\;0}$  \\     
                      &$X_{0^-}\to\Sigma_c^{-}\Sigma_c^{++} $    &$X_{0^-}\to \eta_c a_0(980)$    \\ 
                      &$X_{0^-}\to\Xi_c^+\Xi_c^0 $                       &$X_{0^-}\to J/\psi a_1(1260)$\\\hline
$1^-$            &$X_{1^-}\to\Sigma_c^+\Sigma_c^0 $          &$X_{1^-}\to D^+D^{\ast\;0}$\;,\;$X_{1^-}\to D^{\ast\;+}D^{0}$  \\     
                      &$X_{1^-}\to\Sigma_c^{-}\Sigma_c^{++} $    &$X_{0^-}\to \eta_c a_1(1260)$    \\ 
                      &$X_{1^-}\to\Xi_c^+\Xi_c^0 $                       &$X_{0^-}\to J/\psi a_0(980)$\\\hline
$1^+$             &$X_{1^+}\to\Sigma_c^+\Sigma_c^0 $          &$X_{1^+}\to D^{\ast\;+}D^0$\;,\;$X_{1^+}\to D^{+}D^{\ast\;0}$  \\     
                      &$X_{1^+}\to\Sigma_c^{-}\Sigma_c^{++} $    &$X_{1^+}\to \eta_c\rho(770)$    \\ 
                      &$X_{1^+}\to\Xi_c^+\Xi_c^0 $                       &$X_{1^+}\to J/\psi \pi^+$\\\hline
 \hline
\end{tabular}
\end{center}
\end{table}

It should be noted the masses of the hybrids in our calculation is close in magnitude to the hidden-charm hexaquarks in Refs.~\cite{Wan:2019ake,Wang:2021qmn,Liu:2021gva}, The main difference between the hidden-charm hexaquarks and the tetraquark hybrid states is the branching ratio of $\Xi_c\bar{\Xi}_c$. For the tetraquark hybrid states, the branching ratios of $\Xi_c\bar{\Xi}_c$ are close to the branching ratios of $\Sigma_c\bar{\Sigma}_c$, while due to the Cabibbo-Kobayashi-Maskawa suppression, the branching ratios of $\Xi_c\bar{\Xi}_c$ will be relatively small for the hidden-charm hexaquarks.

\section{Summary}

In summary, we investigate the hidden-charm and -bottom tetraquark hybrid states which consists of two valence quarks and two valence antiquarks together with a valence gluon. We constructed twenty-four currents in configuration of $[\bar{3}_c]_{c q}\otimes[8_c]_{G}\otimes[3_c]_{\bar{c} \bar{q^\prime}}$. In the framework of QCD sum rules, their mass spectra are evaluated, and the numerical results indicate that there may exist 14 hidden-charm and -bottom tetraquark hybrid states, and their masses are tabulated in Tab.~\ref{mass_c} and Tab.~\ref{mass_b}. 

We also analyze their possible production and decay modes of hidden-charm tetraquark hybrid states which are tabulated in Tab.~\ref{production} and Tab.~\ref{decay}, respectively. In those processes, all the parent particles are copiously produced in experiment, and are hopefully measurable in BESIII, BELLEII, PANDA, Super-B, and LHCb experiments.

It's also should be noted that the hybrid in bottom sector are heavier than their charm partner, the production of hidden-bottom
tetraquark hybrids is more difficult. On the other hand, the given production modes in Tab.~\ref{production} are in the processes of $\Upsilon$ decays, which cannot be taken place for hidden-bottom hybrid. Thus, searching the hidden-bottom tetraquark hybrid states are more challenging than their charm partners.

%%%%%%%%%%%%%%%%%%%%%%%%%%%%%%%%%%%%%%%%%%%%%%%%%%%%%%%%%%%%%%%%%%%%%
\vspace{0.7cm} {\bf Acknowledgments}

This work was supported in part by Specific Fund of Fundamental Scientific Research Operating Expenses for Undergraduate Universities in Liaoning Province (No. LJ212410165019 and No. LJKMZ20221431) and Ph.~D. Research Start-up Fund of Liaoning Normal University (No. 2024BSL026).
%%%%%%%%%%%%%%%%%%%%%%%%%%%%%%%%%%%%%%%%%%%%%%%%%%%%%%%%%%%%%%%%%%%%

\begin{widetext}
\appendix
\section{The spectral densities for gluonic tetraquark states}\label{App_A}
\subsection{The spectral densities for $0^+$ gluonic tetraquark states}
For the current shown in Eq. (\ref{current-0+A}),  we obtain the spectral densities as follows:
\begin{eqnarray}
\rho^{pert}_{0^+\;,A} (s) &=& \int^{\alpha_{max}}_{\alpha_{min}} d \alpha \int^{1 - \alpha}_{\beta_{min}} d \beta \frac{g_s^2{\cal F}^5_{\alpha \beta} (\alpha + \beta - 1)^3 \big({\cal F}_{\alpha\beta}+3m_c(\alpha m_u+\beta m_d)\big)}{5\times 3^2 \times 2^{11}\pi^8 \alpha^5 \beta^5}\; , \\
\rho_{0^+\;,A}^{\langle \bar{q} q \rangle}(s) &=& \frac{\langle \bar{q} q \rangle}{\pi^6}
\int^{\alpha_{max}}_{\alpha_{min}} d \alpha \int^{1 - \alpha}_{\beta_{min}} d \beta
\frac{ g_s^2 {\cal F}_{\alpha \beta }^3(\alpha+\beta-1)}{3 \times 2^9 \alpha^4 \beta^4} \big( 2 \alpha\beta{\cal F}_{\alpha \beta} (m_u+m_d)\nonumber\\
&+&4\alpha\beta m_c m_u m_d(\alpha+\beta)-m_c{\cal F}_{\alpha\beta}(\alpha+\beta)(\alpha+\beta-1)\big)\; , \\
\rho_{0^+\;,A}^{\langle G^2 \rangle\;,I}(s) &=& \frac{\langle g_s^2 G^2\rangle}{\pi^6} \int^{\alpha_{max}}_{\alpha_{min}} d \alpha \int^{1 - \alpha}_{\beta_{min}} d \beta \frac{m_c{\cal F}_{\alpha \beta}^2(\alpha+\beta-1)}{3 \times 2^{9} \alpha^3 \beta^3} \big({\cal F}_{\alpha\beta}(\alpha m_u\nonumber\\
&+&\beta m_d)+6\alpha\beta m_c m_u m_d\big)
 \; , \\
 \rho_{0^+\;,A}^{\langle G^2 \rangle\;,II}(s) &=& \frac{g_s^2 \langle g_s^2 G^2\rangle}{\pi^8} \int^{\alpha_{max}}_{\alpha_{min}} d \alpha \int^{1 - \alpha}_{\beta_{min}} d \beta \bigg( \frac{m_c{\cal F}_{\alpha \beta}^2(\alpha+\beta-1)^3}{3^3 \times 2^{12} \alpha^5 \beta^5} \big( \alpha^3{\cal F}_{\alpha\beta}(2 m_c+3 m_u)\nonumber\\
&+&\beta^3 {\cal F}_{\alpha\beta}(2 m_c+3 m_d) + 3m_c^2(\alpha^3+\beta^3)(\alpha m_u+ \beta m_d)\big)+\frac{{\cal F}_{\alpha \beta}^3(\alpha+\beta-1)}{3^3 \times 2^{19} \alpha^4 \beta^4} \nonumber\\
&\times&\big( {\cal F}_{\alpha\beta}(1-5\alpha+4\alpha^2-5\beta+4\beta^2+14\alpha\beta)+ 4m_c(\alpha+\beta-1)(\beta m_d(49\alpha+4\beta\nonumber\\
&-&1) +\alpha m_u(49\beta+4\alpha-1)) \big)\bigg)
 \; , \\
\rho_{0^+\;,A}^{\langle \bar{q} G q \rangle}(s) &=& \frac{\langle g_s \bar{q} \sigma \cdot G q \rangle}{\pi^6} \int_{\alpha_{min}}^{\alpha_{max}}  \int_{\beta_{min}}^{1 - \alpha} d \beta  \frac{g_s^2 {\cal F}^2_{\alpha\beta}}{3^2 \times 2^{13} \alpha^3 \beta^3}\big( -64\alpha\beta {\cal F}_{\alpha\beta}(m_u+m_d) \nonumber\\
&-&96\alpha\beta(\alpha+\beta)m_c m_u m_d + m_c {\cal F}_{\alpha\beta} (\alpha+\beta-1)(94\alpha+94\beta+1)\big)\; ,\\
\rho_{0^+\;,A}^{\langle \bar{q} q\rangle^2}(s) &=& \frac{\langle \bar{q} q\rangle^2}{\pi^4} \int_{\alpha_{min}}^{\alpha_{max}} d \alpha \bigg{\{}\int_{\beta_{min}}^{1 - \alpha} d \beta \big[ -\frac{g_s^2 m_c {\cal F}_{\alpha\beta}^2(\alpha m_d+\beta m_u) }{3 \times 2^{5} \alpha^2\beta^2}  \big] \nonumber\\
&+&\big[\frac{g_s^2 {\cal H}_\alpha^2m_u m_d}{3\times 2^5\alpha(\alpha-1)}\big]\bigg{\}}\; , \\
\rho_{0^+\;,A}^{\langle G^3 \rangle\;,I}(s) &=& \frac{\langle g_s^3 G^3 \rangle}{\pi^6} \int_{\alpha_{min}}^{\alpha_{max}} d\alpha \int_{\beta_{min}}^{1 - \alpha} d \beta - \frac{{\cal F}_{\alpha\beta}}{3 \times 2^{11} \alpha^3 \beta^3} \big[ 2 \alpha\beta {\cal F}_{\alpha\beta}^2 -12\alpha\beta m_u m_d m_c^2 (\alpha^2-\beta^2\nonumber\\
&-&\alpha\beta+\beta-\alpha)  - 3m_c{\cal F}_{\alpha\beta} \big(\alpha m_u(\alpha^2-\beta^2
-3\alpha\beta+\beta-\alpha)+\beta m_d(\alpha^2-\beta^2\nonumber\\
&-&\alpha\beta+\beta-\alpha)\big)   \big] \; , \\
\rho_{0^+\;,A}^{\langle G^3 \rangle\;,II}(s) &=& \frac{\langle g_s^3 G^3 \rangle}{\pi^8} \int_{\alpha_{min}}^{\alpha_{max}} d\alpha \int_{\beta_{min}}^{1 - \alpha} d \beta  \frac{g_s^2 {\cal F}_{\alpha\beta}(\alpha+\beta-1)^3}{3^3 \times 2^{14} \alpha^5 \beta^5} \big[2{\cal F}_{\alpha\beta}^2(\alpha^3+\beta^3)\nonumber\\
&+&12m_c^3(\alpha m_u+\beta m_d)(\alpha^4+\beta^4)+3m_c {\cal F}_{\alpha\beta}\big(6\alpha^4 m_u+6\beta^4 m_d \nonumber\\
&+&\alpha\beta(\alpha^2 m_d +\beta^2 m_u)+4m_c(\alpha^4+\beta^4)\big)  \big] \; , \\
\rho_{0^+\;,A}^{\langle \bar{q} q\rangle\langle \bar{q} G q \rangle}(s) &=& \frac{g_s^2\langle \bar{q} q\rangle \langle g_s \bar{q} \sigma \cdot G q \rangle}{\pi^4} \int_{\alpha_{min}}^{\alpha_{max}} d \alpha \bigg{\{}\int_{\beta_{min}}^{1 - \alpha} d \beta \big[ -\frac{ m_c {\cal F}_{\alpha\beta}( m_u+ m_d) }{3^2 \times 2^{11} \alpha\beta}  \big] \nonumber\\
&+&\frac{1}{3^2\times 2^6\alpha(\alpha-1)}\big[-12{\cal H}_\alpha \alpha(\alpha-1) m_u m_d +4\alpha (\alpha-1)m_c^2 m_u m_d \nonumber\\
&+&5m_c {\cal H}_\alpha (m_u - \alpha m_u+\alpha m_d) \big]\bigg{\}}\; , 
\end{eqnarray}
where, the subscript $I$ and $II$ for the gluon condensate come from dynamic gluon and quarks, respectively. Here,
\begin{eqnarray}
{\cal F}_{\alpha \beta} &=& (\alpha + \beta) m_c^2 - \alpha \beta s \; , {\cal H}_\alpha  = m_c^2 - \alpha (1 - \alpha) s \; , \\
\alpha_{min} &=& \left(1 - \sqrt{1 - 4 m_c^2/s} \right) / 2, \; , \alpha_{max} = \left(1 + \sqrt{1 - 4 m_c^2 / s} \right) / 2  \; , \\
\beta_{min} &=& \alpha m_c^2 /(s \alpha - m_c^2).
\end{eqnarray}

For the current shown in Eq. (\ref{current-0+B}), we obtain the spectral densities as follows:
\begin{eqnarray}
\rho^{pert}_{0^+\;,B} (s) &=& \int^{\alpha_{max}}_{\alpha_{min}} d \alpha \int^{1 - \alpha}_{\beta_{min}} d \beta \frac{g_s^2{\cal F}^5_{\alpha \beta} (\alpha + \beta - 1)^3 \big({\cal F}_{\alpha\beta}-3m_c(\alpha m_u+\beta m_d)\big)}{5\times 3^2 \times 2^{11}\pi^8 \alpha^5 \beta^5}\; ,  \\
\rho_{0^+\;,B}^{\langle \bar{q} q \rangle}(s) &=& \frac{\langle \bar{q} q \rangle}{\pi^6}
\int^{\alpha_{max}}_{\alpha_{min}} d \alpha \int^{1 - \alpha}_{\beta_{min}} d \beta
\frac{ g_s^2 {\cal F}_{\alpha \beta }^3(\alpha+\beta-1)}{3 \times 2^9 \alpha^4 \beta^4} \big( 2 \alpha\beta{\cal F}_{\alpha \beta} (m_u+m_d)\nonumber\\
&-&4\alpha\beta m_c m_u m_d(\alpha+\beta)+m_c{\cal F}_{\alpha\beta}(\alpha+\beta)(\alpha+\beta-1)\big)\; , \\
\rho_{0^+\;,B}^{\langle G^2 \rangle\;,I}(s) &=& \frac{\langle g_s^2 G^2\rangle}{\pi^6} \int^{\alpha_{max}}_{\alpha_{min}} d \alpha \int^{1 - \alpha}_{\beta_{min}} d \beta - \frac{m_c{\cal F}_{\alpha \beta}^2(\alpha+\beta-1)}{3 \times 2^{9} \alpha^3 \beta^3} \big({\cal F}_{\alpha\beta}(\alpha m_u\nonumber\\
&+&\beta m_d)-6\alpha\beta m_c m_u m_d\big)
 \; , \\
 \rho_{0^+\;,B}^{\langle G^2 \rangle\;,II}(s) &=& \frac{g_s^2 \langle g_s^2 G^2\rangle}{\pi^8} \int^{\alpha_{max}}_{\alpha_{min}} d \alpha \int^{1 - \alpha}_{\beta_{min}} d \beta \bigg(- \frac{m_c{\cal F}_{\alpha \beta}^2(\alpha+\beta-1)^3}{3^3 \times 2^{12} \alpha^5 \beta^5} \big( \alpha^3{\cal F}_{\alpha\beta}(3 m_u\nonumber\\
 &-&2 m_c)+\beta^3 {\cal F}_{\alpha\beta}(3 m_d-2 m_c) + 3m_c^2(\alpha^3+\beta^3)(\alpha m_u+ \beta m_d)\big)\nonumber\\
&+&\frac{{\cal F}_{\alpha \beta}^3(\alpha+\beta-1)}{3^3 \times 2^{19} \alpha^4 \beta^4} \big( -{\cal F}_{\alpha\beta}(1-5\alpha+4\alpha^2-5\beta+4\beta^2+14\alpha\beta)\nonumber\\
&+& 4m_c(\alpha+\beta-1)(\beta m_d(49\alpha+4\beta-1) +\alpha m_u(49\beta+4\alpha-1)) \big)\bigg)
 \; , \\
\rho_{0^+\;,B}^{\langle \bar{q} G q \rangle}(s) &=& -\frac{\langle g_s \bar{q} \sigma \cdot G q \rangle}{\pi^6} \int_{\alpha_{min}}^{\alpha_{max}}  \int_{\beta_{min}}^{1 - \alpha} d \beta  \frac{g_s^2 {\cal F}^2_{\alpha\beta}}{3^2 \times 2^{13} \alpha^3 \beta^3}\big( 64\alpha\beta {\cal F}_{\alpha\beta}(m_u+m_d) \nonumber\\
&-&96\alpha\beta(\alpha+\beta)m_c m_u m_d + m_c {\cal F}_{\alpha\beta}(\alpha+\beta-1)(94\alpha+94\beta+1) \big)\; ,\\
\rho_{0^+\;,B}^{\langle \bar{q} q\rangle^2}(s) &=& \frac{\langle \bar{q} q\rangle^2}{\pi^4} \int_{\alpha_{min}}^{\alpha_{max}} d \alpha \bigg{\{}\int_{\beta_{min}}^{1 - \alpha} d \beta \big[ \frac{g_s^2 m_c {\cal F}_{\alpha\beta}^2(\alpha m_d+\beta m_u) }{3 \times 2^{5} \alpha^2\beta^2}  \big] \nonumber\\
&+&\big[\frac{g_s^2 {\cal H}_\alpha^2m_u m_d}{3\times 2^5\alpha(\alpha-1)}\big]\bigg{\}}\; , \\
\rho_{0^+\;,B}^{\langle G^3 \rangle\;,I}(s) &=& \frac{\langle g_s^3 G^3 \rangle}{\pi^6} \int_{\alpha_{min}}^{\alpha_{max}} d\alpha \int_{\beta_{min}}^{1 - \alpha} d \beta - \frac{{\cal F}_{\alpha\beta}}{3 \times 2^{11} \alpha^3 \beta^3} \big[ 2 \alpha\beta {\cal F}_{\alpha\beta}^2 \nonumber\\
&-&12\alpha\beta m_u m_d m_c^2 (\alpha^2-\beta^2-\alpha\beta+\beta-\alpha)  + 3m_c{\cal F}_{\alpha\beta} \big(\alpha m_u(\alpha^2-\beta^2\nonumber\\
&-&3\alpha\beta+\beta-\alpha)+\beta m_d(\alpha^2-\beta^2-\alpha\beta+\beta-\alpha)\big)   \big] \; , \\
\rho_{0^+\;,B}^{\langle G^3 \rangle\;,II}(s) &=& \frac{\langle g_s^3 G^3 \rangle}{\pi^8} \int_{\alpha_{min}}^{\alpha_{max}} d\alpha \int_{\beta_{min}}^{1 - \alpha} d \beta - \frac{g_s^2 {\cal F}_{\alpha\beta}(\alpha+\beta-1)^3}{3^3 \times 2^{14} \alpha^5 \beta^5} \big[-2{\cal F}_{\alpha\beta}^2(\alpha^3+\beta^3)\nonumber\\
&+&12m_c^3(\alpha m_u+\beta m_d)(\alpha^4+\beta^4)+3m_c {\cal F}_{\alpha\beta}\big(6\alpha^4 m_u+6\beta^4 m_d \nonumber\\
&+&\alpha\beta(\alpha^2 m_d +\beta^2 m_u)-4m_c(\alpha^4+\beta^4)\big)  \big] \; , \\
\rho_{0^+\;,B}^{\langle \bar{q} q\rangle\langle \bar{q} G q \rangle}(s) &=& \frac{g_s^2\langle \bar{q} q\rangle \langle g_s \bar{q} \sigma \cdot G q \rangle}{\pi^4} \int_{\alpha_{min}}^{\alpha_{max}} d \alpha \bigg{\{}\int_{\beta_{min}}^{1 - \alpha} d \beta \big[ \frac{ m_c {\cal F}_{\alpha\beta}( m_u+ m_d) }{3^2 \times 2^{11} \alpha\beta}  \big] \nonumber\\
&+&\frac{1}{3^2\times 2^6\alpha(\alpha-1)}\big[-12{\cal H}_\alpha \alpha(\alpha-1) m_u m_d +4\alpha (\alpha-1)m_c^2 m_u m_d \nonumber\\
&-&5m_c {\cal H}_\alpha (m_u - \alpha m_u+\alpha m_d) \big]\bigg{\}}\; .
\end{eqnarray}

For the current showned in Eq. (\ref{current-0+C}), we obtain the spectral densities as follows:
\begin{eqnarray}
\rho^{pert}_{0^+\;,C} (s) &=& \int^{\alpha_{max}}_{\alpha_{min}} d \alpha \int^{1 - \alpha}_{\beta_{min}} d \beta \frac{g_s^2{\cal F}^5_{\alpha \beta} (\alpha + \beta - 1)^3 \big({\cal F}_{\alpha\beta}+3m_c(\alpha m_u-\beta m_d)\big)}{5\times 3^2 \times 2^{12}\pi^8 \alpha^5 \beta^5}\; , \\
\rho_{0^+\;,C}^{\langle \bar{q} q \rangle}(s) &=& \frac{\langle \bar{q} q \rangle}{\pi^6}
\int^{\alpha_{max}}_{\alpha_{min}} d \alpha \int^{1 - \alpha}_{\beta_{min}} d \beta
\frac{ g_s^2 {\cal F}_{\alpha \beta }^3(\alpha+\beta-1)}{3 \times 2^{10} \alpha^4 \beta^4} \big( 2 \alpha\beta{\cal F}_{\alpha \beta} (m_u+m_d)\nonumber\\
&+&4\alpha\beta m_c m_u m_d(\alpha-\beta)-m_c{\cal F}_{\alpha\beta}(\alpha-\beta)(\alpha+\beta-1)\big)\; , \\
\rho_{0^+\;,C}^{\langle G^2 \rangle\;,I}(s) &=& \frac{\langle g_s^2 G^2\rangle}{\pi^6} \int^{\alpha_{max}}_{\alpha_{min}} d \alpha \int^{1 - \alpha}_{\beta_{min}} d \beta \frac{m_c{\cal F}_{\alpha \beta}^2(\alpha+\beta-1)}{3 \times 2^{10} \alpha^3 \beta^3} \big({\cal F}_{\alpha\beta}(-\alpha m_u\nonumber\\
&+&\beta m_d)+6\alpha\beta m_c m_u m_d\big)
 \; , \\
 \rho_{0^+\;,C}^{\langle G^2 \rangle\;,II}(s) &=& \frac{g_s^2 \langle g_s^2 G^2\rangle}{\pi^8} \int^{\alpha_{max}}_{\alpha_{min}} d \alpha \int^{1 - \alpha}_{\beta_{min}} d \beta \bigg( \frac{m_c{\cal F}_{\alpha \beta}^2(\alpha+\beta-1)^3}{3^3 \times 2^{13} \alpha^5 \beta^5} \big( \alpha^3{\cal F}_{\alpha\beta}(2 m_c+3 m_u)\nonumber\\
&+&\beta^3 {\cal F}_{\alpha\beta}(2 m_c-3 m_d) + 3m_c^2(\alpha^3+\beta^3)(\alpha m_u- \beta m_d)\big)+\frac{{\cal F}_{\alpha \beta}^3(\alpha+\beta-1)}{3^3 \times 2^{20} \alpha^4 \beta^4} \nonumber\\
&\times&\big( {\cal F}_{\alpha\beta}(1-5\alpha+4\alpha^2-5\beta+4\beta^2+14\alpha\beta)+ 4m_c(\alpha+\beta-1)(-\beta m_d(49\alpha\nonumber\\
&+&4\beta-1) +\alpha m_u(49\beta+4\alpha-1)) \big)\bigg)
 \; , \\
\rho_{0^+\;,C}^{\langle \bar{q} G q \rangle}(s) &=& \frac{\langle g_s \bar{q} \sigma \cdot G q \rangle}{\pi^6} \int_{\alpha_{min}}^{\alpha_{max}}  \int_{\beta_{min}}^{1 - \alpha} d \beta  \frac{g_s^2 {\cal F}^2_{\alpha\beta}}{3^2 \times 2^{14} \alpha^3 \beta^3}\big( -64\alpha\beta {\cal F}_{\alpha\beta}(m_u+m_d) \nonumber\\
&-&96\alpha\beta(\alpha-\beta)m_c m_u m_d + 95 m_c {\cal F}_{\alpha\beta}(\alpha^2-\beta^2-\alpha+\beta) \big)\; ,\\
\rho_{0^+\;,C}^{\langle \bar{q} q\rangle^2}(s) &=& \frac{\langle \bar{q} q\rangle^2}{\pi^4} \int_{\alpha_{min}}^{\alpha_{max}} d \alpha \bigg{\{}\int_{\beta_{min}}^{1 - \alpha} d \beta \big[ -\frac{g_s^2 m_c {\cal F}_{\alpha\beta}^2(\alpha m_d-\beta m_u) }{3 \times 2^{6} \alpha^2\beta^2}  \big] \nonumber\\
&+&\big[\frac{g_s^2 {\cal H}_\alpha^2m_u m_d}{3\times 2^6\alpha(\alpha-1)}\big]\bigg{\}}\; , \\
\rho_{0^+\;,C}^{\langle G^3 \rangle\;,I}(s) &=& \frac{\langle g_s^3 G^3 \rangle}{\pi^6} \int_{\alpha_{min}}^{\alpha_{max}} d\alpha \int_{\beta_{min}}^{1 - \alpha} d \beta - \frac{{\cal F}_{\alpha\beta}}{3 \times 2^{11} \alpha^3 \beta^3} \big[  \alpha\beta {\cal F}_{\alpha\beta}^2 -6\alpha\beta m_u m_d m_c^2 (4\alpha^2-4\beta^2\nonumber\\
&-&\alpha\beta+4\beta-4\alpha)  + 6m_c{\cal F}_{\alpha\beta} \big(\alpha m_u(\alpha^2-\beta^2
-\alpha\beta+\beta-\alpha)-\beta m_d(\alpha^2-\beta^2\nonumber\\
&+&\alpha\beta+\beta-\alpha)\big)   \big] \; , \\
\rho_{0^+\;,C}^{\langle G^3 \rangle\;,II}(s) &=& \frac{\langle g_s^3 G^3 \rangle}{\pi^8} \int_{\alpha_{min}}^{\alpha_{max}} d\alpha \int_{\beta_{min}}^{1 - \alpha} d \beta  \frac{g_s^2 {\cal F}_{\alpha\beta}(\alpha+\beta-1)^3}{3^3 \times 2^{15} \alpha^5 \beta^5} \big[2{\cal F}_{\alpha\beta}^2(\alpha^3+\beta^3)\nonumber\\
&+&12m_c^3(\alpha m_u-\beta m_d)(\alpha^4+\beta^4)+3m_c {\cal F}_{\alpha\beta}\big(6\alpha^4 m_u-6\beta^4 m_d \nonumber\\
&-&\alpha\beta(\alpha^2 m_d -\beta^2 m_u)+4m_c(\alpha^4+\beta^4)\big)  \big] \; , \\
\rho_{0^+\;,C}^{\langle \bar{q} q\rangle\langle \bar{q} G q \rangle}(s) &=& \frac{g_s^2\langle \bar{q} q\rangle \langle g_s \bar{q} \sigma \cdot G q \rangle}{\pi^4} \int_{\alpha_{min}}^{\alpha_{max}} d \alpha \bigg{\{}\int_{\beta_{min}}^{1 - \alpha} d \beta \big[ \frac{g_s^2 m_c {\cal F}_{\alpha\beta}( m_u- m_d) }{3^2 \times 2^{12} \alpha\beta}  \big] \nonumber\\
&+&\frac{1}{3^2\times 2^7\alpha(\alpha-1)}\big[-12{\cal H}_\alpha \alpha(\alpha-1) m_u m_d +4\alpha (\alpha-1)m_c^2 m_u m_d \nonumber\\
&+&5m_c {\cal H}_\alpha (-m_u + \alpha m_u+\alpha m_d) \big]\bigg{\}}\; .
\end{eqnarray}

For the current showned in Eq. (\ref{current-0+D}), considering the symmetries between $j_{0^+}^C$ and $j_{0^+}^D$, we obtain the spectral densities $\rho_{0^+}^D(s)=\rho_{0^+}^C(s)(m_u\to m_d, m_d\to m_u)$.

\subsection{The spectral densities for $0^-$ gluonic tetraquark states}

For the current shown in Eq. (\ref{current-0-A}),  we obtain the spectral densities as follows:
\begin{eqnarray}
\rho^{pert}_{0^-\;,A} (s) &=& \rho^{pert}_{0^+\;,A} (s)\; , \\
\rho_{0^-\;,A}^{\langle \bar{q} q \rangle}(s) &=& \rho_{0^+\;,A}^{\langle \bar{q} q \rangle}(s)\; , \\
\rho_{0^-\;,A}^{\langle G^2 \rangle\;,I}(s) &=&-\rho_{0^+\;,A}^{\langle G^2 \rangle\;,I}(s)
 \; , \\
 \rho_{0^-\;,A}^{\langle G^2 \rangle\;,II}(s) &=& \rho_{0^+\;,A}^{\langle G^2 \rangle\;,II}(s)
 \; , \\
\rho_{0^-\;,A}^{\langle \bar{q} G q \rangle}(s) &=&\rho_{0^+\;,A}^{\langle \bar{q} G q \rangle}(s)\; ,\\
\rho_{0^-\;,A}^{\langle \bar{q} q\rangle^2}(s) &=& \rho_{0^+\;,A}^{\langle \bar{q} q\rangle^2}(s)\; , \\
\rho_{0^-\;,A}^{\langle G^3 \rangle\;,I}(s) &=& \frac{\langle g_s^3 G^3 \rangle}{\pi^6} \int_{\alpha_{min}}^{\alpha_{max}} d\alpha \int_{\beta_{min}}^{1 - \alpha} d \beta - \frac{{\cal F}_{\alpha\beta}}{3 \times 2^{10} \alpha^3 \beta^3} \big[  \alpha\beta {\cal F}_{\alpha\beta}^2 \nonumber\\
&+&6\alpha\beta m_u m_d m_c^2 (4\alpha^2-4\beta^2-\alpha\beta+\beta-\alpha)  + 6m_c{\cal F}_{\alpha\beta} \big(\alpha m_u(\alpha^2\nonumber\\
&-&\beta^2-\alpha\beta+\beta-\alpha)+\beta m_d(\alpha^2-\beta^2+\alpha\beta+\beta-\alpha)\big)   \big] \; , \\
\rho_{0^-\;,A}^{\langle G^3 \rangle\;,II}(s) &=& \rho_{0^+\;,A}^{\langle G^3 \rangle\;,II}(s) \; , \\
\rho_{0^-\;,A}^{\langle \bar{q} q\rangle\langle \bar{q} G q \rangle}(s) &=&\rho_{0^+\;,A}^{\langle \bar{q} q\rangle\langle \bar{q} G q \rangle}(s)\; .
\end{eqnarray}

For the currents shown in Eq. (\ref{current-0-B}), we obtain the spectral densities as follows:
\begin{eqnarray}
\rho^{pert}_{0^-\;,B} (s) &=& \rho^{pert}_{0^+\;,B} (s) \; , \\
\rho_{0^-\;,B}^{\langle \bar{q} q \rangle}(s) &=& \rho_{0^+\;,B}^{\langle \bar{q} q \rangle}(s) \; , \\
\rho_{0^-\;,B}^{\langle G^2 \rangle\;,I}(s) &=& -\rho_{0^+\;,B}^{\langle G^2 \rangle\;,I}(s)
 \; , \\
 \rho_{0^-\;,B}^{\langle G^2 \rangle\;,II}(s) &=&  \rho_{0^+\;,B}^{\langle G^2 \rangle\;,II}(s)
 \; , \\
\rho_{0^-\;,B}^{\langle \bar{q} G q \rangle}(s) &=& \rho_{0^+\;,B}^{\langle \bar{q} G q \rangle}(s) \; ,\\
\rho_{0^-\;,B}^{\langle \bar{q} q\rangle^2}(s) &=& \rho_{0^+\;,B}^{\langle \bar{q} q\rangle^2}(s) \; , \\
\rho_{0^-\;,B}^{\langle G^3 \rangle\;,I}(s) &=& \frac{\langle g_s^3 G^3 \rangle}{\pi^6} \int_{\alpha_{min}}^{\alpha_{max}} d\alpha \int_{\beta_{min}}^{1 - \alpha} d \beta - \frac{{\cal F}_{\alpha\beta}}{3 \times 2^{10} \alpha^3 \beta^3} \big[  \alpha\beta {\cal F}_{\alpha\beta}^2 +6\alpha\beta m_u m_d m_c^2 (4\alpha^2-4\beta^2\nonumber\\
&-&\alpha\beta+4\beta-4\alpha)  - 6m_c{\cal F}_{\alpha\beta} \big(\alpha m_u(\alpha^2-\beta^2
-\alpha\beta+\beta-\alpha)+\beta m_d(\alpha^2-\beta^2\nonumber\\
&+&\alpha\beta+\beta-\alpha)\big)   \big] \; , \\
\rho_{0^-\;,B}^{\langle G^3 \rangle\;,II}(s) &=& \rho_{0^+\;,B}^{\langle G^3 \rangle\;,II}(s)  \; , \\
\rho_{0^-\;,B}^{\langle \bar{q} q\rangle\langle \bar{q} G q \rangle}(s) &=& \rho_{0^+\;,B}^{\langle \bar{q} q\rangle\langle \bar{q} G q \rangle}(s)\; .
\end{eqnarray}

For the current showned in Eq. (\ref{current-0-C}), we obtain the spectral densities as follows:
\begin{eqnarray}
\rho^{pert}_{0^-\;,C} (s) &=&\rho^{pert}_{0^+\;,C} (s) \; ,  \\
\rho_{0^-\;,C}^{\langle \bar{q} q \rangle}(s) &=& \rho_{0^+\;,C}^{\langle \bar{q} q \rangle}(s)\; , \\
\rho_{0^-\;,C}^{\langle G^2 \rangle\;,I}(s) &=&-\rho_{0^+\;,C}^{\langle G^2 \rangle\;,I}(s)
 \; , \\
 \rho_{0^-\;,C}^{\langle G^2 \rangle\;,II}(s) &=& \rho_{0^+\;,C}^{\langle G^2 \rangle\;,II}(s) 
 \; , \\
\rho_{0^-\;,C}^{\langle \bar{q} G q \rangle}(s) &=&\rho_{0^+\;,C}^{\langle \bar{q} G q \rangle}(s)\; ,\\
\rho_{0^-\;,C}^{\langle \bar{q} q\rangle^2}(s) &=& \rho_{0^+\;,C}^{\langle \bar{q} q\rangle^2}(s)\; , \\
\rho_{0^-\;,C}^{\langle G^3 \rangle\;,I}(s) &=& \frac{\langle g_s^3 G^3 \rangle}{\pi^6} \int_{\alpha_{min}}^{\alpha_{max}} d\alpha \int_{\beta_{min}}^{1 - \alpha} d \beta - \frac{{\cal F}_{\alpha\beta}}{3 \times 2^{12} \alpha^3 \beta^3} \big[ 2 \alpha\beta {\cal F}_{\alpha\beta}^2 +12\alpha\beta m_u m_d m_c^2 (\alpha^2-\beta^2\nonumber\\
&-&\alpha\beta+\beta-\alpha)  + 3m_c{\cal F}_{\alpha\beta} \big(\alpha m_u(\alpha^2-\beta^2
-3\alpha\beta+\beta-\alpha)-\beta m_d(\alpha^2-\beta^2\nonumber\\
&-&\alpha\beta+\beta-\alpha)\big)   \big] \; , \\
\rho_{0^-\;,C}^{\langle G^3 \rangle\;,II}(s) &=& \rho_{0^+\;,C}^{\langle G^3 \rangle\;,II}(s) \; , \\
\rho_{0^-\;,C}^{\langle \bar{q} q\rangle\langle \bar{q} G q \rangle}(s) &=& \rho_{0^+\;,C}^{\langle \bar{q} q\rangle\langle \bar{q} G q \rangle}(s)\; .
\end{eqnarray}

For the current showned in Eq. (\ref{current-0-D}), considering the symmetries between $j_{0^-}^C$ and $j_{0^-}^D$, we obtain the spectral densities $\rho_{0^-}^D(s)=\rho_{0^-}^C(s)(m_u\to m_d, m_d\to m_u)$.

\subsection{The spectral densities for $1^-$ gluonic tetraquark states}

For the current shown in Eq. (\ref{current-1-A}),  we obtain the spectral densities as follows:
\begin{eqnarray}
\rho^{pert}_{1^-\;,A} (s) &=& \int^{\alpha_{max}}_{\alpha_{min}} d \alpha \int^{1 - \alpha}_{\beta_{min}} d \beta \frac{g_s^2{\cal F}^3_{\alpha \beta} (\alpha + \beta - 1)^3}{5\times 3^3 \times 2^{16}\pi^8 \alpha^5 \beta^5} \bigg{\{} 160\alpha\beta m_c^4 m_u m_d (\alpha+\beta)(\alpha\nonumber\\
&+&\beta-1) -3 {\cal F}^3_{\alpha\beta}(13\alpha+13\beta-17) -2m_c {\cal F}^2_{\alpha\beta}\big( -72\beta m_d +44\alpha\beta m_d\nonumber\\
&+&44 \beta^2 m_d- 18m_c (\alpha+\beta)(\alpha+\beta-1)+3\alpha m_u(11\alpha+11\beta+9) \big)\nonumber\\
&+&20m_c^2{\cal F}_{\alpha\beta}\big( \alpha\beta m_u m_d(7-9\alpha-9\beta)\nonumber\\
&+&m_c(\alpha+\beta)(\alpha+\beta-1)(3\alpha m_u +4\beta m_d) \big)  \bigg{\}}\; ,  \\
\rho_{1^-\;,A}^{\langle \bar{q} q \rangle}(s) &=& \frac{\langle \bar{q} q \rangle}{\pi^6}
\int^{\alpha_{max}}_{\alpha_{min}} d \alpha \int^{1 - \alpha}_{\beta_{min}} d \beta
\frac{ g_s^2 {\cal F}_{\alpha \beta }^2(\alpha+\beta-1)}{3^4 \times 2^{12} \alpha^4 \beta^4} \bigg{\{}-24m_c^3\alpha\beta(\alpha+\beta)(\alpha\nonumber\\
&+&\beta-1)\big(4m_c(m_u+m_d)(\alpha+\beta-1)-3m_um_d(3\alpha+4\beta)\big)\nonumber\\
&+&4m_c{\cal F}_{\alpha\beta}\big( -4m_c^2(\alpha+\beta)(3\alpha+4\beta)(\alpha+\beta-1)^2 +2\alpha\beta (\alpha+\beta\nonumber\\
&-&1)(41\alpha+41\beta-11)m_c(m_u+m_d) -3\alpha\beta m_u m_d(21\alpha^2+9\alpha\nonumber\\
&+&28\beta^2+49\alpha\beta-42\beta)\big)+3{\cal F}^2_{\alpha\beta}\big(-9\alpha\beta(m_u+m_d)(9\alpha\nonumber\\
&+&9\beta-11)+2m_c(\alpha+\beta-1)(9\alpha^2+12\beta^2+6\alpha-19\beta+21\alpha\beta)  \big)        \bigg{\}}\; , \\
\rho_{1^-\;,A}^{\langle G^2 \rangle\;,I}(s) &=& \frac{\langle g_s^2 G^2\rangle}{\pi^6} \int^{\alpha_{max}}_{\alpha_{min}} d \alpha \int^{1 - \alpha}_{\beta_{min}} d \beta -\frac{{\cal F}_{\alpha \beta}^2(\alpha+\beta-1)}{3^3 \times 2^{14} \alpha^3 \beta^3} \bigg{\{}({\cal F}_{\alpha\beta}^2(1+9\alpha+9\beta)\nonumber\\
&+&4m_c{\cal F}_{\alpha\beta}\big(\alpha m_u (33-7\alpha-7\beta)-2m_c(\alpha+\beta)(\alpha+\beta-1)\nonumber\\
&+&6\beta m_d(7\alpha+7\beta-6)\big) 
+24m_c^2\big( 27\alpha\beta m_u m_d\nonumber\\
&+&m_c(\alpha+\beta)(\alpha+\beta-1)(\alpha m_u-6\beta m_d)   \big)\bigg{\}}\; , \\
 \rho_{1^-\;,A}^{\langle G^2 \rangle\;,II}(s) &=& \frac{g_s^2 \langle g_s^2 G^2\rangle}{\pi^8} \int^{\alpha_{max}}_{\alpha_{min}} d \alpha \int^{1 - \alpha}_{\beta_{min}} d \beta \bigg{\{}\frac{m_c(\alpha+\beta-1)^3}{3^4 \times 2^{16} \alpha^5 \beta^5} \bigg( 8m_c^5 m_u m_d \alpha \beta (\alpha+\beta)^2\nonumber\\
 &\times&(\alpha+\beta-1)(\alpha^2-\alpha\beta+\beta^2)-{\cal F}_{\alpha\beta}^3\big( 28\beta^3m_d(\alpha+\beta-2)+3\alpha^3 m_u (7\alpha\nonumber\\
 &+&7\beta+13) + 3m_c(\alpha+\beta)(7\alpha+7\beta-11)(\alpha^2-\alpha\beta +\beta^2) \big)+4m_c^3 (\alpha+\beta)\nonumber\\
 &\times&{\cal F}_{\alpha\beta}\big( m_c(\alpha+\beta)(\alpha+\beta-1)(3\alpha m_u+4\beta m_d)(\alpha^2-\alpha \beta+\beta^2)\nonumber\\
 &+&m_u m_d \alpha \beta(3\alpha^3+3\beta^3-5\alpha^2-5\beta^2+6\alpha^2\beta+6\alpha\beta^2-\alpha\beta) \big)\nonumber\\
 &+&m_c{\cal F}_{\alpha\beta}^2\big(-6\alpha\beta m_u m_d(\alpha^2+\beta^2)(5\alpha+5\beta-3) +18 m_c^2(\alpha+\beta\nonumber\\
 &-&1)(\alpha+\beta)^2(\alpha^2-\alpha\beta+\beta^2) + m_c (\alpha+\beta)(4\beta m_d(12\alpha^2-5\alpha^3+6\alpha\beta^2\nonumber\\
 &-&12\alpha\beta+\beta^3+6\beta^2)+3\alpha m_u(\alpha^3+15\alpha\beta-5\beta^3-15\beta^2-21\alpha^2
+6\alpha^2\beta))    \big) \bigg)\nonumber\\
&+&\frac{{\cal F}_{\alpha \beta}(\alpha+\beta-1)}{3^5 \times 2^{21} \alpha^4 \beta^4}\bigg( 48 m_c^4 m_u m_d \alpha \beta (\alpha+\beta)(\alpha+\beta-1)^3 +12m_c^2\nonumber\\
 &\times&{\cal F}_{\alpha\beta}(\alpha+\beta-1)^2\big(\alpha\beta m_u m_d(3-5\alpha-5\beta )+m_c(\alpha+\beta)(\alpha m_u (5\alpha-31\beta\nonumber\\
 &-&1)-2 \beta m_d(63\alpha-5\beta+1) )  \big)-2m_c {\cal F}^2_{\alpha\beta} (\alpha+\beta-1)\big(\alpha m_u(35\alpha^2 - 217\beta^2\nonumber\\
 &-&182\alpha\beta +38\alpha+1478\beta-13 ) +2\beta m_d (35\beta^2 -441\alpha^2-406\alpha\beta -34\beta\nonumber\\
&+&772\alpha+5  )+4m_c (\alpha+\beta)(139\alpha^2 -21\beta^2 +106\alpha\beta -138\alpha +22\beta-1) \big)\nonumber\\
&+&{\cal F}^3_{\alpha\beta}\big( 1251\alpha^3+7\alpha^2(315\beta-419)-(\beta-1)(7+50\beta+189\beta^2)\nonumber\\
&+&\alpha(1675-2698\beta+765\beta^2)    \big) \bigg)\bigg{\}}
 \; , \\
\rho_{1^-\;,A}^{\langle \bar{q} G q \rangle}(s) &=& \frac{\langle g_s \bar{q} \sigma \cdot G q \rangle}{\pi^6} \int_{\alpha_{min}}^{\alpha_{max}}  \int_{\beta_{min}}^{1 - \alpha} d \beta  \frac{g_s^2 {\cal F}_{\alpha\beta}}{3^4 \times 2^{15} \alpha^3 \beta^4}\bigg{\{} -m_c(\alpha+\beta-1)\bigg( {\cal F}_{\alpha\beta}^2\big( 16\alpha\nonumber\\
&\times&(7\alpha^2+\alpha-8) +\beta(210\alpha^2+21\alpha-1)+77\alpha\beta^2+20\beta^2-21\beta^3 \big) +8\beta m_c^3(\alpha\nonumber\\
&+&\beta)(\alpha+\beta-1)^2 (\alpha m_u-16\alpha m_d+\beta m_d) -m_c{\cal F}_{\alpha\beta}(\alpha+\beta-1)\big(\beta(10\alpha+10\beta\nonumber\\
&-&7)(\alpha m_u-16\alpha m_d+\beta m_d)+6m_c(\alpha+\beta)(16\alpha^2-16\alpha-3\beta^2+14\alpha\beta+\beta)      \big)  \bigg) \nonumber\\
&-&48\beta\bigg( -8m_c^3\alpha\beta(\alpha+\beta)(\alpha+\beta-1) \big( 3m_c(m_u+m_d)(\alpha+\beta-1)-m_um_d(3\alpha\nonumber\\
&+&4\beta) \big)+{\cal F}_{\alpha\beta}^2 \big( 6\alpha\beta(m_u+m_d)(8-7\alpha-7\beta)+m_c(\alpha+\beta-1)(21\alpha^2+9\alpha\nonumber\\
&+&49\alpha\beta-42\beta+28\beta^2)    \big) -2m_c {\cal F}_{\alpha\beta}\big( 3m_c^2(\alpha+\beta)(\alpha+\beta-1)^2(3\alpha+4\beta) \nonumber\\
&-&3m_c(m_u+m_d)\alpha\beta(\alpha+\beta-1)(11\alpha+11\beta-4) +\alpha\beta m_u m_d (-27\beta\nonumber\\
&+&5(\alpha+\beta)(3\alpha+4\beta))\big)\bigg)     \bigg{\}}\; ,\\
\rho_{1^-\;,A}^{\langle \bar{q} q\rangle^2}(s) &=& \frac{\langle \bar{q} q\rangle^2}{\pi^4} \int_{\alpha_{min}}^{\alpha_{max}} d \alpha \bigg{\{}\int_{\beta_{min}}^{1 - \alpha} d \beta \big[ -\frac{g_s^2 {\cal F}_{\alpha\beta} }{3^3 \times 2^{8} \alpha^2\beta^2} \big( -4m_c^2(\alpha+\beta)(9\alpha\beta m_u m_d +4m_c^2(\alpha\nonumber\\
&+&\beta-1)^2-2m_c(\alpha+\beta-1)(3\alpha m_d+4\beta m_u))+{\cal F}_{\alpha\beta}\big(54m_cm_u\beta+45m_um_d\alpha\beta\nonumber\\
&+&4m_c^2(\alpha+\beta-1)(5\alpha+5\beta-4)-10m_c(\alpha+\beta)(3\alpha m_d+4\beta m_u)  \big)  \big] \nonumber\\
&+&\frac{g_s^2 {\cal H}_\alpha^2m_u m_d}{3\times 2^8\alpha(\alpha-1)}\bigg{\}}\; , \\
\rho_{1^-\;,A}^{\langle G^3 \rangle\;,I}(s) &=& \frac{\langle g_s^3 G^3 \rangle}{\pi^6} \int_{\alpha_{min}}^{\alpha_{max}} d\alpha \int_{\beta_{min}}^{1 - \alpha} d \beta  \frac{1}{3^3 \times 2^{14} \alpha^3 \beta^3} \bigg{\{} -1152m_c^4m_dm_u\alpha^2\beta^2(\alpha+\beta)(\alpha\nonumber\\
&+&\beta-1) +\beta {\cal F}_{\alpha\beta}^3\big( 357\alpha^2+(\beta-1)(7\beta+3)+ 4\alpha(91\beta-62) \big)+12 m_c^2{\cal F}_{\alpha\beta}\nonumber\\ 
&\times&\big( - m_c  (\alpha+\beta)(\alpha+\beta-1)(\alpha m_u(2\alpha^2-2\alpha-\beta^2+\beta+47\alpha\beta) +2\beta m_d(3\beta^2\nonumber\\
&-&3\beta-2\alpha^2+2\alpha-3\alpha\beta)) +3\alpha\beta m_u m_d (9\beta^2-9\beta+48\alpha^2\beta-6\alpha^2+6\alpha\nonumber\\
&-&13\alpha\beta+48\alpha\beta^2)  \big)  +3m_c{\cal F}^2_{\alpha\beta}\big( -2\beta m_c(\alpha+\beta)(\alpha+\beta-1)(51\alpha+\beta-1)  \nonumber\\
&+&2\beta m_d (-10\alpha^3+18\alpha^2-25\alpha^2\beta+3\beta(\beta-1)(5\beta-4)+9\alpha\beta-8\alpha)\nonumber\\
&+&\alpha m_u(10\alpha^3-\beta(\beta-1)(5\beta-31) +245\alpha^2\beta\nonumber\\
&-&36\alpha^2+26\alpha-108\alpha\beta+230\alpha\beta^2   )    \big) \bigg{\}} \; , \\
\rho_{1^-\;,A}^{\langle G^3 \rangle\;,II}(s) &=& \frac{\langle g_s^3 G^3 \rangle}{\pi^8} \int_{\alpha_{min}}^{\alpha_{max}} d\alpha \int_{\beta_{min}}^{1 - \alpha} d \beta  \frac{g_s^2 (\alpha+\beta-1)^3}{3^4 \times 2^{18} \alpha^5 \beta^5} \bigg{\{} -3{\cal F}_{\alpha\beta}^3 (\alpha+\beta)(7\alpha+7\beta\nonumber\\
&-&1)(\alpha^2+\beta^2-\alpha\beta) -m_c {\cal F}^2_{\alpha\beta}\big(15m_u\alpha(\alpha+\beta+3) (6\alpha^3+\beta^3) +4\beta m_d\nonumber\\
&\times&(5\alpha+5\beta-12) (\alpha^3+6\beta^3)+18m_c(4\alpha^5+\alpha^3\beta-\alpha^3\beta^2-\alpha^2\beta^3-8\beta^4\nonumber\\
&+&4\beta^5-8\alpha^4+3\alpha^4\beta+\alpha\beta^3+3\alpha\beta^4) \big)+4m_c^2{\cal F}_{\alpha\beta}\big( -6m_um_d\alpha\beta(\alpha\nonumber\\
&+&\beta)(3\alpha+3\beta-1)(\alpha^2-\alpha\beta+\beta^2)+18m_c^2(\alpha+\beta-1)(\alpha+\beta)(\alpha^4+\beta^4)\nonumber\\
&+&4\beta m_cm_d(9\alpha^4-2\alpha^5-\alpha^3\beta-\alpha^4\beta+\alpha^3\beta^2-6\alpha\beta^3+6\alpha^2\beta^3+4\beta^4\nonumber\\
&+&9\alpha\beta^4+3\beta^5  ) +3\alpha m_um_c(3\alpha^5-6\alpha^3\beta+6\alpha^3\beta^2+\alpha^2\beta^3-\alpha\beta^3-\alpha\beta^4\nonumber\\
&-&18\beta^4-2\beta^5-23\alpha^4+9\alpha^4\beta) \big)+8m_c^4\big( m_c(\alpha+\beta-1)(\alpha+\beta)(\alpha^4+\beta^4)(3\alpha m_u\nonumber\\
&+&4\beta m_d)+\alpha\beta m_u m_d ( 5\alpha^5+6\alpha^3\beta(\beta-1)+6\alpha^2\beta^3 +\beta^4(5\beta-7)+ \alpha^4(11\beta\nonumber\\
&-&7)+\alpha\beta^3(11\beta-6) )       \big)   \bigg{\}} \; , \\
\rho_{1^-\;,A}^{\langle \bar{q} q\rangle\langle \bar{q} G q \rangle}(s) &=& \frac{g_s^2\langle \bar{q} q\rangle \langle g_s \bar{q} \sigma \cdot G q \rangle}{\pi^4} \int_{\alpha_{min}}^{\alpha_{max}} d \alpha \bigg{\{}\int_{\beta_{min}}^{1 - \alpha} d \beta \big[ \frac{ m_c }{3^4 \times 2^{12} \alpha^2\beta^2}\big( 2m_c^2(\alpha+\beta\nonumber\\
&-&1)(\alpha+\beta) (-2m_u\alpha\beta-\alpha m_d(\beta-48\alpha)-2m_c(15\alpha-\beta)(\alpha+\beta-1))  \big)\nonumber\\
&-&{\cal F}_{\alpha\beta}(m_u\alpha\beta(5-6\alpha-6\beta) -2m_c(15\alpha-\beta)(\alpha+\beta-1) (3\alpha+3\beta-2)\nonumber\\
&+&\alpha m_d(48\alpha-\beta)(3\alpha+3\beta+2))  \big] +\frac{1}{3^4\times 2^8\alpha(\alpha-1)}\big[-108{\cal H}_\alpha \alpha(\alpha\nonumber\\
&-&1) m_u m_d +54\alpha (\alpha-1)m_c^2 m_u m_d -5m_c {\cal H}_\alpha (-7m_u + 7\alpha m_u+15\alpha m_d) \big]\bigg{\}}\; .
\end{eqnarray}

For the currents shown in Eq. (\ref{current-1-B}), we obtain the spectral densities as follows:
\begin{eqnarray}
\rho^{pert}_{1^-\;,B} (s) &=& \int^{\alpha_{max}}_{\alpha_{min}} d \alpha \int^{1 - \alpha}_{\beta_{min}} d \beta -\frac{g_s^2{\cal F}^3_{\alpha \beta} (\alpha + \beta - 1)^3}{5\times 3^3 \times 2^{16}\pi^8 \alpha^5 \beta^5} \bigg{\{} 160\alpha\beta m_c^4 m_u m_d (\alpha+\beta)(\alpha+\beta-1) \nonumber\\
&-&3 {\cal F}^3_{\alpha\beta}(13\alpha+13\beta-17) +2m_c {\cal F}^2_{\alpha\beta}\big( -72\beta m_d +44\alpha\beta m_d+44 \beta^2 m_d\nonumber\\
&+& 18m_c (\alpha+\beta)(\alpha+\beta-1)+3\alpha m_u(11\alpha+11\beta+9) \big)-20m_c^2{\cal F}_{\alpha\beta}\big( \alpha\beta m_u m_d(-7\nonumber\\
&+&9\alpha+9\beta)+m_c(\alpha+\beta)(\alpha+\beta-1)(3\alpha m_u +4\beta m_d) \big)  \bigg{\}}\; ,  \\
\rho_{1^-\;,B}^{\langle \bar{q} q \rangle}(s) &=& \frac{\langle \bar{q} q \rangle}{\pi^6}
\int^{\alpha_{max}}_{\alpha_{min}} d \alpha \int^{1 - \alpha}_{\beta_{min}} d \beta
\frac{ g_s^2 {\cal F}_{\alpha \beta }^2(\alpha+\beta-1)}{3^4 \times 2^{12} \alpha^4 \beta^4} \bigg{\{}24m_c^3\alpha\beta(\alpha+\beta)(\alpha+\beta-1)\nonumber\\
&\times&\big(4m_c(m_u+m_d)(\alpha+\beta-1)+3m_um_d(3\alpha+4\beta)\big)-4m_c{\cal F}_{\alpha\beta}\big( 4m_c^2(\alpha+\beta)\nonumber\\
&\times&(3\alpha+4\beta)(\alpha+\beta-1)^2 +2\alpha\beta (\alpha+\beta-1)(41\alpha+41\beta-11)m_c(m_u+m_d) \nonumber\\
&+&3\alpha\beta m_u m_d(21\alpha^2+9\alpha+28\beta^2+49\alpha\beta-42\beta)\big)+3{\cal F}^2_{\alpha\beta}\big(9\alpha\beta(m_u+m_d)(9\alpha\nonumber\\
&+&9\beta-11)+2m_c(\alpha+\beta-1)(9\alpha^2+12\beta^2+6\alpha-19\beta+21\alpha\beta)  \big)        \bigg{\}}\; , \\
\rho_{1^-\;,B}^{\langle G^2 \rangle\;,I}(s) &=& \frac{\langle g_s^2 G^2\rangle}{\pi^6} \int^{\alpha_{max}}_{\alpha_{min}} d \alpha \int^{1 - \alpha}_{\beta_{min}} d \beta \frac{{\cal F}_{\alpha \beta}^2(\alpha+\beta-1)}{3^3 \times 2^{14} \alpha^3 \beta^3} \bigg{\{}({\cal F}_{\alpha\beta}^2(1+9\alpha+9\beta)\nonumber\\
&+&4m_c{\cal F}_{\alpha\beta}\big(\alpha m_u (-33+7\alpha+7\beta)-2m_c(\alpha+\beta)(\alpha+\beta-1)-6\beta m_d(7\alpha+7\beta-6)\big)   \nonumber\\
&+&24m_c^2\big( 27\alpha\beta m_u m_d-m_c(\alpha+\beta)(\alpha+\beta-1)(\alpha m_u-6\beta m_d)   \big)\bigg{\}}\; , \\
 \rho_{1^-\;,B}^{\langle G^2 \rangle\;,II}(s) &=& \frac{g_s^2 \langle g_s^2 G^2\rangle}{\pi^8} \int^{\alpha_{max}}_{\alpha_{min}} d \alpha \int^{1 - \alpha}_{\beta_{min}} d \beta \bigg{\{}\frac{m_c(\alpha+\beta-1)^3}{3^4 \times 2^{16} \alpha^5 \beta^5} \bigg( 8m_c^5 m_u m_d \alpha \beta (\alpha+\beta)^2\nonumber\\
 &\times&(\alpha+\beta-1)(\alpha^2-\alpha\beta+\beta^2)+{\cal F}_{\alpha\beta}^3\big( 28\beta^3m_d(\alpha+\beta-2)+3\alpha^3 m_u (7\alpha\nonumber\\
 &+&7\beta+13) - 3m_c(\alpha+\beta)(7\alpha+7\beta-11)(\alpha^2-\alpha\beta +\beta^2) \big)-4m_c^3 (\alpha+\beta)\nonumber\\
 &\times&{\cal F}_{\alpha\beta}\big( m_c(\alpha+\beta)(\alpha+\beta-1)(3\alpha m_u+4\beta m_d)(\alpha^2-\alpha \beta+\beta^2)-m_u m_d \alpha \beta(3\alpha^3\nonumber\\
 &+&3\beta^3-5\alpha^2-5\beta^2+6\alpha^2\beta+6\alpha\beta^2-\alpha\beta) \big)+m_c{\cal F}_{\alpha\beta}^2\big(-6\alpha\beta m_u m_d(\alpha^2+\beta^2)(5\alpha\nonumber\\
 &+&5\beta-3) +18 m_c^2(\alpha+\beta-1)(\alpha+\beta)^2(\alpha^2-\alpha\beta+\beta^2) - m_c (\alpha+\beta)(4\beta m_d(12\alpha^2\nonumber\\
 &-&5\alpha^3+6\alpha\beta^2-12\alpha\beta+\beta^3+6\beta^2)+3\alpha m_u(\alpha^3+15\alpha\beta-5\beta^3-15\beta^2-21\alpha^2\nonumber\\
 &+&6\alpha^2\beta))    \big) \bigg)-\frac{{\cal F}_{\alpha \beta}(\alpha+\beta-1)}{3^5 \times 2^{21} \alpha^4 \beta^4}\bigg( 48 m_c^4 m_u m_d \alpha \beta (\alpha+\beta)(\alpha+\beta-1)^3 -12m_c^2\nonumber\\
 &\times&{\cal F}_{\alpha\beta}(\alpha+\beta-1)^2\big(\alpha\beta m_u m_d(-3+5\alpha+5\beta )+m_c(\alpha+\beta)(\alpha m_u (5\alpha-31\beta-1)\nonumber\\
 &-&2 \beta m_d(63\alpha-5\beta+1) )  \big)-2m_c {\cal F}^2_{\alpha\beta} (\alpha+\beta-1)\big(-\alpha m_u(35\alpha^2 - 217\beta^2\nonumber\\
 &-&182\alpha\beta +38\alpha+1478\beta-13 ) -2\beta m_d (35\beta^2 -441\alpha^2-406\alpha\beta -34\beta\nonumber\\
&+&772\alpha+5  )+4m_c (\alpha+\beta)(139\alpha^2 -21\beta^2 +106\alpha\beta -138\alpha +22\beta-1) \big)\nonumber\\
&+&{\cal F}^3_{\alpha\beta}\big( 1251\alpha^3+7\alpha^2(315\beta-419)-(\beta-1)(7+50\beta+189\beta^2)\nonumber\\
&+&\alpha(1675-2698\beta+765\beta^2)    \big) \bigg)\bigg{\}}
 \; , \\
\rho_{1^-\;,B}^{\langle \bar{q} G q \rangle}(s) &=& \frac{\langle g_s \bar{q} \sigma \cdot G q \rangle}{\pi^6} \int_{\alpha_{min}}^{\alpha_{max}}  \int_{\beta_{min}}^{1 - \alpha} d \beta  \frac{g_s^2 {\cal F}_{\alpha\beta}}{3^4 \times 2^{15} \alpha^3 \beta^4}\bigg{\{} -m_c(\alpha+\beta-1)\bigg( {\cal F}_{\alpha\beta}^2\big( 16\alpha\nonumber\\
&\times&(7\alpha^2+\alpha-8) +\beta(210\alpha^2+21\alpha-1)+77\alpha\beta^2+20\beta^2-21\beta^3 \big) -8\beta m_c^3(\alpha\nonumber\\
&+&\beta)(\alpha+\beta-1)^2 (\alpha m_u-16\alpha m_d+\beta m_d) -m_c{\cal F}_{\alpha\beta}(\alpha+\beta-1)\big(\beta(10\alpha+10\beta\nonumber\\
&-&7)(\alpha m_u-16\alpha m_d+\beta m_d)+6m_c(\alpha+\beta)(16\alpha^2-16\alpha-3\beta^2+14\alpha\beta+\beta)      \big)  \bigg) \nonumber\\
&-&48\beta\bigg( 8m_c^3\alpha\beta(\alpha+\beta)(\alpha+\beta-1) \big( 3m_c(m_u+m_d)(\alpha+\beta-1)+m_um_d(3\alpha\nonumber\\
&+&4\beta) \big)+{\cal F}_{\alpha\beta}^2 \big( 6\alpha\beta(m_u+m_d)(-8+7\alpha+7\beta)+m_c(\alpha+\beta-1)(21\alpha^2+9\alpha\nonumber\\
&+&49\alpha\beta-42\beta+28\beta^2)    \big) -2m_c {\cal F}_{\alpha\beta}\big( 3m_c^2(\alpha+\beta)(\alpha+\beta-1)^2(3\alpha+4\beta) \nonumber\\
&+&3m_c(m_u+m_d)\alpha\beta(\alpha+\beta-1)(11\alpha+11\beta-4) +\alpha\beta m_u m_d (-27\beta\nonumber\\
&+&5(\alpha+\beta)(3\alpha+4\beta))\big)\bigg)     \bigg{\}}\; ,\\
\rho_{1^-\;,B}^{\langle \bar{q} q\rangle^2}(s) &=& \frac{\langle \bar{q} q\rangle^2}{\pi^4} \int_{\alpha_{min}}^{\alpha_{max}} d \alpha \bigg{\{}\int_{\beta_{min}}^{1 - \alpha} d \beta \big[ -\frac{g_s^2 {\cal F}_{\alpha\beta} }{3^3 \times 2^{8} \alpha^2\beta^2} \big( 4m_c^2(\alpha+\beta)(9\alpha\beta m_u m_d +4m_c^2(\alpha\nonumber\\
&+&\beta-1)^2+2m_c(\alpha+\beta-1)(3\alpha m_d+4\beta m_u))+{\cal F}_{\alpha\beta}\big(54m_cm_u\beta-45m_um_d\alpha\beta\nonumber\\
&-&4m_c^2(\alpha+\beta-1)(5\alpha+5\beta-4)-10m_c(\alpha+\beta)(3\alpha m_d+4\beta m_u)  \big)  \big] \nonumber\\
&-&\frac{g_s^2 {\cal H}_\alpha^2m_u m_d}{3\times 2^8\alpha(\alpha-1)}\bigg{\}}\; , \\
\rho_{1^-\;,B}^{\langle G^3 \rangle\;,I}(s) &=& \frac{\langle g_s^3 G^3 \rangle}{\pi^6} \int_{\alpha_{min}}^{\alpha_{max}} d\alpha \int_{\beta_{min}}^{1 - \alpha} d \beta  \frac{1}{3^3 \times 2^{14} \alpha^3 \beta^3} \bigg{\{} 1152m_c^4m_dm_u\alpha^2\beta^2(\alpha+\beta)(\alpha\nonumber\\
&+&\beta-1) +\beta {\cal F}_{\alpha\beta}^3\big(- 357\alpha^2+4\alpha(62-91\beta) +\beta(4-7\beta)+3 \big)+12 m_c^2{\cal F}_{\alpha\beta}\nonumber\\ 
&\times&\big( - m_c  (\alpha+\beta)(\alpha+\beta-1)(\alpha m_u(2\alpha^2-2\alpha-\beta^2+\beta+47\alpha\beta) +2\beta m_d(3\beta^2\nonumber\\
&-&3\beta-2\alpha^2+2\alpha-3\alpha\beta)) -3\alpha\beta m_u m_d (9\beta^2-9\beta+48\alpha^2\beta-6\alpha^2+6\alpha\nonumber\\
&-&13\alpha\beta+48\alpha\beta^2)  \big)  +3m_c{\cal F}^2_{\alpha\beta}\big( 2\beta m_c(\alpha+\beta)(\alpha+\beta-1)(51\alpha+\beta-1)  \nonumber\\
&+&2\beta m_d (-10\alpha^3+18\alpha^2-25\alpha^2\beta+3\beta(\beta-1)(5\beta-4)+9\alpha\beta-8\alpha)\nonumber\\
&+&\alpha m_u(10\alpha^3-\beta(\beta-1)(5\beta-31) +245\alpha^2\beta\nonumber\\
&-&36\alpha^2+26\alpha-108\alpha\beta+230\alpha\beta^2   )    \big) \bigg{\}} \; , \\
\rho_{1^-\;,B}^{\langle G^3 \rangle\;,II}(s) &=& \frac{\langle g_s^3 G^3 \rangle}{\pi^8} \int_{\alpha_{min}}^{\alpha_{max}} d\alpha \int_{\beta_{min}}^{1 - \alpha} d \beta  \frac{g_s^2 (\alpha+\beta-1)^3}{3^4 \times 2^{18} \alpha^5 \beta^5} \bigg{\{} 3{\cal F}_{\alpha\beta}^3 (\alpha+\beta)(7\alpha+7\beta\nonumber\\
&-&11)(\alpha^2+\beta^2-\alpha\beta) -m_c {\cal F}^2_{\alpha\beta}\big(15m_u\alpha(\alpha+\beta+3) (6\alpha^3+\beta^3) +4\beta m_d\nonumber\\
&\times&(5\alpha+5\beta-12) (\alpha^3+6\beta^3)-18m_c(4\alpha^5+\alpha^3\beta-\alpha^3\beta^2-\alpha^2\beta^3-8\beta^4\nonumber\\
&+&4\beta^5-8\alpha^4+3\alpha^4\beta+\alpha\beta^3+3\alpha\beta^4) \big)-4m_c^2{\cal F}_{\alpha\beta}\big( -6m_um_d\alpha\beta(\alpha\nonumber\\
&+&\beta)(3\alpha+3\beta-1)(\alpha^2-\alpha\beta+\beta^2)+18m_c^2(\alpha+\beta-1)(\alpha+\beta)(\alpha^4+\beta^4)\nonumber\\
&-&4\beta m_cm_d(9\alpha^4-2\alpha^5-\alpha^3\beta-\alpha^4\beta+\alpha^3\beta^2-6\alpha\beta^3+6\alpha^2\beta^3+4\beta^4\nonumber\\
&+&9\alpha\beta^4+3\beta^5  ) +3\alpha m_um_c(3\alpha^5-6\alpha^3\beta+6\alpha^3\beta^2+\alpha^2\beta^3-\alpha\beta^3-\alpha\beta^4\nonumber\\
&-&18\beta^4-2\beta^5-23\alpha^4+9\alpha^4\beta) \big)+8m_c^4\big( m_c(\alpha+\beta-1)(\alpha+\beta)(\alpha^4+\beta^4)(3\alpha m_u\nonumber\\
&+&4\beta m_d)-\alpha\beta m_u m_d ( 5\alpha^5+6\alpha^3\beta(\beta-1)+6\alpha^2\beta^3 +\beta^4(5\beta-7) +\alpha^4(11\beta\nonumber\\
&-&7)+\alpha\beta^3(11\beta-6) )       \big)   \bigg{\}} \; , \\
\rho_{1^-\;,B}^{\langle \bar{q} q\rangle\langle \bar{q} G q \rangle}(s) &=& \frac{g_s^2\langle \bar{q} q\rangle \langle g_s \bar{q} \sigma \cdot G q \rangle}{\pi^4} \int_{\alpha_{min}}^{\alpha_{max}} d \alpha \bigg{\{}\int_{\beta_{min}}^{1 - \alpha} d \beta \big[ -\frac{ m_c }{3^4 \times 2^{12} \alpha^2\beta^2}\big( 2m_c^2(\alpha+\beta\nonumber\\
&-&1)(\alpha+\beta) (2m_u\alpha\beta+\alpha m_d(\beta-48\alpha)-2m_c(15\alpha-\beta)(\alpha+\beta-1))  \big)\nonumber\\
&+&{\cal F}_{\alpha\beta}(m_u\alpha\beta(5-6\alpha-6\beta) +2m_c(15\alpha-\beta)(\alpha+\beta-1) (3\alpha+3\beta-2)\nonumber\\
&+&\alpha m_d(48\alpha-\beta)(3\alpha+3\beta+2))  \big] +\frac{1}{3^4\times 2^8\alpha(\alpha-1)}\big[108{\cal H}_\alpha \alpha(\alpha\nonumber\\
&-&1) m_u m_d -54\alpha (\alpha-1)m_c^2 m_u m_d -5m_c {\cal H}_\alpha (-7m_u + 7\alpha m_u+15\alpha m_d) \big]\bigg{\}}\; .
\end{eqnarray}

For the current showned in Eq. (\ref{current-1-C}), we obtain the spectral densities as follows:
\begin{eqnarray}
\rho^{pert}_{1^-\;,C} (s) &=& \int^{\alpha_{max}}_{\alpha_{min}} d \alpha \int^{1 - \alpha}_{\beta_{min}} d \beta -\frac{g_s^2{\cal F}^3_{\alpha \beta} (\alpha + \beta - 1)^3}{5\times 3^3 \times 2^{16}\pi^8 \alpha^5 \beta^5} \bigg{\{} 160\alpha\beta m_c^4 m_u m_d (\alpha+\beta)(\alpha+\beta-1) \nonumber\\
&+&3 {\cal F}^3_{\alpha\beta}(13\alpha+13\beta-17) +2m_c {\cal F}^2_{\alpha\beta}\big( 72\beta m_d -44\alpha\beta m_d-44 \beta^2 m_d\nonumber\\
&-& 18m_c (\alpha+\beta)(\alpha+\beta-1)+3\alpha m_u(11\alpha+11\beta+9) \big)-20m_c^2{\cal F}_{\alpha\beta}\big( \alpha\beta m_u m_d(-7\nonumber\\
&+&9\alpha+9\beta)+m_c(\alpha+\beta)(\alpha+\beta-1)(3\alpha m_u -4\beta m_d) \big)  \bigg{\}}\; ,  \\
\rho_{1^-\;,C}^{\langle \bar{q} q \rangle}(s) &=& \frac{\langle \bar{q} q \rangle}{\pi^6}
\int^{\alpha_{max}}_{\alpha_{min}} d \alpha \int^{1 - \alpha}_{\beta_{min}} d \beta
\frac{ g_s^2 {\cal F}_{\alpha \beta }^2(\alpha+\beta-1)}{3^4 \times 2^{12} \alpha^4 \beta^4} \bigg{\{}24m_c^3\alpha\beta(\alpha+\beta)(\alpha+\beta-1)\nonumber\\
&\times&\big(4m_c(m_u+m_d)(\alpha+\beta-1)+3m_um_d(3\alpha-4\beta)\big)-4m_c{\cal F}_{\alpha\beta}\big( 4m_c^2(\alpha+\beta)\nonumber\\
&\times&(3\alpha-4\beta)(\alpha+\beta-1)^2 -2\alpha\beta (\alpha+\beta-1)(13\alpha+13\beta+11)m_c(m_u+m_d) \nonumber\\
&+&3\alpha\beta m_u m_d(21\alpha^2+9\alpha-28\beta^2-7\alpha\beta+42\beta)\big)-3{\cal F}^2_{\alpha\beta}\big(9\alpha\beta(m_u+m_d)(9\alpha\nonumber\\
&+&9\beta-11)-2m_c(\alpha+\beta-1)(9\alpha^2-12\beta^2+6\alpha+19\beta-3\alpha\beta)  \big)        \bigg{\}}\; , \\
\rho_{1^-\;,C}^{\langle G^2 \rangle\;,I}(s) &=& \frac{\langle g_s^2 G^2\rangle}{\pi^6} \int^{\alpha_{max}}_{\alpha_{min}} d \alpha \int^{1 - \alpha}_{\beta_{min}} d \beta \frac{{\cal F}_{\alpha \beta}^2(\alpha+\beta-1)}{3^3 \times 2^{14} \alpha^3 \beta^3} \bigg{\{}({\cal F}_{\alpha\beta}^2(1+9\alpha+9\beta)\nonumber\\
&+&4m_c{\cal F}_{\alpha\beta}\big(\alpha m_u (33-7\alpha-7\beta)-2m_c(\alpha+\beta)(\alpha+\beta-1)-6\beta m_d(7\alpha+7\beta-6)\big)   \nonumber\\
&-&24m_c^2\big( 27\alpha\beta m_u m_d-m_c(\alpha+\beta)(\alpha+\beta-1)(\alpha m_u+6\beta m_d)   \big)\bigg{\}}\; , \\
 \rho_{1^-\;,C}^{\langle G^2 \rangle\;,II}(s) &=& \frac{g_s^2 \langle g_s^2 G^2\rangle}{\pi^8} \int^{\alpha_{max}}_{\alpha_{min}} d \alpha \int^{1 - \alpha}_{\beta_{min}} d \beta \bigg{\{}-\frac{m_c(\alpha+\beta-1)^3}{3^4 \times 2^{16} \alpha^5 \beta^5} \bigg( 8m_c^5 m_u m_d \alpha \beta (\alpha+\beta)^2\nonumber\\
 &\times&(\alpha+\beta-1)(\alpha^2-\alpha\beta+\beta^2)+{\cal F}_{\alpha\beta}^3\big(- 28\beta^3m_d(\alpha+\beta-2)+3\alpha^3 m_u (7\alpha\nonumber\\
 &+&7\beta+13) + 3m_c(\alpha+\beta)(7\alpha+7\beta-11)(\alpha^2-\alpha\beta +\beta^2) \big)-4m_c^3 (\alpha+\beta)\nonumber\\
 &\times&{\cal F}_{\alpha\beta}\big( m_c(\alpha+\beta)(\alpha+\beta-1)(3\alpha m_u-4\beta m_d)(\alpha^2-\alpha \beta+\beta^2)-m_u m_d \alpha \beta(3\alpha^3\nonumber\\
 &+&3\beta^3-5\alpha^2-5\beta^2+6\alpha^2\beta+6\alpha\beta^2-\alpha\beta) \big)+m_c{\cal F}_{\alpha\beta}^2\big(-6\alpha\beta m_u m_d(\alpha^2+\beta^2)(5\alpha\nonumber\\
 &+&5\beta-3) -18 m_c^2(\alpha+\beta-1)(\alpha+\beta)^2(\alpha^2-\alpha\beta+\beta^2) + m_c (\alpha+\beta)(4\beta m_d(12\alpha^2\nonumber\\
 &-&5\alpha^3+6\alpha\beta^2-12\alpha\beta+\beta^3+6\beta^2)+3\alpha m_u(\alpha^3+15\alpha\beta-5\beta^3-15\beta^2-21\alpha^2\nonumber\\
 &+&6\alpha^2\beta))    \big) \bigg)-\frac{{\cal F}_{\alpha \beta}(\alpha+\beta-1)}{3^5 \times 2^{21} \alpha^4 \beta^4}\bigg( 48 m_c^4 m_u m_d \alpha \beta (\alpha+\beta)(\alpha+\beta-1)^3 -12m_c^2\nonumber\\
 &\times&{\cal F}_{\alpha\beta}(\alpha+\beta-1)^2\big(\alpha\beta m_u m_d(-3+5\alpha+5\beta )+m_c(\alpha+\beta)(\alpha m_u (5\alpha-31\beta-1)\nonumber\\
 &+&2 \beta m_d(63\alpha-5\beta+1) )  \big)+2m_c {\cal F}^2_{\alpha\beta} (\alpha+\beta-1)\big(\alpha m_u(35\alpha^2 - 217\beta^2\nonumber\\
 &-&182\alpha\beta +38\alpha+1478\beta-13 ) -2\beta m_d (35\beta^2 -441\alpha^2-406\alpha\beta -34\beta\nonumber\\
&+&772\alpha+5  )+4m_c (\alpha+\beta)(139\alpha^2 -21\beta^2 +106\alpha\beta -138\alpha +22\beta-1) \big)\nonumber\\
&+&{\cal F}^3_{\alpha\beta}\big( 1251\alpha^3+7\alpha^2(315\beta-419)-(\beta-1)(7+50\beta+189\beta^2)\nonumber\\
&+&\alpha(1675-2698\beta+765\beta^2)    \big) \bigg)\bigg{\}}
 \; , \\
\rho_{1^-\;,C}^{\langle \bar{q} G q \rangle}(s) &=& \frac{\langle g_s \bar{q} \sigma \cdot G q \rangle}{\pi^6} \int_{\alpha_{min}}^{\alpha_{max}}  \int_{\beta_{min}}^{1 - \alpha} d \beta  \frac{g_s^2 {\cal F}_{\alpha\beta}}{3^5 \times 2^{15} \alpha^3 \beta^4}\bigg{\{} -m_c(\alpha+\beta-1)\bigg( {\cal F}_{\alpha\beta}^2\big( 48\alpha\nonumber\\
&\times&(7\alpha^2+\alpha-8) +\beta(658\alpha^2+13\alpha+19)+371\alpha\beta^2-62\beta^2+49\beta^3 \big) -24\beta m_c^3(\alpha\nonumber\\
&+&\beta)(\alpha+\beta-1)^2 (\alpha m_u-16\alpha m_d+\beta m_d) -3m_c{\cal F}_{\alpha\beta}(\alpha+\beta-1)\big(-\beta(10\alpha+10\beta\nonumber\\
&-&7)(\alpha m_u-16\alpha m_d+\beta m_d)+2m_c(\alpha+\beta)(48\alpha^2-48\alpha+73\beta^2+46\alpha\beta-\beta)      \big)  \bigg) \nonumber\\
&-&144\beta\bigg( 8m_c^3\alpha\beta(\alpha+\beta)(\alpha+\beta-1) \big( 3m_c(m_u+m_d)(\alpha+\beta-1)+m_um_d(3\alpha\nonumber\\
&-&4\beta) \big)+{\cal F}_{\alpha\beta}^2 \big( 6\alpha\beta(m_u+m_d)(8-7\alpha-7\beta)+m_c(\alpha+\beta-1)(21\alpha^2+9\alpha\nonumber\\
&-&7\alpha\beta+42\beta-28\beta^2)    \big) -2m_c {\cal F}_{\alpha\beta}\big( 3m_c^2(\alpha+\beta)(\alpha+\beta-1)^2(3\alpha-4\beta) \nonumber\\
&-&3m_c(m_u+m_d)\alpha\beta(\alpha+\beta-1)(\alpha+\beta+4) -\alpha\beta m_u m_d (-27\beta\nonumber\\
&+&5(\alpha+\beta)(-3\alpha+4\beta))\big)\bigg)     \bigg{\}}\; ,\\
\rho_{1^-\;,C}^{\langle \bar{q} q\rangle^2}(s) &=& \frac{\langle \bar{q} q\rangle^2}{\pi^4} \int_{\alpha_{min}}^{\alpha_{max}} d \alpha \bigg{\{}\int_{\beta_{min}}^{1 - \alpha} d \beta \big[ \frac{g_s^2 {\cal F}_{\alpha\beta} }{3^3 \times 2^{8} \alpha^2\beta^2} \big( 4m_c^2(\alpha+\beta)(9\alpha\beta m_u m_d -4m_c^2(\alpha\nonumber\\
&+&\beta-1)^2-2m_c(\alpha+\beta-1)(3\alpha m_d-4\beta m_u))+{\cal F}_{\alpha\beta}\big(54m_cm_u\beta-45m_um_d\alpha\beta\nonumber\\
&+&4m_c^2(\alpha+\beta-1)(5\alpha+5\beta-4)+10m_c(\alpha+\beta)(3\alpha m_d-4\beta m_u)  \big)  \big] \nonumber\\
&+&\frac{g_s^2 {\cal H}_\alpha^2m_u m_d}{3\times 2^8\alpha(\alpha-1)}\bigg{\}}\; , \\
\rho_{1^-\;,C}^{\langle G^3 \rangle\;,I}(s) &=& \frac{\langle g_s^3 G^3 \rangle}{\pi^6} \int_{\alpha_{min}}^{\alpha_{max}} d\alpha \int_{\beta_{min}}^{1 - \alpha} d \beta  \frac{1}{3^3 \times 2^{12} \alpha^3 \beta^2} \bigg{\{}  -6m_c^2 (3\alpha-\beta\nonumber\\
&+&1)(\alpha+\beta)(\alpha+\beta-1)+{\cal F}_{\alpha\beta}(3+21\alpha^2-7\beta^2+4\beta-22\alpha+14\alpha\beta) \bigg{\}} \; , \\
\rho_{1^-\;,C}^{\langle G^3 \rangle\;,II}(s) &=& \frac{\langle g_s^3 G^3 \rangle}{\pi^8} \int_{\alpha_{min}}^{\alpha_{max}} d\alpha \int_{\beta_{min}}^{1 - \alpha} d \beta  \frac{g_s^2 (\alpha+\beta-1)^3}{3^4 \times 2^{18} \alpha^5 \beta^5} \bigg{\{} -3{\cal F}_{\alpha\beta}^3 (\alpha+\beta)(7\alpha+7\beta\nonumber\\
&-&11)(\alpha^2+\beta^2-\alpha\beta) +72{\cal F}_{\alpha\beta}m_c^4(\alpha^4+\beta^4)(\alpha+\beta)(\alpha+\beta-1)\nonumber\\
&-&18{\cal F}_{\alpha\beta}^2 m_c^2\big(4\alpha^5-\alpha^3\beta(\beta-1)-\alpha^2\beta^3+4\beta^4(\beta-2) \nonumber\\
&+&\alpha^4(3\beta-8) +\alpha\beta^3(3\beta+1)\big) \bigg{\}} \; , \\
\rho_{1^-\;,C}^{\langle \bar{q} q\rangle\langle \bar{q} G q \rangle}(s) &=& \frac{g_s^2\langle \bar{q} q\rangle \langle g_s \bar{q} \sigma \cdot G q \rangle}{\pi^4} \int_{\alpha_{min}}^{\alpha_{max}} d \alpha \bigg{\{}\int_{\beta_{min}}^{1 - \alpha} d \beta \big[ -\frac{ m_c }{3^4 \times 2^{12} \alpha^2\beta^2}\big( 2m_c^2(\alpha+\beta\nonumber\\
&-&1)(\alpha+\beta) (-2m_u\alpha\beta+\alpha m_d(\beta-48\alpha)-2m_c(15\alpha-\beta)(\alpha+\beta-1))  \big)\nonumber\\
&+&{\cal F}_{\alpha\beta}(m_u\alpha\beta(6\alpha+6\beta-5) +2m_c(15\alpha-\beta)(\alpha+\beta-1) (3\alpha+3\beta-2)\nonumber\\
&+&\alpha m_d(48\alpha-\beta)(3\alpha+3\beta+2))  \big] +\frac{1}{3^4\times 2^8\alpha(\alpha-1)}\big[-108{\cal H}_\alpha \alpha(\alpha\nonumber\\
&-&1) m_u m_d +54\alpha (\alpha-1)m_c^2 m_u m_d -5m_c {\cal H}_\alpha (7m_u - 7\alpha m_u+15\alpha m_d) \big]\bigg{\}}\; .
\end{eqnarray}

For the current showned in Eq. (\ref{current-1-D}), we obtain the spectral densities as follows:
\begin{eqnarray}
\rho^{pert}_{1^-\;,D} (s) &=& \int^{\alpha_{max}}_{\alpha_{min}} d \alpha \int^{1 - \alpha}_{\beta_{min}} d \beta \frac{g_s^2{\cal F}^3_{\alpha \beta} (\alpha + \beta - 1)^3}{5\times 3^3 \times 2^{16}\pi^8 \alpha^5 \beta^5} \bigg{\{} 160\alpha\beta m_c^4 m_u m_d (\alpha+\beta)(\alpha+\beta-1) \nonumber\\
&+&3 {\cal F}^3_{\alpha\beta}(13\alpha+13\beta-17) -2m_c {\cal F}^2_{\alpha\beta}\big( 72\beta m_d -44\alpha\beta m_d-44 \beta^2 m_d\nonumber\\
&+& 18m_c (\alpha+\beta)(\alpha+\beta-1)+3\alpha m_u(11\alpha+11\beta+9) \big)+20m_c^2{\cal F}_{\alpha\beta}\big( \alpha\beta m_u m_d(7\nonumber\\
&-&9\alpha-9\beta)+m_c(\alpha+\beta)(\alpha+\beta-1)(3\alpha m_u -4\beta m_d) \big)  \bigg{\}}\; ,  \\
\rho_{1^-\;,D}^{\langle \bar{q} q \rangle}(s) &=& \frac{\langle \bar{q} q \rangle}{\pi^6}
\int^{\alpha_{max}}_{\alpha_{min}} d \alpha \int^{1 - \alpha}_{\beta_{min}} d \beta
\frac{ g_s^2 {\cal F}_{\alpha \beta }^2(\alpha+\beta-1)}{3^4 \times 2^{12} \alpha^4 \beta^4} \bigg{\{}24m_c^3\alpha\beta(\alpha+\beta)(\alpha+\beta-1)\nonumber\\
&\times&\big(4m_c(m_u+m_d)(1-\alpha-\beta)+3m_um_d(3\alpha-4\beta)\big)-4m_c{\cal F}_{\alpha\beta}\big( 4m_c^2(\alpha+\beta)\nonumber\\
&\times&(3\alpha-4\beta)(\alpha+\beta-1)^2 +2\alpha\beta (\alpha+\beta-1)(13\alpha+13\beta+11)m_c(m_u+m_d) \nonumber\\
&+&3\alpha\beta m_u m_d(21\alpha^2+9\alpha-28\beta^2-7\alpha\beta+42\beta)\big)+3{\cal F}^2_{\alpha\beta}\big(9\alpha\beta(m_u+m_d)(9\alpha\nonumber\\
&+&9\beta-11)+2m_c(\alpha+\beta-1)(9\alpha^2-12\beta^2+6\alpha+19\beta-3\alpha\beta)  \big)        \bigg{\}}\; , \\
\rho_{1^-\;,D}^{\langle G^2 \rangle\;,I}(s) &=& \frac{\langle g_s^2 G^2\rangle}{\pi^6} \int^{\alpha_{max}}_{\alpha_{min}} d \alpha \int^{1 - \alpha}_{\beta_{min}} d \beta -\frac{{\cal F}_{\alpha \beta}^2(\alpha+\beta-1)}{3^3 \times 2^{14} \alpha^3 \beta^3} \bigg{\{}({\cal F}_{\alpha\beta}^2(1+9\alpha+9\beta)\nonumber\\
&+&4m_c{\cal F}_{\alpha\beta}\big(\alpha m_u (7\alpha+7\beta-33)-2m_c(\alpha+\beta)(\alpha+\beta-1)+6\beta m_d(7\alpha+7\beta-6)\big)   \nonumber\\
&-&24m_c^2\big( 27\alpha\beta m_u m_d+m_c(\alpha+\beta)(\alpha+\beta-1)(\alpha m_u+6\beta m_d)   \big)\bigg{\}}\; , \\
 \rho_{1^-\;,D}^{\langle G^2 \rangle\;,II}(s) &=& \frac{g_s^2 \langle g_s^2 G^2\rangle}{\pi^8} \int^{\alpha_{max}}_{\alpha_{min}} d \alpha \int^{1 - \alpha}_{\beta_{min}} d \beta \bigg{\{}\frac{m_c(\alpha+\beta-1)^3}{3^4 \times 2^{16} \alpha^5 \beta^5} \bigg( 8m_c^5 m_u m_d \alpha \beta (\alpha+\beta)^2\nonumber\\
 &\times&(\alpha+\beta-1)(\alpha^2-\alpha\beta+\beta^2)+{\cal F}_{\alpha\beta}^3\big(28\beta^3m_d(\alpha+\beta-2)-3\alpha^3 m_u (7\alpha\nonumber\\
 &+&7\beta+13) + 3m_c(\alpha+\beta)(7\alpha+7\beta-11)(\alpha^2-\alpha\beta +\beta^2) \big)+4m_c^3 (\alpha+\beta)\nonumber\\
 &\times&{\cal F}_{\alpha\beta}\big( m_c(\alpha+\beta)(\alpha+\beta-1)(3\alpha m_u-4\beta m_d)(\alpha^2-\alpha \beta+\beta^2)+m_u m_d \alpha \beta(3\alpha^3\nonumber\\
 &+&3\beta^3-5\alpha^2-5\beta^2+6\alpha^2\beta+6\alpha\beta^2-\alpha\beta) \big)-m_c{\cal F}_{\alpha\beta}^2\big(6\alpha\beta m_u m_d(\alpha^2+\beta^2)(5\alpha\nonumber\\
 &+&5\beta-3) +18 m_c^2(\alpha+\beta-1)(\alpha+\beta)^2(\alpha^2-\alpha\beta+\beta^2) + m_c (\alpha+\beta)(4\beta m_d(12\alpha^2\nonumber\\
 &-&5\alpha^3+6\alpha\beta^2-12\alpha\beta+\beta^3+6\beta^2)-3\alpha m_u(\alpha^3+15\alpha\beta-5\beta^3-15\beta^2-21\alpha^2\nonumber\\
 &+&6\alpha^2\beta))    \big) \bigg)-\frac{{\cal F}_{\alpha \beta}(\alpha+\beta-1)}{3^5 \times 2^{21} \alpha^4 \beta^4}\bigg( -48 m_c^4 m_u m_d \alpha \beta (\alpha+\beta)(\alpha+\beta-1)^3 -12m_c^2\nonumber\\
 &\times&{\cal F}_{\alpha\beta}(\alpha+\beta-1)^2\big(\alpha\beta m_u m_d(3-5\alpha-5\beta )+m_c(\alpha+\beta)(\alpha m_u (5\alpha-31\beta-1)\nonumber\\
 &+&2 \beta m_d(63\alpha-5\beta+1) )  \big)-2m_c {\cal F}^2_{\alpha\beta} (\alpha+\beta-1)\big(\alpha m_u(-35\alpha^2 + 217\beta^2\nonumber\\
 &+&182\alpha\beta -38\alpha-1478\beta+13 ) +2\beta m_d (35\beta^2 -441\alpha^2-406\alpha\beta -34\beta\nonumber\\
&+&772\alpha+5  )+4m_c (\alpha+\beta)(139\alpha^2 -21\beta^2 +106\alpha\beta -138\alpha +22\beta-1) \big)\nonumber\\
&+&{\cal F}^3_{\alpha\beta}\big( 1251\alpha^3+7\alpha^2(315\beta-419)-(\beta-1)(7+50\beta+189\beta^2)\nonumber\\
&+&\alpha(1675-2698\beta+765\beta^2)    \big) \bigg)\bigg{\}}
 \; , \\
\rho_{1^-\;,D}^{\langle \bar{q} G q \rangle}(s) &=& \frac{\langle g_s \bar{q} \sigma \cdot G q \rangle}{\pi^6} \int_{\alpha_{min}}^{\alpha_{max}}  \int_{\beta_{min}}^{1 - \alpha} d \beta  \frac{g_s^2 {\cal F}_{\alpha\beta}}{3^5 \times 2^{15} \alpha^3 \beta^4}\bigg{\{} -m_c(\alpha+\beta-1)\bigg( {\cal F}_{\alpha\beta}^2\big( 48\alpha\nonumber\\
&\times&(7\alpha^2+\alpha-8) +\beta(658\alpha^2+13\alpha+19)+371\alpha\beta^2-62\beta^2+49\beta^3 \big) +24\beta m_c^3(\alpha\nonumber\\
&+&\beta)(\alpha+\beta-1)^2 (\alpha m_u-16\alpha m_d+\beta m_d) -3m_c{\cal F}_{\alpha\beta}(\alpha+\beta-1)\big(\beta(10\alpha+10\beta\nonumber\\
&-&7)(\alpha m_u-16\alpha m_d+\beta m_d)+2m_c(\alpha+\beta)(48\alpha^2-48\alpha+7\beta^2+46\alpha\beta-\beta)      \big)  \bigg) \nonumber\\
&+&144\beta\bigg( 8m_c^3\alpha\beta(\alpha+\beta)(\alpha+\beta-1) \big( 3m_c(m_u+m_d)(\alpha+\beta-1)+m_um_d(3\alpha\nonumber\\
&-&4\beta) \big)+{\cal F}_{\alpha\beta}^2 \big( 6\alpha\beta(m_u+m_d)(8-7\alpha-7\beta)-m_c(\alpha+\beta-1)(21\alpha^2+9\alpha\nonumber\\
&-&7\alpha\beta+42\beta-28\beta^2)    \big) +2m_c {\cal F}_{\alpha\beta}\big( 3m_c^2(\alpha+\beta)(\alpha+\beta-1)^2(3\alpha-4\beta) \nonumber\\
&+&3m_c(m_u+m_d)\alpha\beta(\alpha+\beta-1)(\alpha+\beta+4) +\alpha\beta m_u m_d (-27\beta\nonumber\\
&+&5(\alpha+\beta)(3\alpha-4\beta))\big)\bigg)     \bigg{\}}\; ,\\
\rho_{1^-\;,D}^{\langle \bar{q} q\rangle^2}(s) &=& \frac{\langle \bar{q} q\rangle^2}{\pi^4} \int_{\alpha_{min}}^{\alpha_{max}} d \alpha \bigg{\{}\int_{\beta_{min}}^{1 - \alpha} d \beta \big[ \frac{g_s^2 {\cal F}_{\alpha\beta} }{3^3 \times 2^{8} \alpha^2\beta^2} \big( 4m_c^2(\alpha+\beta)(-9\alpha\beta m_u m_d +4m_c^2(\alpha\nonumber\\
&+&\beta-1)^2-2m_c(\alpha+\beta-1)(3\alpha m_d-4\beta m_u))-{\cal F}_{\alpha\beta}\big(-54m_cm_u\beta-45m_um_d\alpha\beta\nonumber\\
&+&4m_c^2(\alpha+\beta-1)(5\alpha+5\beta-4)-10m_c(\alpha+\beta)(3\alpha m_d-4\beta m_u)  \big)  \big] \nonumber\\
&-&\frac{g_s^2 {\cal H}_\alpha^2m_u m_d}{3\times 2^8\alpha(\alpha-1)}\bigg{\}}\; , \\
\rho_{1^-\;,D}^{\langle G^3 \rangle\;,I}(s) &=& \frac{\langle g_s^3 G^3 \rangle}{\pi^6} \int_{\alpha_{min}}^{\alpha_{max}} d\alpha \int_{\beta_{min}}^{1 - \alpha} d \beta  \frac{1}{3^3 \times 2^{12} \alpha^3 \beta^2} \bigg{\{}  6m_c^2 (9\alpha+\beta\nonumber\\
&-&1)(\alpha+\beta)(\alpha+\beta-1)+{\cal F}_{\alpha\beta}(3-63\alpha^2-7\beta^2+4\beta+38\alpha-70\alpha\beta) \bigg{\}} \; , \\
\rho_{1^-\;,D}^{\langle G^3 \rangle\;,II}(s) &=& \frac{\langle g_s^3 G^3 \rangle}{\pi^8} \int_{\alpha_{min}}^{\alpha_{max}} d\alpha \int_{\beta_{min}}^{1 - \alpha} d \beta  \frac{g_s^2 (\alpha+\beta-1)^3}{3^4 \times 2^{18} \alpha^5 \beta^5} \bigg{\{} 3{\cal F}_{\alpha\beta}^3 (\alpha+\beta)(7\alpha+7\beta\nonumber\\
&-&11)(\alpha^2+\beta^2-\alpha\beta) -72{\cal F}_{\alpha\beta}m_c^4(\alpha^4+\beta^4)(\alpha+\beta)(\alpha+\beta-1)\nonumber\\
&+&18{\cal F}_{\alpha\beta}^2 m_c^2\big(4\alpha^5-\alpha^3\beta(\beta-1)-\alpha^2\beta^3+4\beta^4(\beta-2) \nonumber\\
&+&\alpha^4(3\beta-8) +\alpha\beta^3(3\beta+1)\big) \bigg{\}} \; , \\
\rho_{1^-\;,D}^{\langle \bar{q} q\rangle\langle \bar{q} G q \rangle}(s) &=& \frac{g_s^2\langle \bar{q} q\rangle \langle g_s \bar{q} \sigma \cdot G q \rangle}{\pi^4} \int_{\alpha_{min}}^{\alpha_{max}} d \alpha \bigg{\{}\int_{\beta_{min}}^{1 - \alpha} d \beta \big[ -\frac{ m_c }{3^4 \times 2^{12} \alpha^2\beta^2}\big( 2m_c^2(\alpha+\beta\nonumber\\
&-&1)(\alpha+\beta) (-2m_u\alpha\beta+\alpha m_d(\beta-48\alpha)+2m_c(15\alpha-\beta)(\alpha+\beta-1))  \big)\nonumber\\
&+&{\cal F}_{\alpha\beta}(m_u\alpha\beta(6\alpha+6\beta-5) -2m_c(15\alpha-\beta)(\alpha+\beta-1) (3\alpha+3\beta-2)\nonumber\\
&+&\alpha m_d(48\alpha-\beta)(3\alpha+3\beta+2))  \big] +\frac{1}{3^4\times 2^8\alpha(\alpha-1)}\big[108{\cal H}_\alpha \alpha(\alpha\nonumber\\
&-&1) m_u m_d -54\alpha (\alpha-1)m_c^2 m_u m_d -5m_c {\cal H}_\alpha (7m_u - 7\alpha m_u+15\alpha m_d) \big]\bigg{\}}\; .
\end{eqnarray}

For the currents showned in Eqs. (\ref{current-1-E})-(\ref{current-1-H}), considering the symmetries between $j_{1^-}^A$ and $j_{1^-}^E$, $j_{1^-}^B$ and $j_{1^-}^F$, $j_{1^-}^C$ and $j_{1^-}^G$, and $j_{1^-}^D$ and $j_{1^-}^H$, we obtain the spectral densities $\rho_{1^-}^A(s)=\rho_{1^-}^E(s)(m_u\to m_d, m_d\to m_u)$, $\rho_{1^-}^B(s)=\rho_{1^-}^F(s)(m_u\to m_d, m_d\to m_u)$, $\rho_{1^-}^C(s)=\rho_{1^-}^G(s)(m_u\to m_d, m_d\to m_u)$, and $\rho_{1^-}^D(s)=\rho_{1^-}^H(s)(m_u\to m_d, m_d\to m_u)$, respectively.

\subsection{The spectral densities for $1^+$ gluonic tetraquark states}

For the current shown in Eq. (\ref{current-1+A}),  we obtain the spectral densities as follows:
\begin{eqnarray}
\rho^{pert}_{1^+\;,A} (s) &=& \rho^{pert}_{1^-\;,A} (s) \; ,  \\
\rho_{1^+\;,A}^{\langle \bar{q} q \rangle}(s) &=& \rho_{1^-\;,A}^{\langle \bar{q} q \rangle}(s)\; , \\
\rho_{1^+\;,A}^{\langle G^2 \rangle\;,I}(s) &=&-\rho_{1^-\;,A}^{\langle G^2 \rangle\;,I}(s) \; , \\
 \rho_{1^+\;,A}^{\langle G^2 \rangle\;,II}(s) &=&  \rho_{1^-\;,A}^{\langle G^2 \rangle\;,II}(s)
 \; , \\
\rho_{1^+\;,A}^{\langle \bar{q} G q \rangle}(s) &=&\rho_{1^-\;,A}^{\langle \bar{q} G q \rangle}(s) \; ,\\
\rho_{1^+\;,A}^{\langle \bar{q} q\rangle^2}(s) &=& \rho_{1^-\;,A}^{\langle \bar{q} q\rangle^2}(s)\; , \\
\rho_{1^+\;,A}^{\langle G^3 \rangle\;,I}(s) &=& \frac{\langle g_s^3 G^3 \rangle}{\pi^6} \int_{\alpha_{min}}^{\alpha_{max}} d\alpha \int_{\beta_{min}}^{1 - \alpha} d \beta  \frac{1}{3^3 \times 2^{12} \alpha^3 \beta^2} \bigg{\{} -6m_c^2(3\alpha-\beta\nonumber\\
&+&1)(\alpha+\beta-1)(\alpha+\beta) +{\cal F}_{\alpha\beta}\big( 3+21\alpha^2\nonumber\\
&+&(4-7\beta)\beta+2\alpha(7\beta-11) \big) \bigg{\}} \; , \\
\rho_{1^+\;,A}^{\langle G^3 \rangle\;,II}(s) &=&\rho_{1^-\;,A}^{\langle G^3 \rangle\;,II}(s) \; , \\
\rho_{1^+\;,A}^{\langle \bar{q} q\rangle\langle \bar{q} G q \rangle}(s) &=& \rho_{1^-\;,A}^{\langle \bar{q} q\rangle\langle \bar{q} G q \rangle}(s)\; .
\end{eqnarray}

For the currents shown in Eq. (\ref{current-1+B}), we obtain the spectral densities as follows:
\begin{eqnarray}
\rho^{pert}_{1^+\;,B} (s) &=& \rho^{pert}_{1^-\;,B} (s) \; ,  \\
\rho_{1^+\;,B}^{\langle \bar{q} q \rangle}(s) &=&\rho_{1^-\;,B}^{\langle \bar{q} q \rangle}(s)\; , \\
\rho_{1^+\;,B}^{\langle G^2 \rangle\;,I}(s) &=& -\rho_{1^-\;,B}^{\langle G^2 \rangle\;,I}(s)\; , \\
 \rho_{1^+\;,B}^{\langle G^2 \rangle\;,II}(s) &=& \rho_{1^-\;,B}^{\langle G^2 \rangle\;,II}(s)
 \; , \\
\rho_{1^+\;,B}^{\langle \bar{q} G q \rangle}(s) &=& \rho_{1^-\;,B}^{\langle \bar{q} G q \rangle}(s)\; ,\\
\rho_{1^+\;,B}^{\langle \bar{q} q\rangle^2}(s) &=& \rho_{1^-\;,B}^{\langle \bar{q} q\rangle^2}(s)\; , \\
\rho_{1^-\;,B}^{\langle G^3 \rangle\;,I}(s) &=& \frac{\langle g_s^3 G^3 \rangle}{\pi^6} \int_{\alpha_{min}}^{\alpha_{max}} d\alpha \int_{\beta_{min}}^{1 - \alpha} d \beta  \frac{{\cal F}_{\alpha\beta}^2}{3^3 \times 2^{12} \alpha^3 \beta^2} \bigg{\{} 6m_c^2\nonumber\\
&\times&(\alpha+\beta)(\alpha+\beta-1)(9\alpha+\beta-1) +{\cal F}_{\alpha\beta}(3-63\alpha^2\nonumber\\
&+&38\alpha-70\alpha\beta+4\beta-7\beta^2)  \bigg{\}} \; , \\
\rho_{1^+\;,B}^{\langle G^3 \rangle\;,II}(s) &=&\rho_{1^-\;,B}^{\langle G^3 \rangle\;,II}(s) \; , \\
\rho_{1^+\;,B}^{\langle \bar{q} q\rangle\langle \bar{q} G q \rangle}(s) &=& \rho_{1^-\;,B}^{\langle \bar{q} q\rangle\langle \bar{q} G q \rangle}(s)\; .
\end{eqnarray}

For the current showned in Eq. (\ref{current-1+C}), we obtain the spectral densities as follows:
\begin{eqnarray}
\rho^{pert}_{1^+\;,C} (s) &=& \rho^{pert}_{1^-\;,C} (s)\; ,  \\
\rho_{1^+\;,C}^{\langle \bar{q} q \rangle}(s) &=& \rho_{1^-\;,C}^{\langle \bar{q} q \rangle}(s)\; , \\
\rho_{1^+\;,C}^{\langle G^2 \rangle\;,I}(s) &=& -\rho_{1^-\;,C}^{\langle G^2 \rangle\;,I}(s)\; , \\
 \rho_{1^+\;,C}^{\langle G^2 \rangle\;,II}(s) &=& \rho_{1^-\;,C}^{\langle G^2 \rangle\;,II}(s) 
 \; , \\
\rho_{1^+\;,C}^{\langle \bar{q} G q \rangle}(s) &=& \rho_{1^-\;,C}^{\langle \bar{q} G q \rangle}(s)\; ,\\
\rho_{1^+\;,C}^{\langle \bar{q} q\rangle^2}(s) &=& \rho_{1^-\;,C}^{\langle \bar{q} q\rangle^2}(s)\; , \\
\rho_{1^-\;,C}^{\langle G^3 \rangle\;,I}(s) &=& \frac{\langle g_s^3 G^3 \rangle}{\pi^6} \int_{\alpha_{min}}^{\alpha_{max}} d\alpha \int_{\beta_{min}}^{1 - \alpha} d \beta  \frac{1}{3^3 \times 2^{12} \alpha^3 \beta^2} \bigg{\{}  -6m_c^2 (51\alpha+\beta\nonumber\\
&-&1)(\alpha+\beta)(\alpha+\beta-1)+{\cal F}_{\alpha\beta}(357\alpha^2+7\beta^2\nonumber\\
&-&4\beta-248\alpha+364\alpha\beta-3) \bigg{\}} \; , \\
\rho_{1^+\;,C}^{\langle G^3 \rangle\;,II}(s) &=& \rho_{1^-\;,C}^{\langle G^3 \rangle\;,II}(s) \; , \\
\rho_{1^+\;,C}^{\langle \bar{q} q\rangle\langle \bar{q} G q \rangle}(s) &=&\rho_{1^-\;,C}^{\langle \bar{q} q\rangle\langle \bar{q} G q \rangle}(s)\; .
\end{eqnarray}

For the current showned in Eq. (\ref{current-1+D}), we obtain the spectral densities as follows:
\begin{eqnarray}
\rho^{pert}_{1^+\;,D} (s) &=&\rho^{pert}_{1^-\;,D} (s)\; ,  \\
\rho_{1^+\;,D}^{\langle \bar{q} q \rangle}(s) &=&\rho_{1^-\;,D}^{\langle \bar{q} q \rangle}(s)\; , \\
\rho_{1^+\;,D}^{\langle G^2 \rangle\;,I}(s) &=&-\rho_{1^-\;,D}^{\langle G^2 \rangle\;,I}(s)\; , \\
 \rho_{1^+\;,D}^{\langle G^2 \rangle\;,II}(s) &=& \rho_{1^-\;,D}^{\langle G^2 \rangle\;,II}(s)
 \; , \\
\rho_{1^+\;,D}^{\langle \bar{q} G q \rangle}(s) &=& \rho_{1^-\;,D}^{\langle \bar{q} G q \rangle}(s)\; ,\\
\rho_{1^+\;,D}^{\langle \bar{q} q\rangle^2}(s) &=& \rho_{1^-\;,D}^{\langle \bar{q} q\rangle^2}(s)\; , \\
\rho_{1^-\;,D}^{\langle G^3 \rangle\;,I}(s) &=& \frac{\langle g_s^3 G^3 \rangle}{\pi^6} \int_{\alpha_{min}}^{\alpha_{max}} d\alpha \int_{\beta_{min}}^{1 - \alpha} d \beta  \frac{1}{3^3 \times 2^{14} \alpha^3 \beta^2} \bigg{\{}  6m_c^2 (51\alpha+\beta\nonumber\\
&-&1)(\alpha+\beta)(\alpha+\beta-1)+{\cal F}_{\alpha\beta}(3-357\alpha^2\nonumber\\
&-&7\beta^2+4\beta+248\alpha-364\alpha\beta) \bigg{\}} \; , \\
\rho_{1^+\;,D}^{\langle G^3 \rangle\;,II}(s) &=& \rho_{1^-\;,D}^{\langle G^3 \rangle\;,II}(s) \; , \\
\rho_{1^+\;,D}^{\langle \bar{q} q\rangle\langle \bar{q} G q \rangle}(s) &=& \rho_{1^-\;,D}^{\langle \bar{q} q\rangle\langle \bar{q} G q \rangle}(s)\; .
\end{eqnarray}

For the currents showned in Eqs. (\ref{current-1+E})-(\ref{current-1+H}), considering the symmetries between $j_{1^+}^A$ and $j_{1^+}^E$, $j_{1^+}^B$ and $j_{1^+}^F$, $j_{1^+}^C$ and $j_{1^+}^G$, and $j_{1^+}^D$ and $j_{1^+}^H$, we obtain the spectral densities $\rho_{1^+}^A(s)=\rho_{1^+}^E(s)(m_u\to m_d, m_d\to m_u)$, $\rho_{1^+}^B(s)=\rho_{1^+}^F(s)(m_u\to m_d, m_d\to m_u)$, $\rho_{1^+}^C(s)=\rho_{1^+}^G(s)(m_u\to m_d, m_d\to m_u)$, and $\rho_{1^+}^D(s)=\rho_{1^+}^H(s)(m_u\to m_d, m_d\to m_u)$, respectively.

\section{The ratios $R^{OPE}$, $R^{PC}$, and the masses $m$ are plopted as functions of Borel parameter $M_B^2$}\label{App_B}

We display the figures of the $R^{OPE}$, $R^{PC}$, and the masses $m$ as functions of Borel parameter $M_B^2$ for hidden-charm and -bottom tetraquark hybrid states below. It should be noted that the differences between the numerical results for $j_{0^+}^C$ and $j_{0^+}^D$ are so tiny that we just plot the case of $j_{0^+}^C$. The same applies to the pairs $(j_{0^-}^C, j_{0^-}^D)$, $(j_{1^-}^B, j_{1^-}^F)$, $(j_{1^-}^C, j_{1^-}^G)$, and $(j_{1^+}^C, j_{1^+}^G)$. 

\begin{figure}
\includegraphics[width=6.8cm]{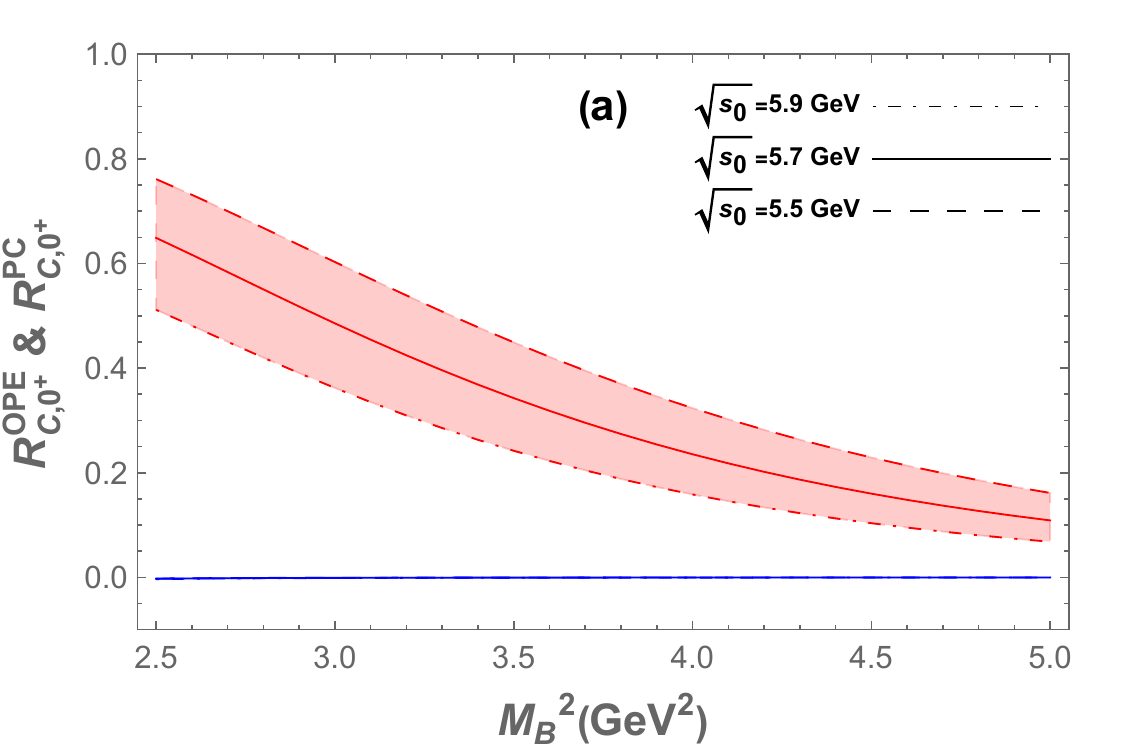}
\includegraphics[width=6.8cm]{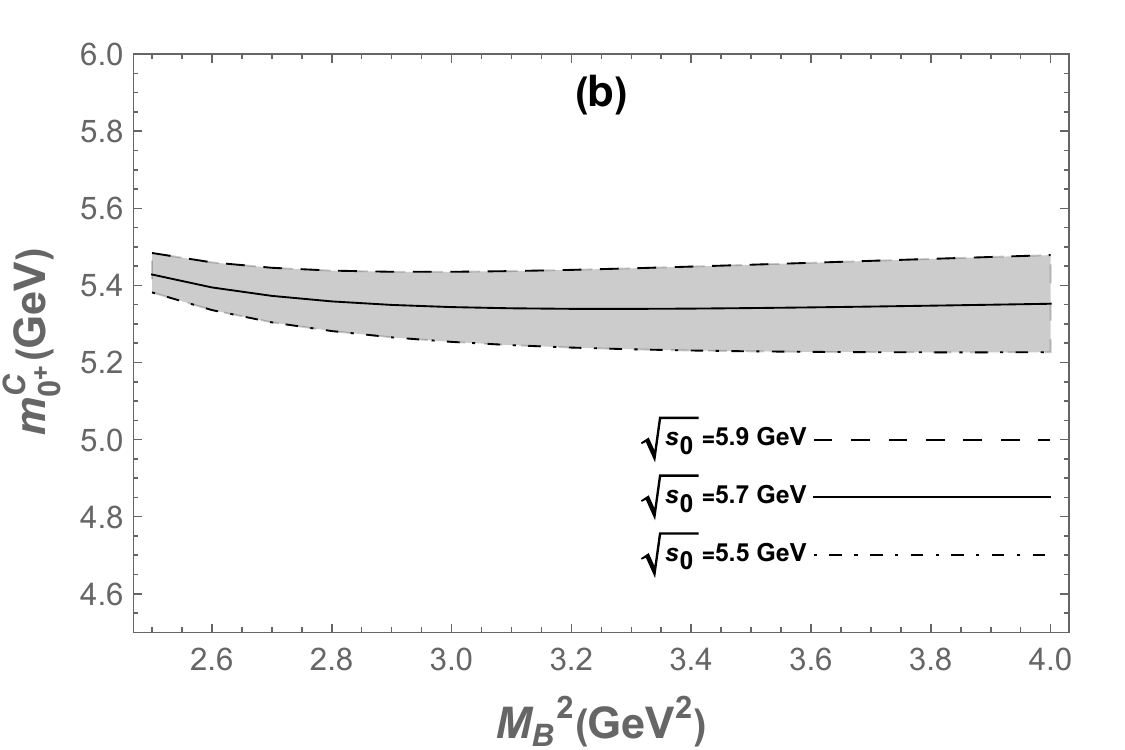}
\caption{Similar captions as in Fig.~\ref{fig0+B}, but for the current in Eq.~(\ref{current-0+C}).} \label{fig0+C}
\end{figure}

\begin{figure}
\includegraphics[width=6.8cm]{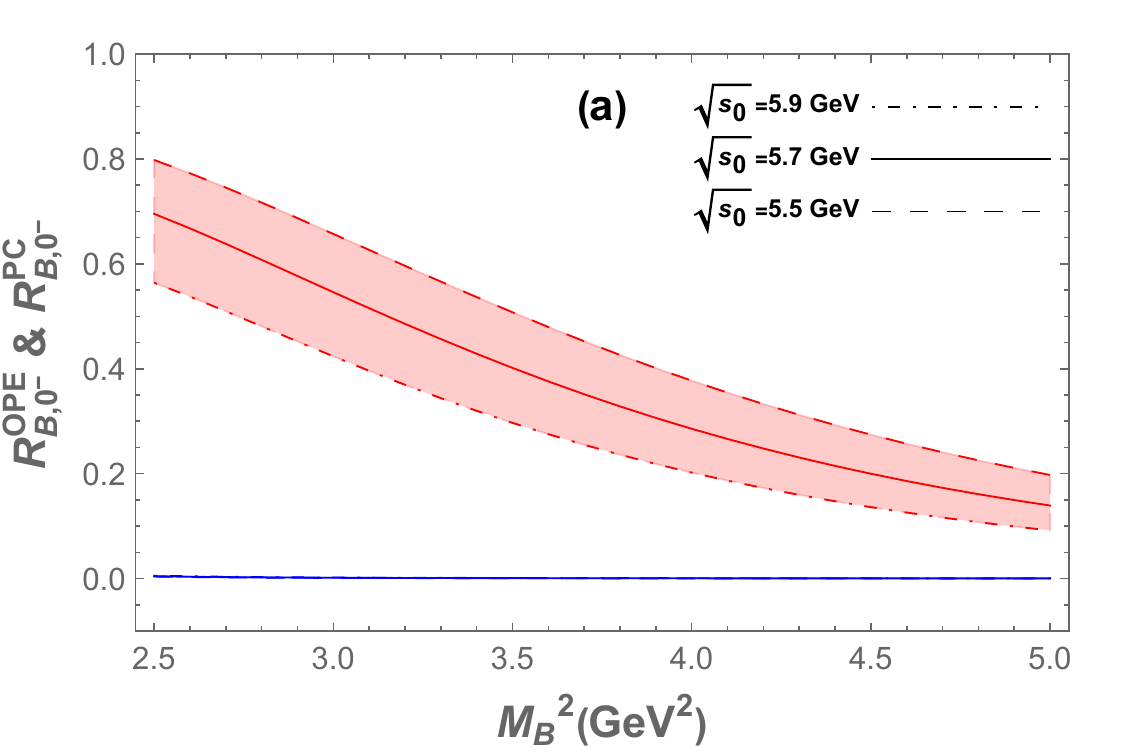}
\includegraphics[width=6.8cm]{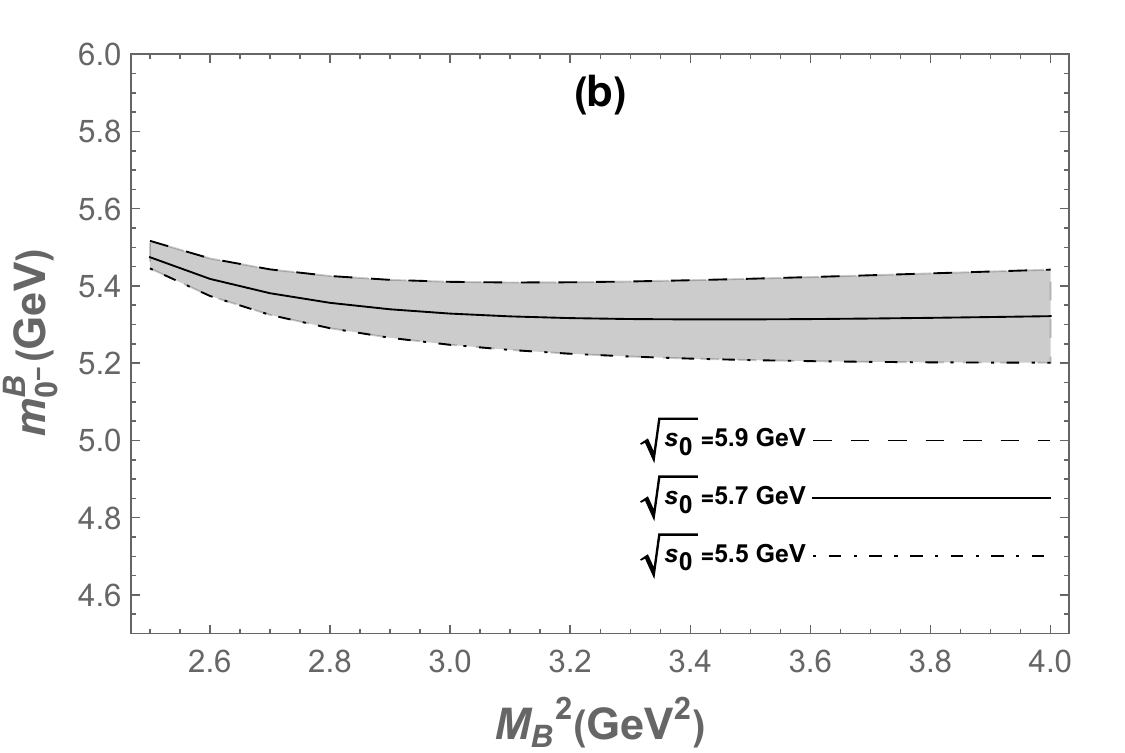}
\caption{Similar captions as in Fig.~\ref{fig0+B}, but for the current in Eq.~(\ref{current-0-B}).} \label{fig0-B}
\end{figure}

\begin{figure}
\includegraphics[width=6.8cm]{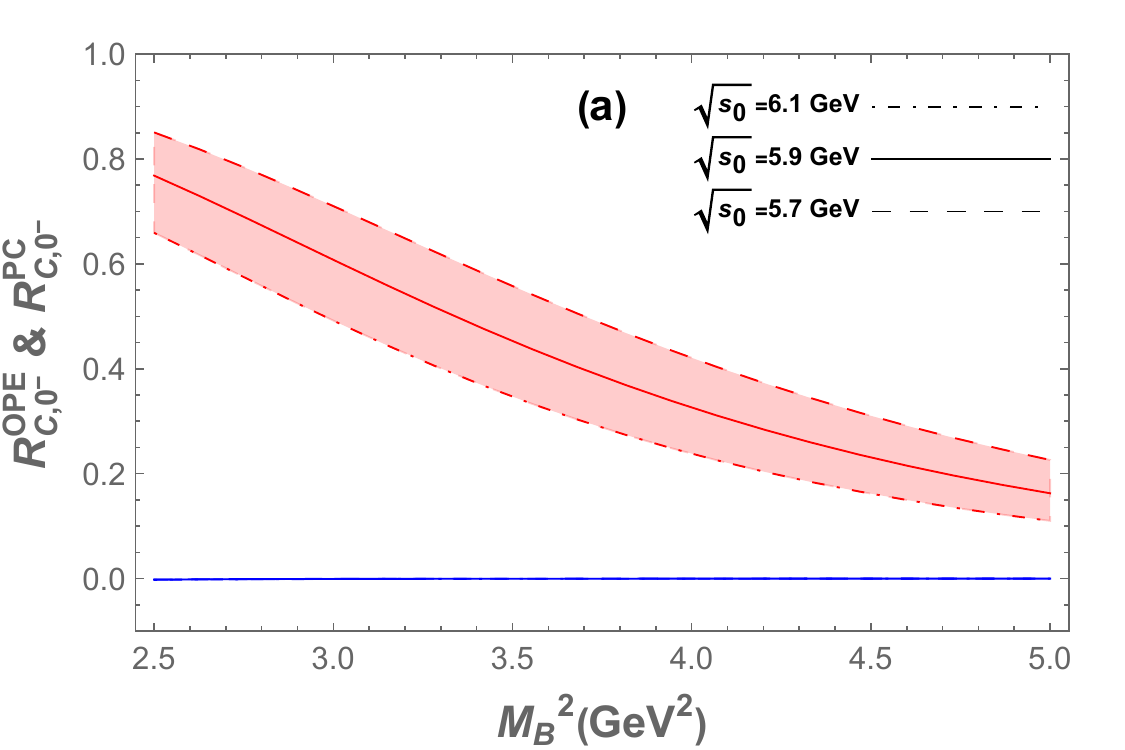}
\includegraphics[width=6.8cm]{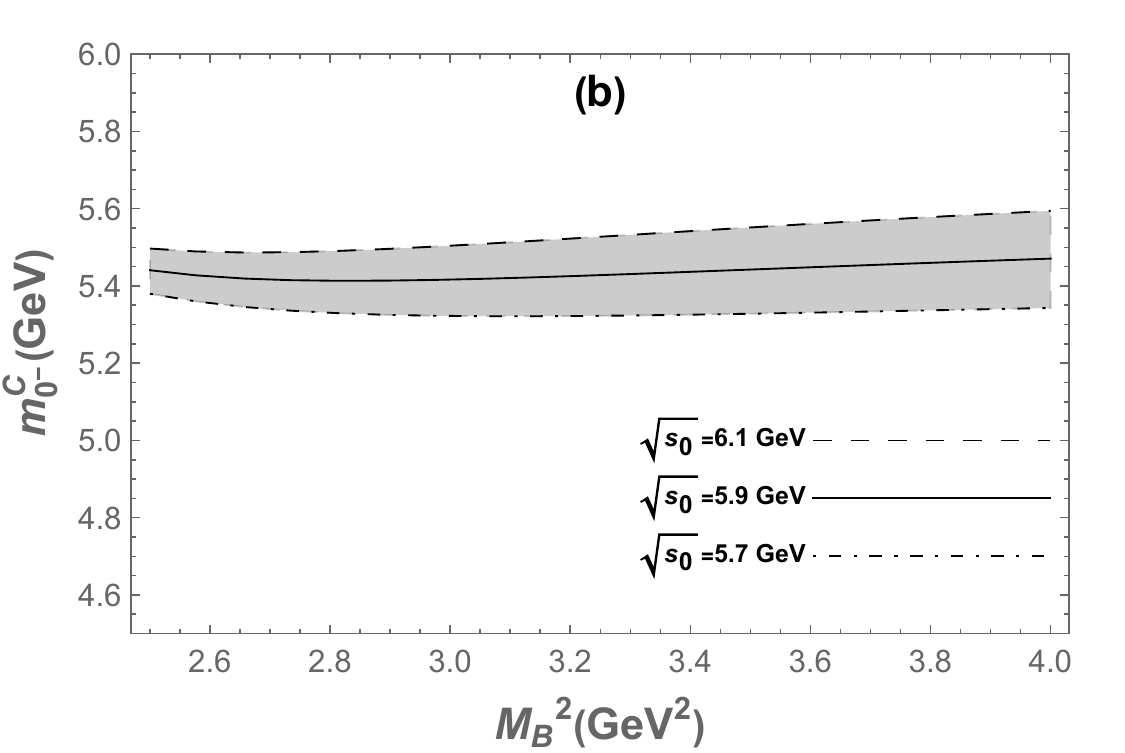}
\caption{Similar captions as in Fig.~\ref{fig0+B}, but for the current in Eq.~(\ref{current-0-C}).} \label{fig0-C}
\end{figure}

\begin{figure}
\includegraphics[width=6.8cm]{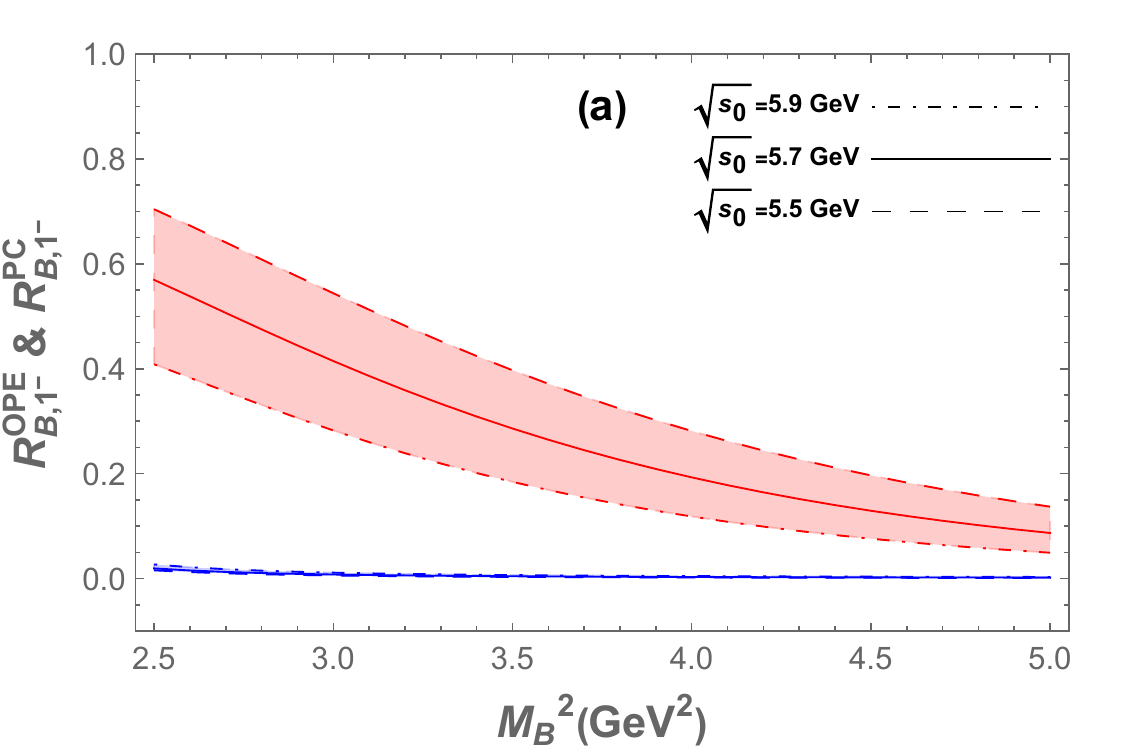}
\includegraphics[width=6.8cm]{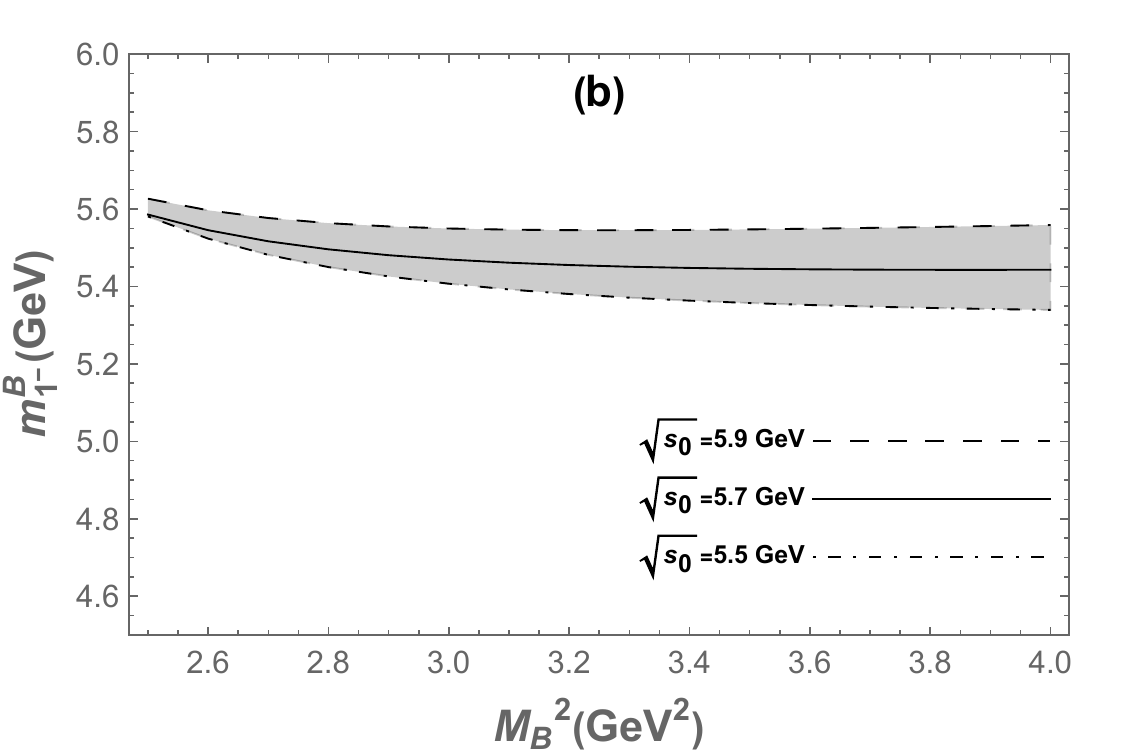}
\caption{Similar captions as in Fig.~\ref{fig0+B}, but for the current in Eq.~(\ref{current-1-B})} \label{fig1-B}
\end{figure}

\begin{figure}
\includegraphics[width=6.8cm]{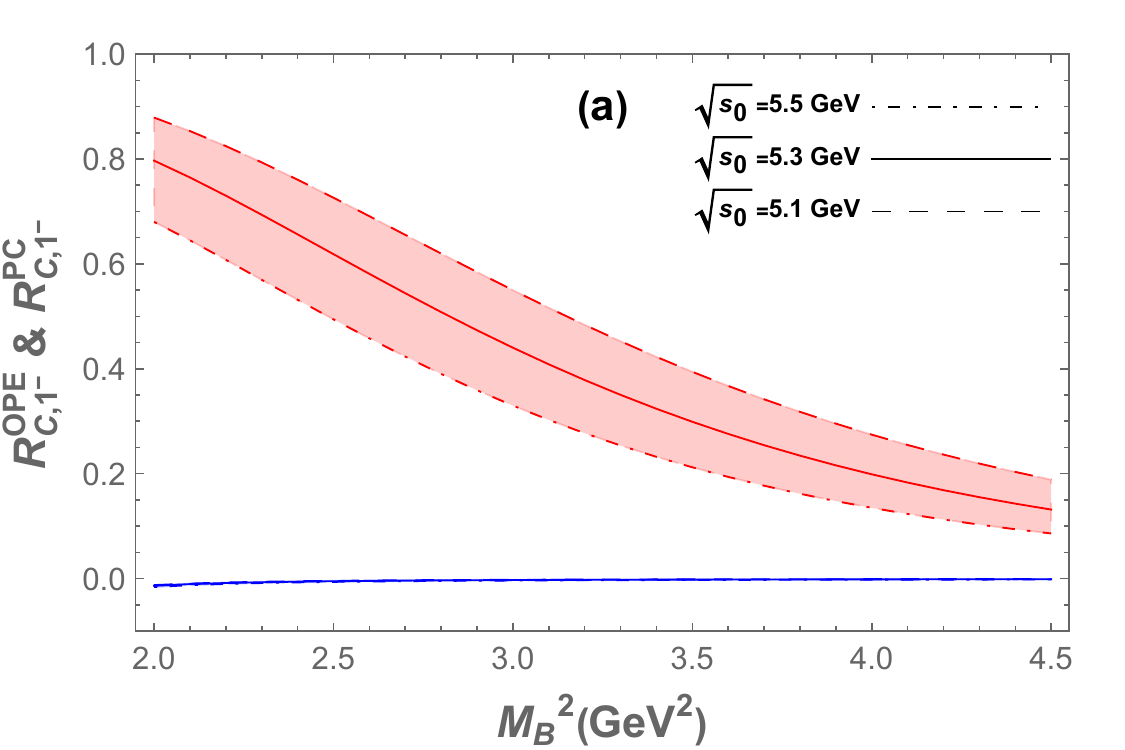}
\includegraphics[width=6.8cm]{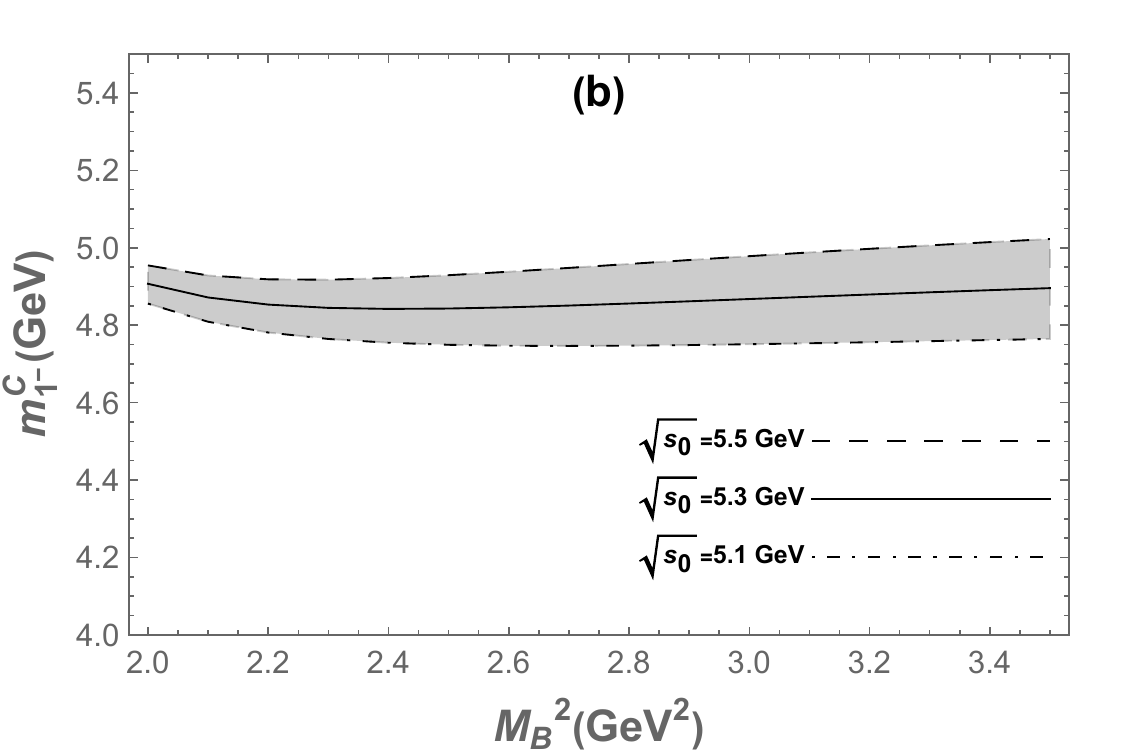}
\caption{Similar captions as in Fig.~\ref{fig0+B}, but for the current in Eq.~(\ref{current-1-C}).} \label{fig1-C}
\end{figure}

\begin{figure}
\includegraphics[width=6.8cm]{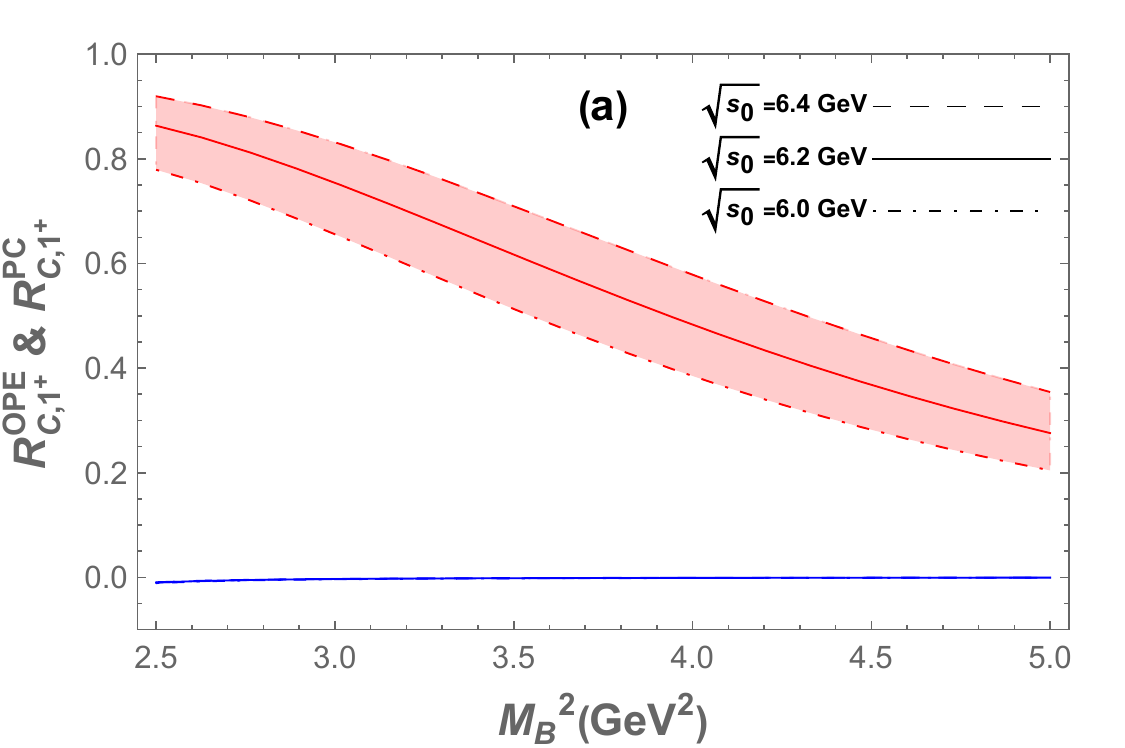}
\includegraphics[width=6.8cm]{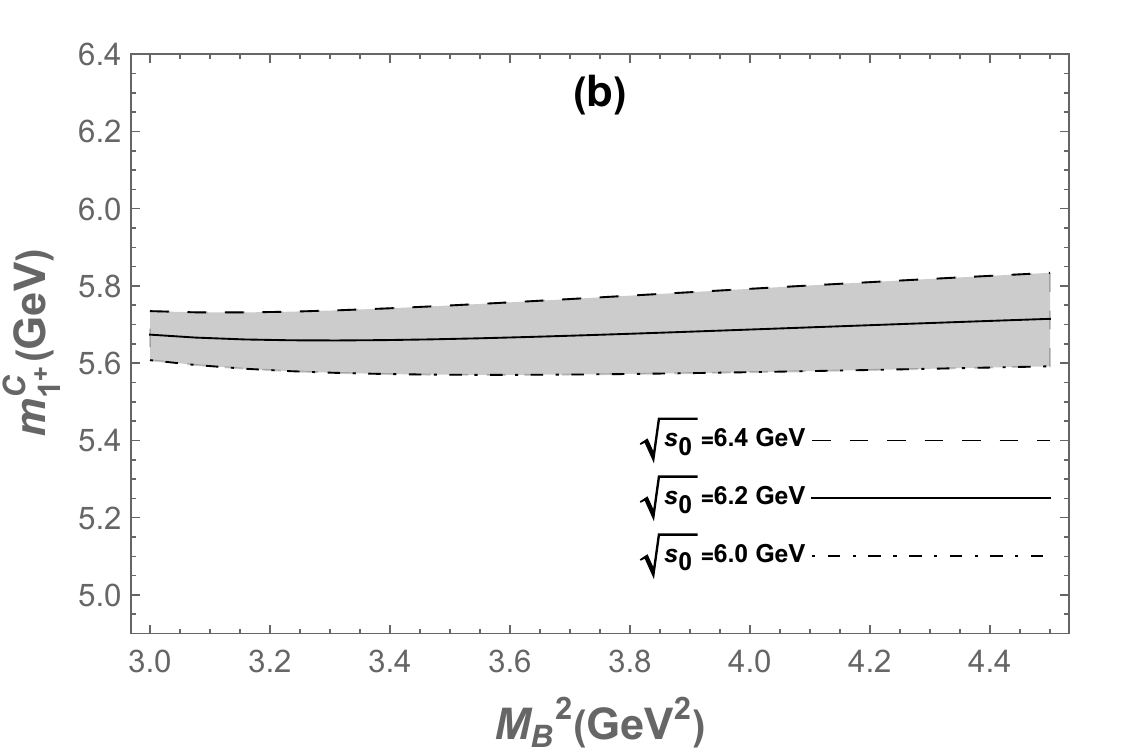}
\caption{Similar captions as in Fig.~\ref{fig0+B}, but for the current in Eq.~(\ref{current-1+C}).} \label{fig1+C}
\end{figure}

\begin{figure}
\includegraphics[width=6.8cm]{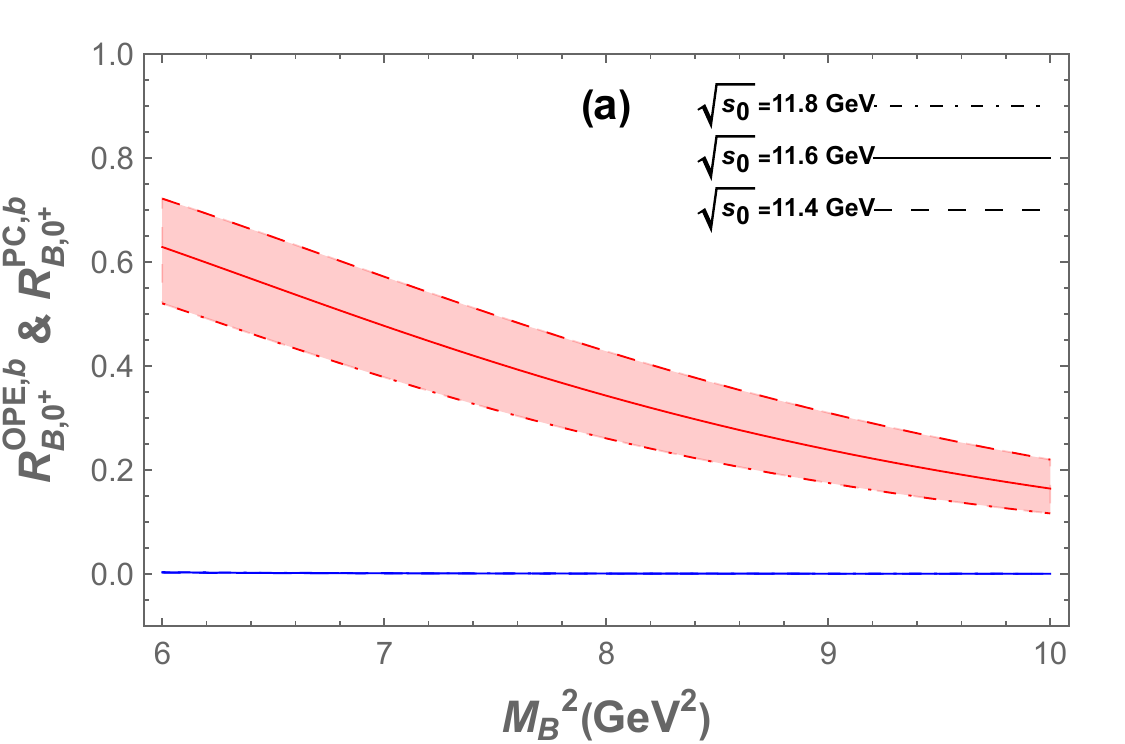}
\includegraphics[width=6.8cm]{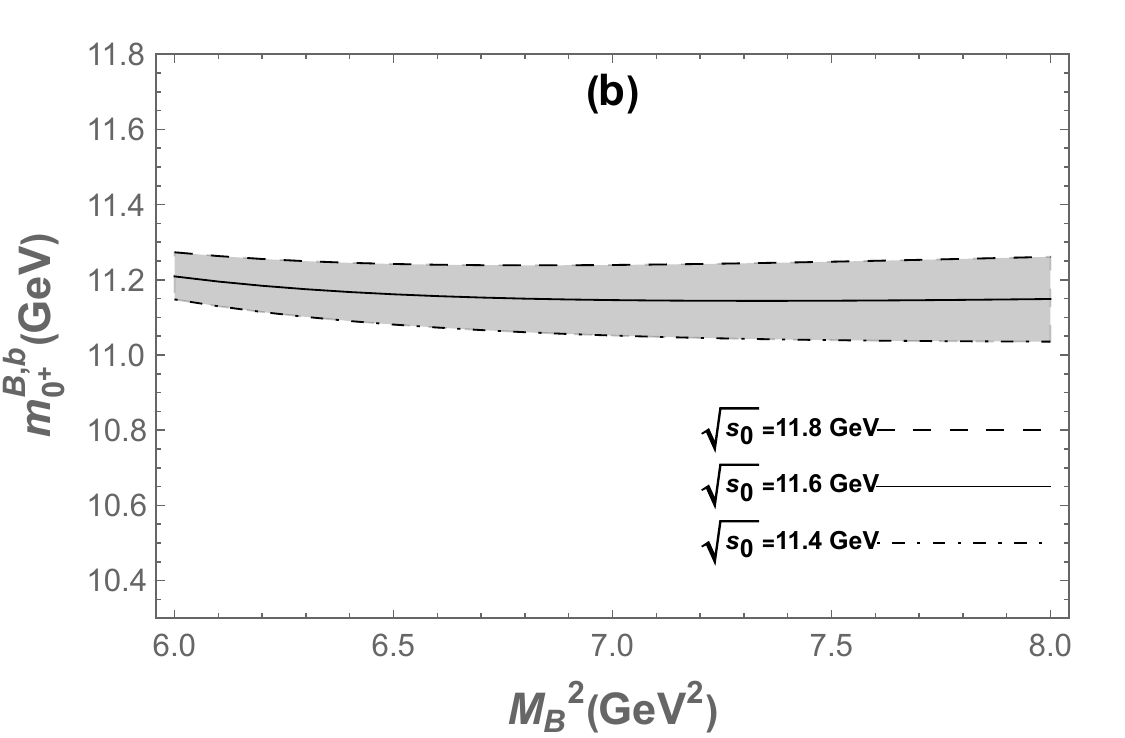}
\caption{Similar captions as in Fig.~\ref{fig0+B}, but for the $b$-sector and for the current in Eq.~(\ref{current-0+B}).} \label{fig0+Bb}
\end{figure}

\begin{figure}
\includegraphics[width=6.8cm]{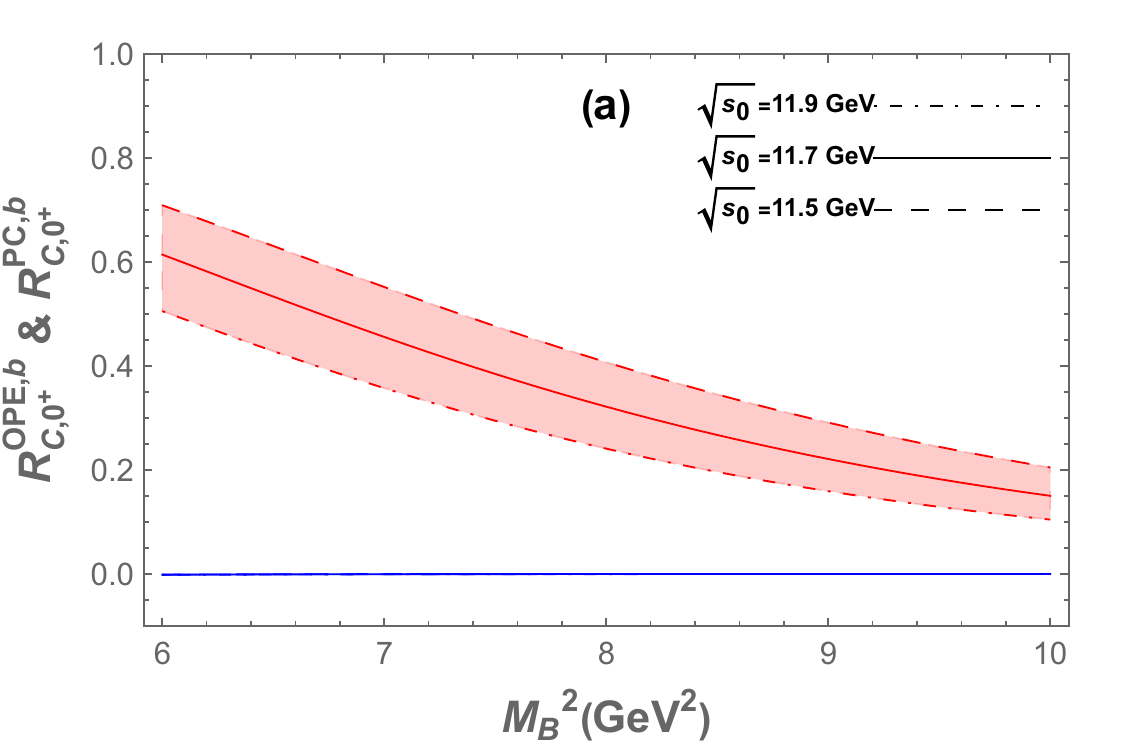}
\includegraphics[width=6.8cm]{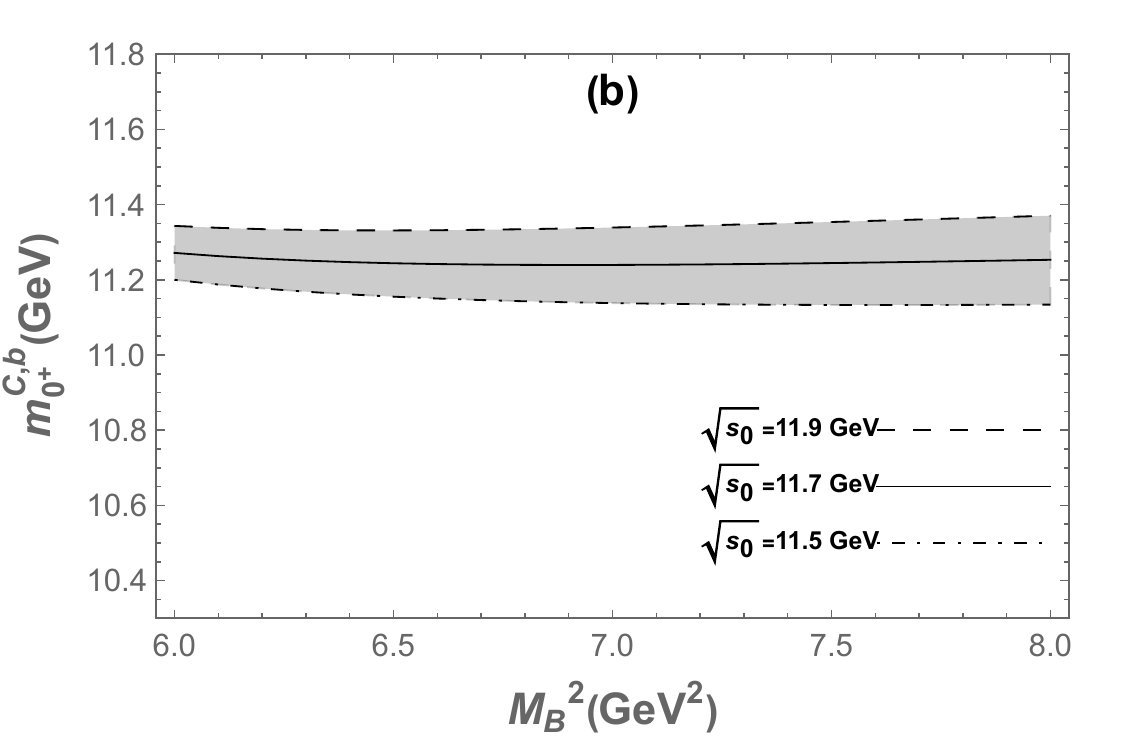}
\caption{Similar captions as in Fig.~\ref{fig0+B}, but for the $b$-sector and for the current in Eq.~(\ref{current-0+C}).} \label{fig0+Cb}
\end{figure}

\begin{figure}
\includegraphics[width=6.8cm]{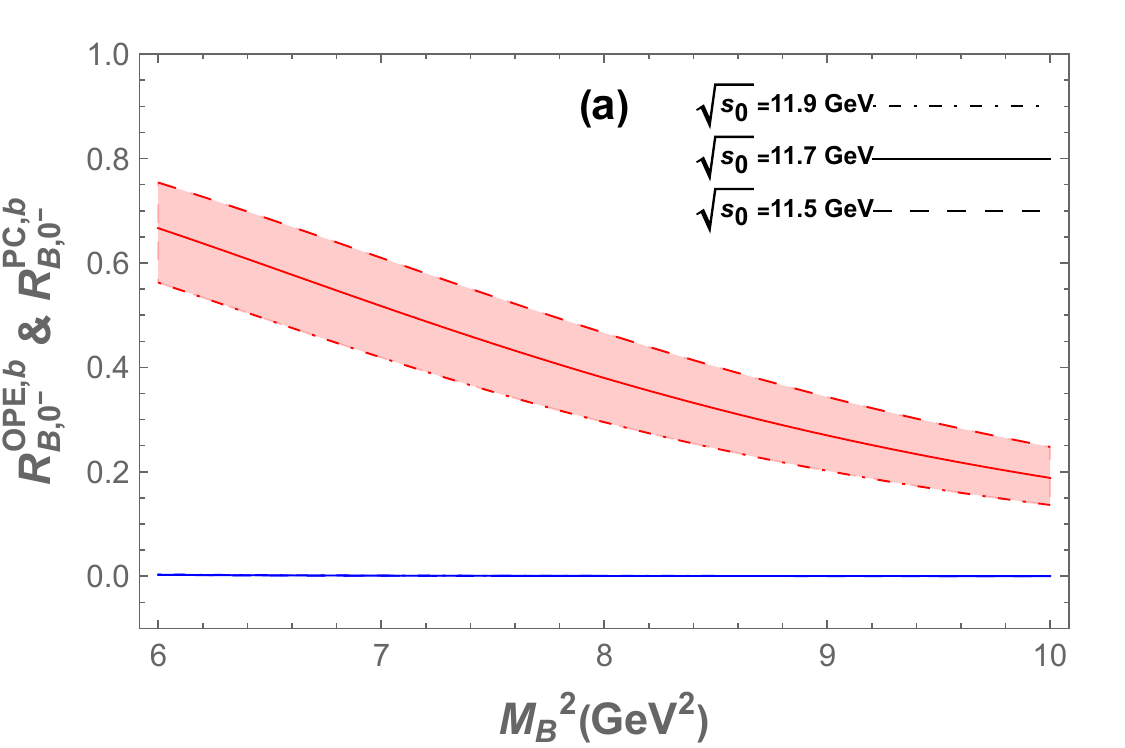}
\includegraphics[width=6.8cm]{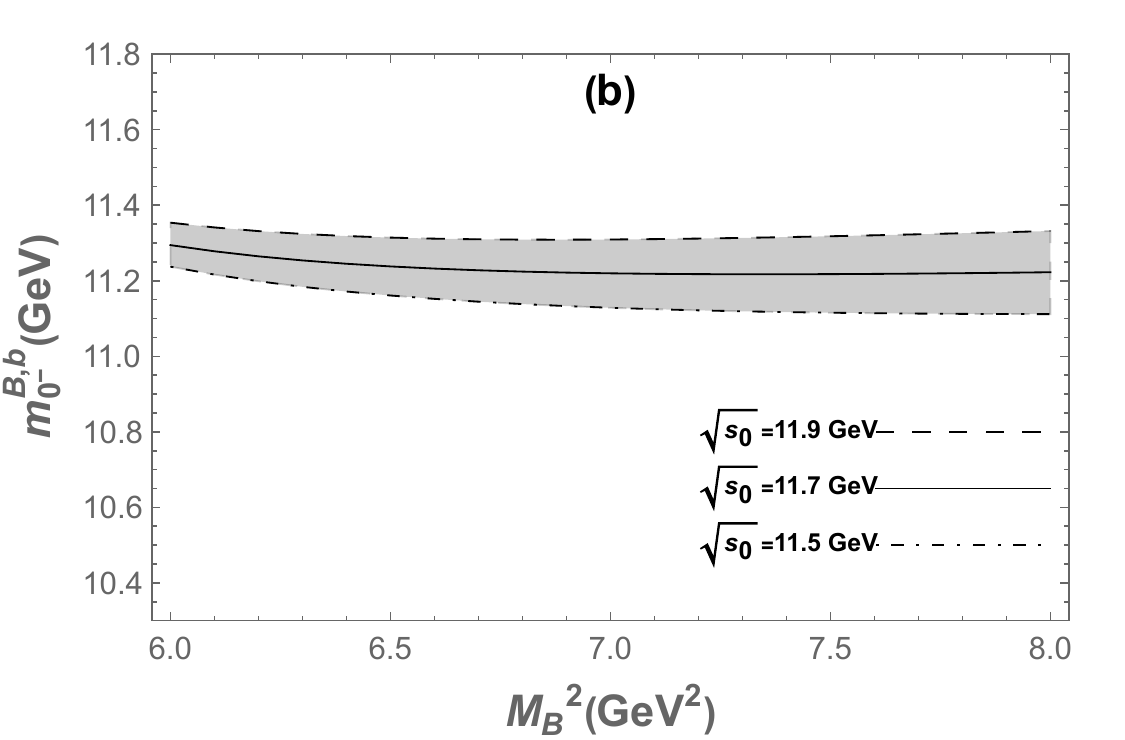}
\caption{Similar captions as in Fig.~\ref{fig0+B}, but for the $b$-sector and for the current in Eq.~(\ref{current-0-B}).} \label{fig0-Bb}
\end{figure}

\begin{figure}
\includegraphics[width=6.8cm]{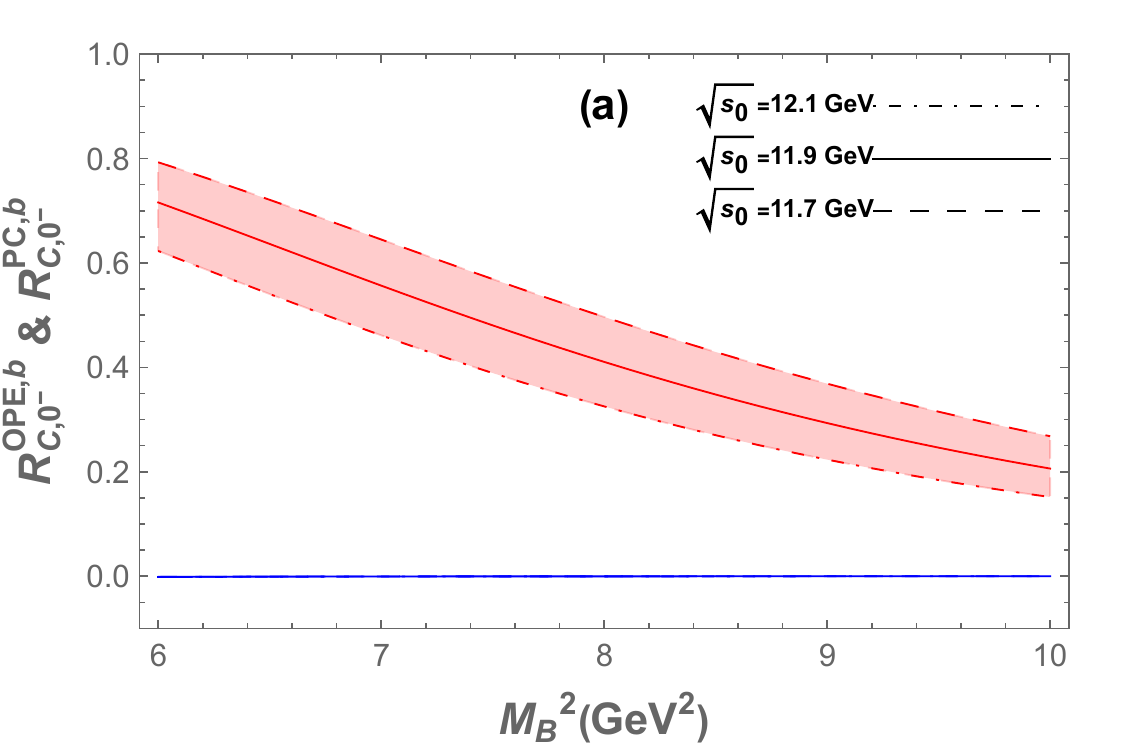}
\includegraphics[width=6.8cm]{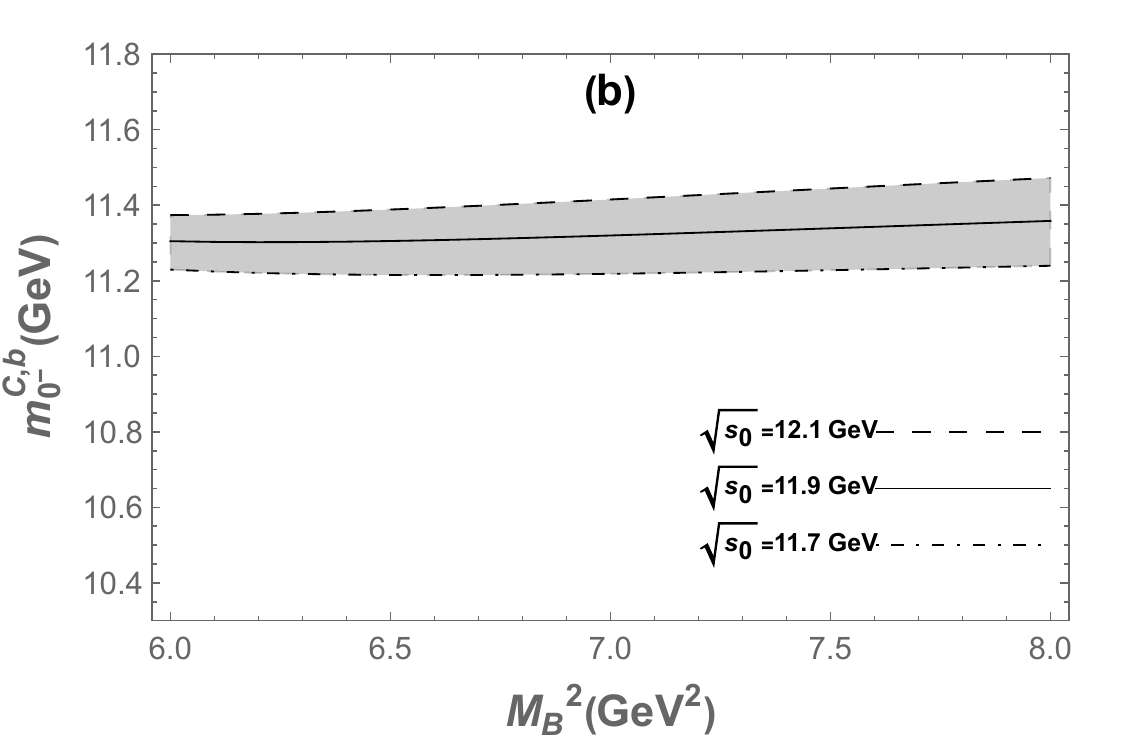}
\caption{ Similar captions as in Fig.~\ref{fig0+B}, but for the $b$-sector and for the current in Eq.~(\ref{current-0-C}).} \label{fig0-Bb}
\end{figure}

\begin{figure}
\includegraphics[width=6.8cm]{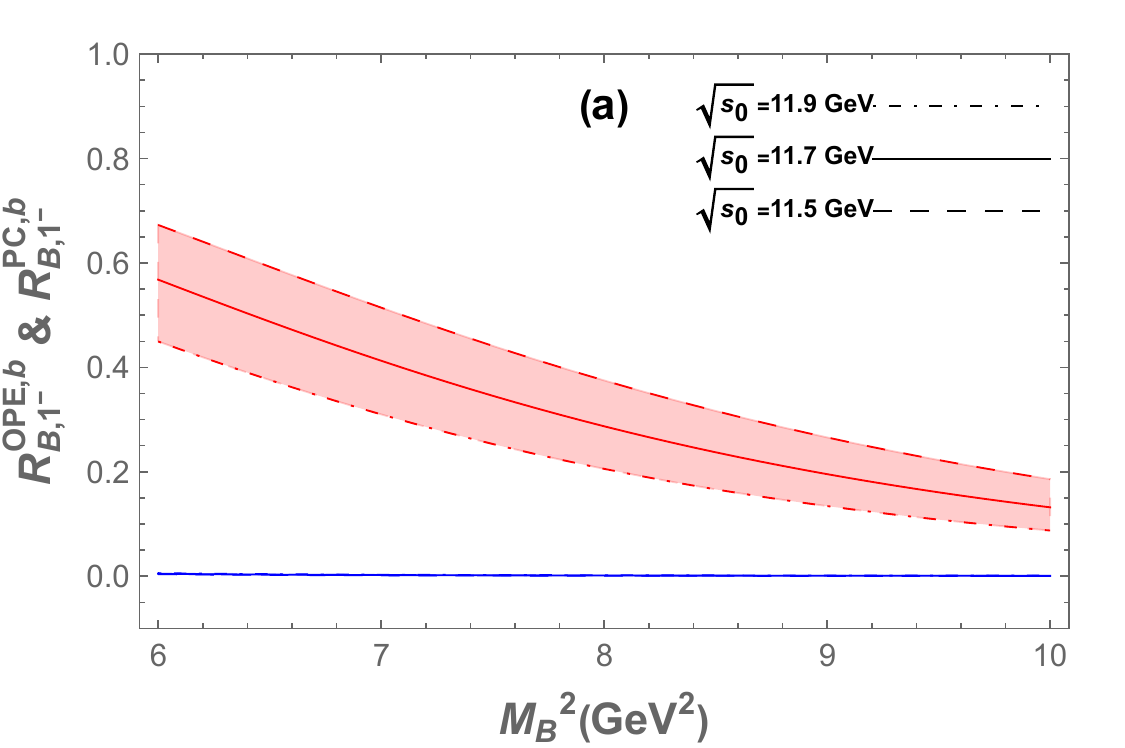}
\includegraphics[width=6.8cm]{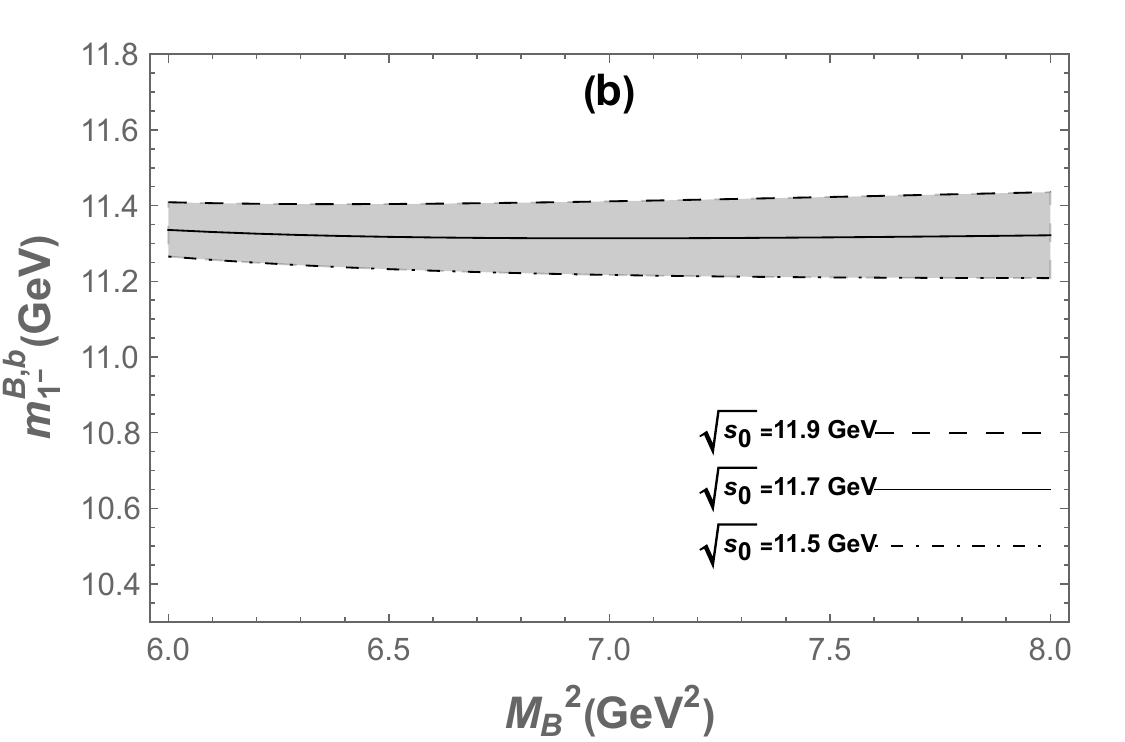}
\caption{Similar captions as in Fig.~\ref{fig0+B}, but for the $b$-sector and for the current in Eq.~(\ref{current-1-B}).} \label{fig1-B}
\end{figure}

\begin{figure}
\includegraphics[width=6.8cm]{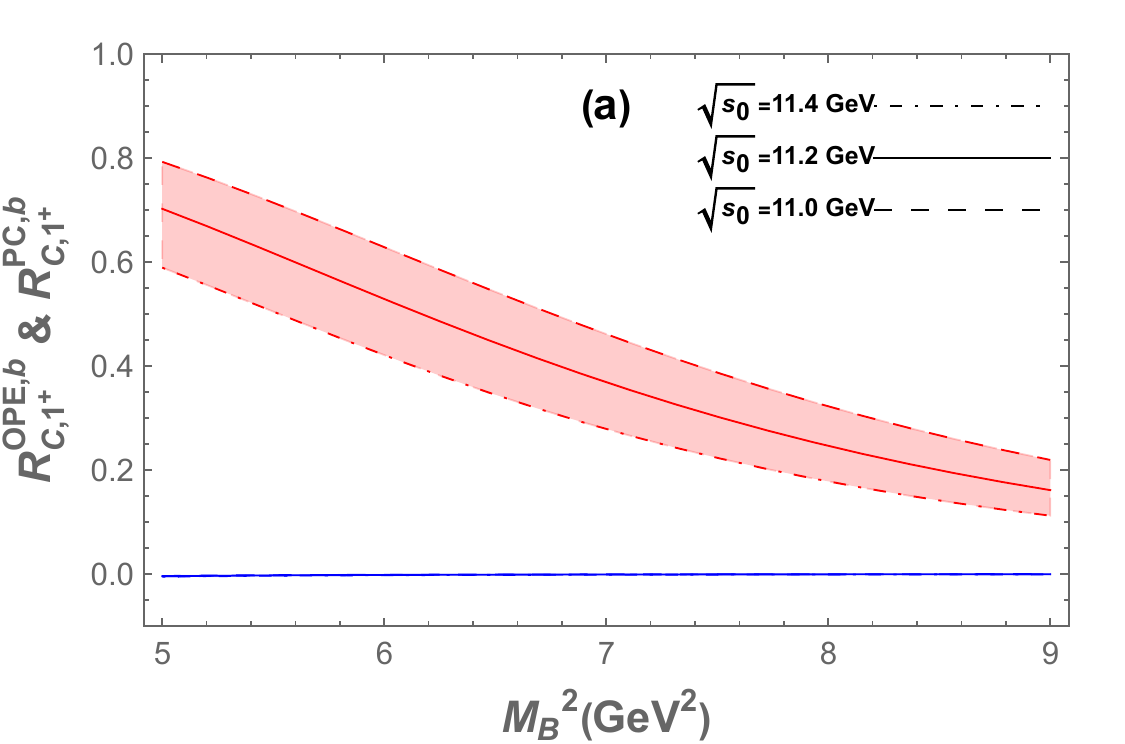}
\includegraphics[width=6.8cm]{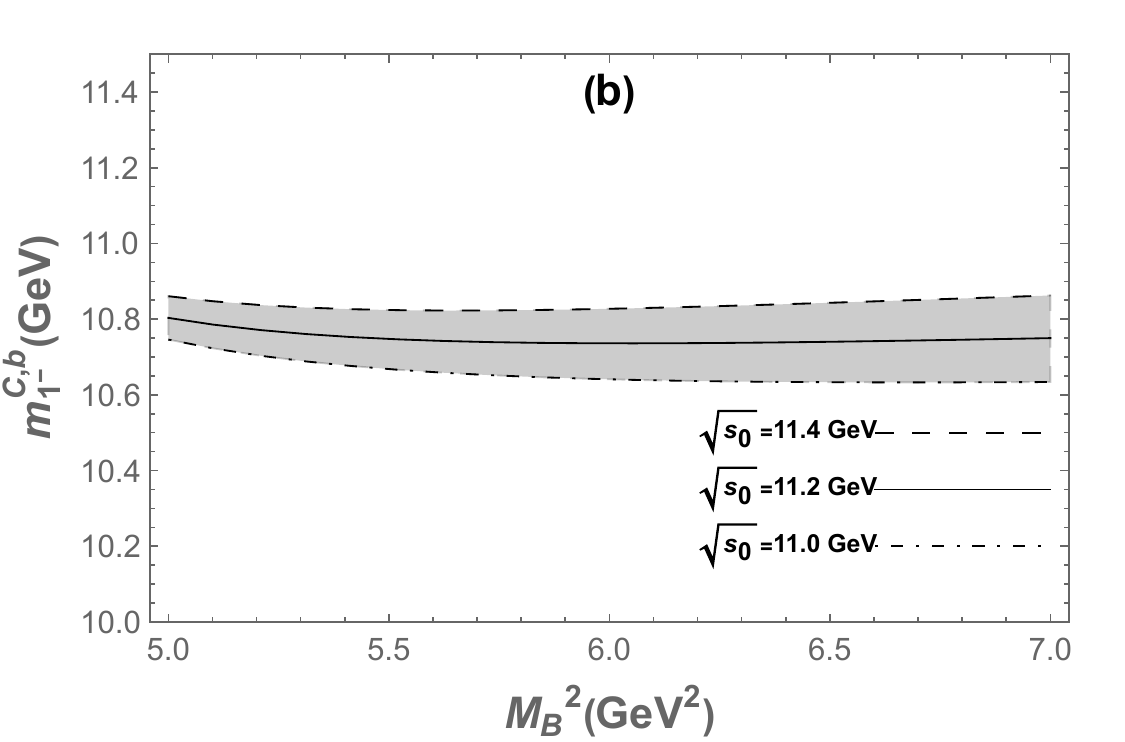}
\caption{ Similar captions as in Fig.~\ref{fig0+B}, but for the $b$-sector and for the current in Eq.~(\ref{current-1-C}).} \label{fig1-Cb}
\end{figure}

\begin{figure}
\includegraphics[width=6.8cm]{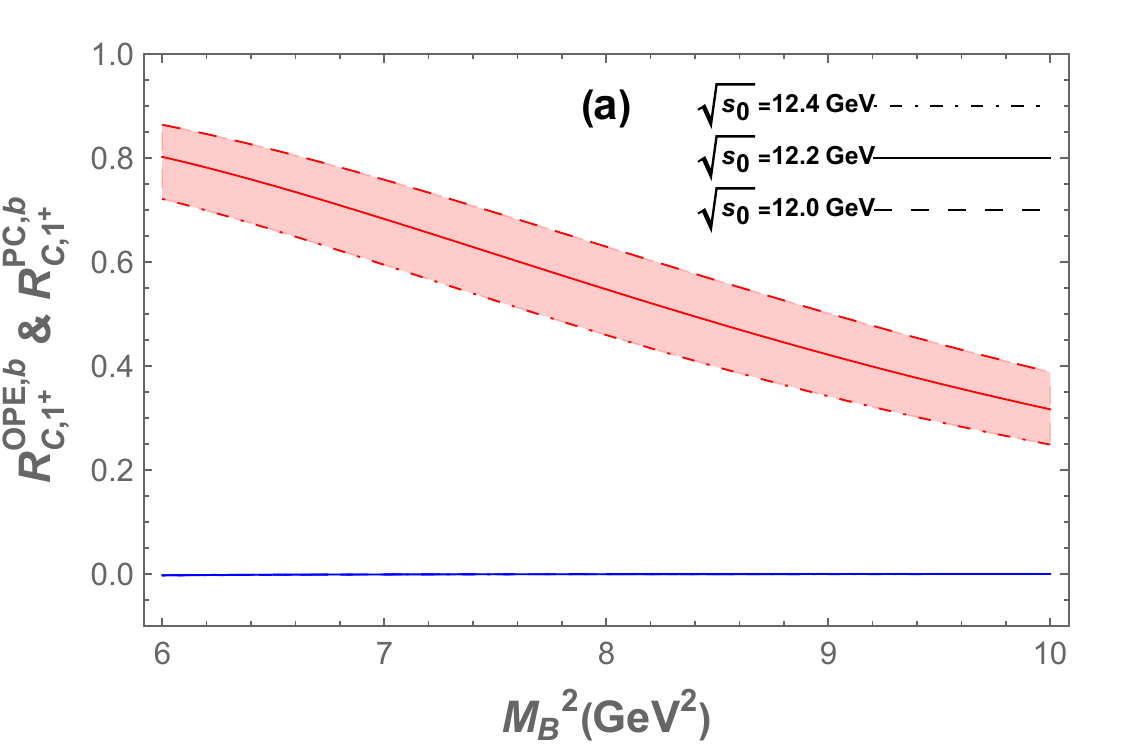}
\includegraphics[width=6.8cm]{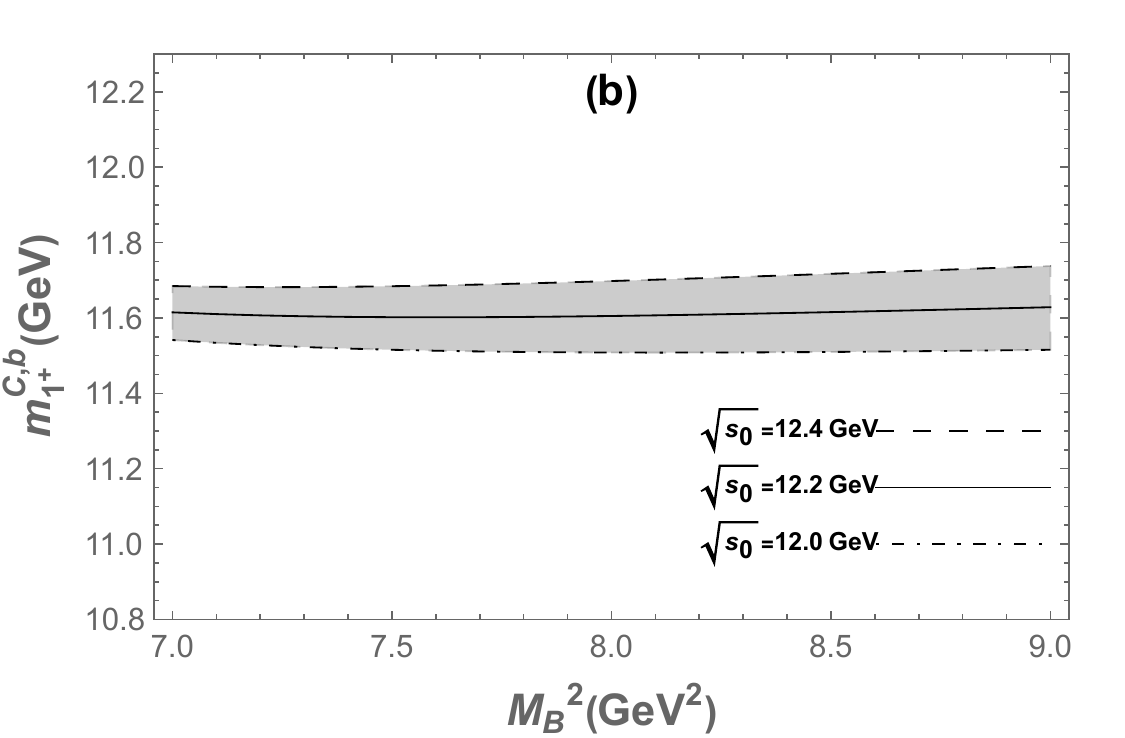}
\caption{Similar captions as in Fig.~\ref{fig0+B}, but for the $b$-sector and for the current in Eq.~(\ref{current-1+C}).} \label{fig1+Cb}
\end{figure}

\end{widetext}

\end{document}